\documentclass[11pt,a4paper]{article}

\usepackage{epsfig}
\usepackage{amsmath}
\usepackage{amssymb}
\usepackage{verbatim}
\usepackage{bm}
\usepackage{dsfont}
\usepackage[footnotesize]{caption}
\usepackage{subfigure}
\usepackage{cancel}
\usepackage{cite}
\usepackage{textcomp}
\usepackage{calc}
\usepackage{geometry}
\usepackage{bbm}
\usepackage{color}

\usepackage{hyphenat}
\begin{document}


\pagenumbering{roman}

\noindent April 2016
\hfill IPMU 16-0051
 
\vskip 2.5cm

\begin{center}
{\LARGE\bf Polonyi Inflation}

\bigskip

{\bf Dynamical Supersymmetry Breaking
and Late-Time\\ \boldmath{$R$} Symmetry Breaking
as the Origin of Cosmic Inflation}

\vskip 2cm

\renewcommand*{\thefootnote}{\fnsymbol{footnote}}

{\large
Kai~Schmitz\,$^{a,\,b,\,\hspace{-0.25mm}}$%
\footnote{Corresponding author. E-mail: kai.schmitz@mpi-hd.mpg.de}
and Tsutomu~T.~Yanagida\,$^{b}$}\\[3mm]
{\it{
$^{a}$ Max Planck Institute for Nuclear Physics (MPIK), 69117 Heidelberg, Germany\\
$^{b}$ Kavli IPMU (WPI), UTIAS, The University of Tokyo, Kashiwa, Chiba 277-8583, Japan}}

\end{center}

\vskip 1cm

\renewcommand*{\thefootnote}{\arabic{footnote}}
\setcounter{footnote}{0}


\begin{abstract}


Spontaneously broken supersymmetry (SUSY) and a vanishingly small
cosmological constant imply that $R$ symmetry must be spontaneously broken at low energies.
Based on this observation, we suppose that, in the sector responsible for low-energy
$R$ symmetry breaking, a discrete $R$ symmetry remains preserved at
high energies and only becomes dynamically broken at relatively late times
in the cosmological evolution, i.e., after the dynamical breaking of SUSY.
Prior to $R$ symmetry breaking, the Universe is then bound to be in a
quasi-de Sitter phase---which offers a dynamical explanation for the
occurrence of cosmic inflation. 
This scenario yields a new perspective on the interplay
between SUSY breaking and inflation, which neatly fits into
the paradigm of high-scale SUSY:
inflation is driven by the SUSY-breaking vacuum energy density, while
the chiral field responsible for SUSY breaking, the Polonyi field,
serves as the inflaton.
Because $R$ symmetry is broken only after inflation,
slow-roll inflation is not spoiled by otherwise dangerous
gravitational corrections in supergravity.
We illustrate our idea by means of a concrete example,
in which both SUSY and $R$ symmetry are broken by strong gauge dynamics and
in which late-time $R$ symmetry breaking is triggered by a
small inflaton field value.
In this model, the scales of inflation and SUSY breaking are
unified;
the inflationary predictions are similar to those of F-term
hybrid inflation in supergravity;
reheating proceeds via gravitino decay at temperatures consistent
with thermal leptogenesis;
and the sparticle mass spectrum follows from pure gravity mediation.
Dark matter consists of thermally produced winos with a mass in the TeV range.


\end{abstract}


\newpage

\tableofcontents

\newpage

\pagenumbering{arabic}


\section{Introduction: Inflation and supersymmetry breaking unified}
\label{sec:introduction}


\subsection{High-scale supersymmetry breaking as the origin of inflation}
\label{subsec:origin}


The paradigm of cosmic inflation~\cite{Guth:1980zm,Linde:1981mu}
is one of the main pillars of modern cosmology.
Not only does inflation account for the vast size of the observable Universe and
its high degree of homogeneity and isotropy on cosmological scales; it also seeds
the post-inflationary formation of structure on galactic scales.
In this sense, inflation is a key aspect of our cosmic past
and part of the reason why our Universe is capable of harboring life.
From the perspective of particle physics, the origin of inflation is,
however, rather unclear.
After decades of model building, there exists a plethora of inflation
models in the literature~\cite{Lyth:1998xn}.
But a consensus about how to embed inflation into
particle physics is out of sight.
In this situation, it seems appropriate to take a step back and ask
ourselves what avenues have been left unexplored so far. 
In particular, we should question our dearly cherished prejudices
and re-examine whether inflation might not be connected to
other high-energy phenomena which, up to now, have been taken to be
mostly unrelated to inflation.
As we are going to demonstrate in this paper, an important example in this
respect might be the interplay between inflation and the
spontaneous breaking of supersymmetry (SUSY).%
\footnote{For related earlier work on the relation between inflation
and supersymmetry breaking, see,
e.g.,~\cite{Randall:1994fr,Riotto:1997iv,Izawa:1997jc,Buchmuller:2000zm}.
\smallskip}


In recent years, the picture of supersymmetry as a solution to the hierarchy problem
has become increasingly challenged by the experimental data.
The null results of SUSY searches at the Large Hadron Collider (LHC)~\cite{Chatrchyan:2014lfa}
and the rather large standard model (SM) Higgs boson mass of
a $125\,\textrm{GeV}$~\cite{Aad:2012tfa} indicate that supersymmetry,
if it exists in nature, must be broken at a high scale~\cite{Okada:1990vk}.
Based on this observation, one could feel tempted to give up on supersymmetry
as an extension of the standard model altogether.
But this would not do justice to supersymmetry's other virtues.
Provided that supersymmetry is broken at a high scale~\cite{Giudice:1998xp,Wells:2003tf},
such as in the minimal framework of pure
gravity mediation (PGM)~\cite{Ibe:2006de,Ibe:2012hu},%
\footnote{For closely related schemes for the mediation of supersymmetry breaking
to the visible sector, see~\cite{Hall:2011jd,Hall:2012zp}.}
it may no longer be responsible
for stabilizing the electroweak scale.
But in this case, supersymmetry is still capable of providing a viable candidate
for dark matter~\cite{Ibe:2012hu,Hall:2012zp,Ibe:2013jya},
ensuring the unification of the SM gauge couplings~\cite{Evans:2015bxa}
and setting the stage for a UV completion of the standard model
in the context of string theory.
In addition, high-scale supersymmetry does not suffer from
a number of phenomenological problems that low-scale realizations
of supersymmetry breaking are plagued with.
A high SUSY breaking scale does away with the cosmological gravitino
problem~\cite{Pagels:1981ke} and reduces the tension with constraints
on flavor-changing neutral currents and $CP$ violation~\cite{Gabbiani:1996hi}.
Moreover, in PGM, the SUSY-breaking (or ``Polonyi'') field is required
to be a non-singlet~\cite{Babu:2015xba}, which solves the
cosmological Polonyi problem~\cite{Coughlan:1983ci}.


In this paper, we will now concentrate our attention to yet another
intriguing feature of supersymmetry which comes into reach, once 
we let go of the notion that supersymmetry's main purpose is
to solve the hierarchy problem in the standard model.
The spontaneous breaking of supersymmetry at a scale
$\Lambda_{\rm SUSY}$ results in a nonzero
contribution to the total vacuum energy density, $\Lambda_{\rm SUSY}^4$.
If we allow $\Lambda_{\rm SUSY}$ to take values as large as, say,
the unification scale, $\Lambda_{\rm GUT} \sim 10^{16}\,\textrm{GeV}$,
this SUSY-breaking vacuum energy density might, in fact, be the origin
of the inflationary phase in the early universe!
Such a connection between inflation and supersymmetry breaking not only
appears economical, but also very natural.


First of all, supersymmetry tends to render inflation 
technically more natural, independent of the scale at which it is broken.
Thanks to the SUSY nonrenormalization theorem~\cite{Grisaru:1979wc}, the 
superpotential $W$ in supersymmetric models of inflation does not receive
any radiative corrections in perturbation theory.
This represents an important advantage in preserving the
required flatness of the inflaton potential.
Besides, all remaining radiative corrections
(which can be collected in an effective K\"ahler potential
$K$ to leading order~\cite{Gaillard:1993es}) scale with
the soft SUSY-breaking mass scale~\cite{Ellis:1982ed}
and are, thus, under theoretical control.
Supersymmetry, therefore, has the ability to stabilize the inflaton
potential against radiative corrections;
and it is, thus, conceivable that supersymmetry's actual importance
may lie in the fact that it is capable of taming the hierarchy among
different mass scales in the inflaton sector rather than in the standard model.
Second of all, the spontaneous breaking of global supersymmetry via nonvanishing
F-terms, i.e., via the O'Raifeartaigh mechanism~\cite{O'Raifeartaigh:1975pr},
always results in a pseudoflat direction in the scalar potential~\cite{Ray:2006wk}.
Together with the constant vacuum energy density $\Lambda_{\rm SUSY}^4$, such a flat
potential for a scalar field is exactly one of the crucial requirements
for the successful realization of an inflationary stage in the early universe.
In principle, the necessary ingredients for inflation are, therefore,
already intrinsic features of every O'Raifeartaigh model.
Inflation may be driven by the SUSY-breaking vacuum energy density
$\Lambda_{\rm SUSY}^4$ and the inflaton field may be identified with
the pseudoflat direction in the scalar potential.


The main obstacle in implementing this idea in realistic models is gravity.
Here, the crucial point is that the vanishingly small value of the cosmological constant (CC)
tells us that we live in a near-Minkowski vacuum with an almost
zero total vacuum energy density, $\left<V\right>\simeq 0$.
Note that, as pointed out by Weinberg, this not a mere observation, but
a necessary condition for a sufficient amount of structure formation
in our Universe, so that it can support life~\cite{Weinberg:1987dv}.
In the context of supergravity (SUGRA)~\cite{Nilles:1983ge}, the fact that
$\left<V\right>\simeq 0$ means that the SUSY-breaking vacuum energy density $\Lambda_{\rm SUSY}^4$
must be balanced, with very high precision, by a nonvanishing
vacuum expectation value (VEV) of the superpotential, $\left<W\right>$,
\begin{align}
\left<V\right> = \left<\left|F\right|\right>^2 - 3\, \exp\left[\frac{\left<K\right>}{M_{\rm Pl}^2}\right]
\frac{\left<\left|W\right|\right>^2}{M_{\rm Pl}^2} \simeq 0  \,, \quad
\left<\left|F\right|\right> = \Lambda_{\rm SUSY}^2 \,, 
\label{eq:Min}
\end{align}
where $M_{\rm Pl} = \left(8\pi\, G\right)^{-1/2} \simeq 2.44 \times 10^{18}\,\textrm{GeV}$
denotes the reduced Planck mass.
If the SUSY breaking scale $\Lambda_{\rm SUSY}$ is indeed of
$\mathcal{O}\left(10^{16}\right)\,\textrm{GeV}$, the requirement of a zero CC
results in a huge VEV of the superpotential, which, in turn, generates dangerously
large SUGRA corrections to the scalar potential.
These corrections then easily spoil the flatness of the potential and render
inflation impossible~\cite{Ovrut:1983my}.


The most attractive and, in fact, only way out of this problem is $R$ symmetry.
Under $R$ symmetry, the superpotential carries charge $2$, so that $\left<W\right> = 0$,
as long as $R$ symmetry is preserved~\cite{Dine:2009swa}.
In other words, by imposing $R$ symmetry, we promote $\left<W\right>$
to the order parameter of spontaneous $R$ symmetry breaking, in
analogy to $\left<F\right>$ which acts as the order parameter
of spontaneous supersymmetry breaking.
In the true vacuum, $R$ symmetry must be broken, so that $\left<W\right> \neq 0$,
in order to satisfy the condition in Eq.~\eqref{eq:Min}.
But this does not necessarily mean that $\left<W\right>$ must be
nonzero during the entire cosmological evolution.
It is conceivable that, at early times,
$R$ symmetry is, in fact, a good symmetry, so that $\left<W\right> \approx 0$,
thereby ``switching off'' the most dangerous SUGRA corrections
to the Polonyi potential.
The main intention of our paper now is to present a minimal dynamical
model, in which this is indeed the case, so that inflation driven by 
the SUSY-breaking vacuum energy density, i.e., \textit{Polonyi inflation},
becomes a viable option.


\subsection{Two avenues towards a vanishing cosmological constant}


Let us now outline our general philosophy in, to some extent, simplified terms.
We suppose that, from the perspective of the low-energy effective
theory, the breaking of supersymmetry and $R$ symmetry
are two distinct dynamical processes, taking place in two different
hidden sectors at two different times, $t_{\rm SUSY}$ and $t_R$.
In particular, we assume that the parameters $t_{\rm SUSY}$ and $t_R$ may be sampled in the
UV theory, so that in some patches of the early universe (or in the landscape
of string vacua~\cite{Susskind:2003kw} for that purpose)
$t_{\rm SUSY} < t_R$ and in other patches $t_{\rm SUSY} > t_R$.
Note that this differentiation equally includes the case of a maximally symmetric
initial state, $\min\left\{t_{\rm SUSY}, t_R\right\} > t_{\rm ini}$, as well
as the possibility that either supersymmetry or $R$ symmetry is already broken from
the very beginning, $\min\left\{t_{\rm SUSY}, t_R\right\} \equiv t_{\rm ini}$.
Our crucial observation is that regions in space where $R$ symmetry is broken
before supersymmetry correspond to bubbles of anti-de Sitter (AdS) space with an
AdS radius $R_{\rm AdS}$ equal to the inverse of the gravitino mass $m_{3/2}$,
while regions in space where supersymmetry is broken before $R$ symmetry correspond
to bubbles of de Sitter (dS) space with a dS Radius $R_{\rm dS}$ equal to the
inverse of the Hubble parameter $H$,
\begin{align}
t_{\rm SUSY} > t_R \quad\Rightarrow\quad & \textrm{AdS bubbles with radius }
R_{\rm AdS} = m_{3/2}^{-1} \,, &
m_{3/2} & = \exp\left[\left<K\right>/2/M_{\rm Pl}^2\right]
\left<\left|W\right|\right>/M_{\rm Pl}^2 \,, \label{eq:bubbles} \\ \nonumber
t_{\rm SUSY} < t_R \quad\Rightarrow\quad & \textrm{\phantom{A}dS bubbles with radius }
R_{\rm dS} = H^{-1}  \,, &
H & = \left<\left|F\right|\right>/\sqrt{3}/M_{\rm Pl} \,.
\end{align}
As long as supersymmetry is unbroken, the AdS bubbles represent open
Friedmann-Lema\^itre-Robertson-Walker (FLRW) universes with a negative
CC and an oscillatory scale factor.
The dS bubbles, on the other hand, turn, as long as $R$ symmetry is unbroken,
into (asymptotically) flat FLRW universes with a positive CC and an
exponentially growing scale factor~\cite{Carroll:1997ar}. 
This is to say that the dS bubbles experience inflation, while the
AdS bubbles remain limited in their spatial extent.
Furthermore, if we suppose that there is no other source of inflation present
in the theory, this means that the AdS bubbles will never develop into habitable
universes.
Intelligent observers and humans can, therefore, only life in regions
where, initially, $R$ symmetry remains unbroken up to a certain high energy scale.
This applies, in particular, to our own observable Universe.
Under the above assumptions, the inflationary period in the early
history of our Universe must have been a consequence of spontaneous
supersymmetry breaking and $R$ symmetry must
have been broken only at late times, i.e., after a sufficient
amount of inflation.
Put differently, we can say that supersymmetry breaking and
Weinberg's argument regarding the size of the CC
\textit{postdict} a period of inflation and late-time $R$
symmetry breaking in our cosmic past.


This conclusion also sheds new light on the role of the CC itself.
The fine-tuning of the CC is now a dynamical process that takes
place only after inflation.
In order to obtain zero CC, we have to require that the gravitino
mass generated during $R$ symmetry breaking matches the inflationary
Hubble rate,%
\footnote{Note that the relation between $m_{3/2}$ and $H_{\rm inf}$ in our
scenario is conceptually quite different from other inflation models,
such as, e.g., the one in~\cite{Randall:1994fr},
where $H_{\rm inf}\simeq m_{3/2}$ on purely phenomenological grounds.
In our case, $H_{\rm inf}\simeq m_{3/2}$ is a property of the low-energy
effective Lagrangian of our Universe that ensures that we live in a vacuum
with an almost zero CC.}
\begin{align}
\left<V\right> \simeq 0 \quad\Rightarrow\quad
m_{3/2} \simeq H_{\rm inf} \,.
\label{eq:m32Hinf}
\end{align}
If $R$ symmetry breaking results in a gravitino mass smaller
than $H_{\rm inf}$, inflation never ends; if it ``overshoots'' and
the gravitino mass is eventually larger than $H_{\rm inf}$, our Universe
becomes AdS.
We, thus, recognize the requirement that inflation must terminate at one 
point or another as part of the reason why the CC in our Universe is fine-tuned.
A certain amount of fine-tuning during late-time
$R$ symmetry breaking is inevitable, as inflation would otherwise not exit
into a near-Minkowski vacuum.
This situation needs to be contrasted with standard scenarios of inflation,
such as chaotic~\cite{Linde:1983gd} or hybrid~\cite{Linde:1991km} inflation,
where the vacuum energy density driving inflation is neither related to
$R$ symmetry breaking nor to low-energy supersymmetry breaking.
These scenarios require an independent reason for the vanishing of the
inflationary vacuum energy density
(e.g., a tuning of the inflaton potential or some kind of waterfall transition),
whereas in our case this reason is already inherent to the fine-tuning
of the CC in the course of spontaneous $R$ symmetry breaking at the end of inflation.


\subsection{Ingredients for a realistic model of Polonyi inflation}
\label{subsec:ingredients}


The above reasoning is just a rough sketch.
To construct a realistic model of Polonyi inflation,
we need to be more specific.
This pertains, first of all, to the kind of $R$ symmetry that we have in mind.
Naively, our first attempt might be to protect $\left<W\right>$
by means of a global $U(1)_R$ symmetry.
On general grounds, quantum gravity is, however, expected to explicitly break
all global symmetries (see~\cite{Banks:2010zn} and references therein),
so that a global $U(1)_R$ does not appear to be a viable possibility.
Meanwhile, gauging a continuous $R$ symmetry is a subtle issue
that easily results in conflicts with anomaly constraints at low energies
(see~\cite{Antoniadis:2014hfa} for a recent discussion).
This leaves us with a \textit{discrete} gauged
(i.e., anomaly-free~\cite{Krauss:1988zc}) $R$ symmetry, $Z_N^R$,
as a unique choice to ensure the vanishing of the superpotential at early times.


Interestingly enough, such a discrete $R$ symmetry readily comes with a number
of other advantages in the context of SUSY phenomenology:
(i) A discrete $R$ symmetry prevents too rapid proton decay
via perilous dimension-5 operators~\cite{Sakai:1981pk};
(ii) it may give rise to an accidental approximate
global Peccei-Quinn symmetry and, thus, help in solving 
the strong $CP$ problem~\cite{Harigaya:2013vja,Harigaya:2015soa};
(iii) and it may account for the approximate global continuous
$R$ symmetry which is required to realize stable~\cite{Nelson:1993nf}
or meta-stable~\cite{Intriligator:2006dd} SUSY-breaking vacua in a large
class of models of dynamical supersymmetry breaking (DSB).
Moreover, if we restrict ourselves to the special case of
a $Z_4^R$ symmetry, $R$ symmetry can also help us in solving the
$\mu$ problem~\cite{Kim:1983dt} in the minimal supersymmetric
standard model (MSSM).
Any discrete $R$ symmetry suppresses the bilinear Higgs mass
term (i.e., the $\mu$ term) in the superpotential.
But only in the case of a $Z_4^R$ symmetry, we are, in addition, allowed
to include a Higgs bilinear term in the K\"ahler potential,
$K \supset H_u H_d$ (see \cite{Harigaya:2015soa} and references
therein for an extended discussion of this point). 
We are then able to generate the $\mu$ term in the course of
$R$ symmetry breaking~\cite{Inoue:1991rk}, which directly
relates the $\mu$ parameter to the gravitino mass.%
\footnote{This solution to the $\mu$ problem is not to be confused
with the Giudice-Masiero mechanism~\cite{Giudice:1988yz}, which
relates the generation of the $\mu$ term to the spontaneous breaking
of supersymmetry rather than to $R$ symmetry breaking.}
It is for this reason that we will assume a discrete
$Z_4^R$ symmetry in the following.


For our purposes, it will not be necessary to specify the origin 
of this $Z_4^R$ symmetry.
But it is interesting to note that orbifold compactifications
of the heterotic string have the ability to yield discrete $R$ symmetries
in the low-energy effective theory.
In this case, the discrete $R$ symmetry at low energies is nothing but
a remnant of the higher-dimensional Lorentz symmetry that survives the
compactification of the internal space (for early as well as more recent work
on this topic, see \cite{Font:1988nc} and \cite{Bizet:2013gf}, respectively).
We also mention that the special case of a discrete $Z_4^R$ symmetry has
received particular attention in the context of orbifold compactifications in
recent years~\cite{Lee:2010gv}.
The assumption of a discrete $Z_4^R$ symmetry in the low-energy effective
theory is, therefore, well motivated and stands theoretically on a sound footing.


Second of all, inflation is more than just a pure dS phase.
It represents a stage of quasi-dS expansion, in the
course of which the Hubble parameter slowly varies.
We, thus, need to specify the dynamics of supersymmetry breaking
more precisely and check whether the corresponding Polonyi potential is, in fact,
suitable for slow-roll inflation.
Here, we shall work within the framework of dynamical supersymmetry
breaking~\cite{Affleck:1983mk}, in which supersymmetry is assumed
to be broken by the dynamics of a strongly coupled SUSY gauge theory.
This gives us the advantage that the SUSY breaking scale $\Lambda_{\rm SUSY}$
is generated dynamically via dimensional transmutation.
As far as supersymmetry breaking is concerned, we will, therefore, not
have to rely on any dimensionful input parameters.
Instead, $\Lambda_{\rm SUSY}$ and, thus, the energy scale of inflation,
$\Lambda_{\rm inf}$, will be controlled by the dynamical scale $\Lambda$
of the SUSY-breaking hidden sector,
\begin{align}
 \Lambda_{\rm SUSY} \equiv \Lambda_{\rm inf} \sim \Lambda \,.
\end{align}
In this sense, our inflation model should be regarded as a variant
of \textit{dynamical inflation}~\cite{Dimopoulos:1997fv,Craig:2008tv,Harigaya:2012pg},
as we assume the energy scale of inflation to be
generated by strong dynamics.
Similarly, our model is closely related to \textit{natural inflation}~\cite{Freese:1990rb},
which treats the inflaton as an axion-like field that is likewise subject to
a scalar potential generated by nonperturbative dynamics (for recent dynamical
implementations of natural inflation in field theory, see~\cite{Harigaya:2014eta}),
as well as to \textit{modulus inflation}~\cite{Thomas:1995dq}, where
one identifies the inflaton as a (composite) modulus in the
effective low-energy regime of strongly coupled gauge theories.
 

In our case, the role of the inflaton is played by the scalar component
of the chiral Polonyi field $\Phi$, which breaks supersymmetry via its nonzero $F$-term.
In global supersymmetry and at the classical level, the scalar Polonyi potential
is exactly flat and, thus, an ideal starting point for the realization of inflation.
At the quantum level and in supergravity, the Polonyi potential, however, receives
corrections, which may or may not spoil the flatness of the potential.
Here, the SUGRA corrections lead, in particular, to the notorious $\eta$
problem~\cite{Dine:1983ys}, which typically requires a parameter
fine-tuning at the level of $1\cdots10\,\%$ or so.
The quantum and gravity corrections to the effective scalar potential
scale with the coupling strengths of the Yukawa
interactions between the Polonyi field and matter fields, $\lambda$, as well as
with the coefficients of the higher-dimensional Polonyi terms in the effective
K\"ahler potential, $\epsilon$, respectively.
In order to assess the prospects of successful Polonyi inflation, one,
therefore, has to study the viability of inflation as a function
of the parameters $\lambda$ and $\epsilon$ and identify those parameter
ranges that lead to consistency with the observational data
on the cosmic microwave background (CMB)~\cite{Ade:2015xua}.


\subsection[Our setup: minimal model based on two strongly coupled $SU(2)$ gauge theories]
{Our setup: minimal model based on two strongly coupled \boldmath{$SU(2)$} gauge theories}
\label{subsec:setup}


To this end, we will present in this paper a \textit{minimal} realization
of the idea of Polonyi inflation.
Our model is based on two strongly coupled
$SU(2)$ hidden gauge sectors (featuring two quark/antiquark
pairs each), which we take to be responsible for the dynamical
breaking of supersymmetry and $R$ symmetry, respectively.
We supplement both sectors with an appropriate number of gauge singlet fields,
so that the SUSY-breaking sector becomes identical 
to the simplest version of the IYIT DSB model~\cite{Izawa:1996pk},
while the $R$-symmetry breaking sector turns into a strongly coupled
SQCD theory with a quantum mechanically deformed moduli space
and field-dependent quark masses~\cite{Seiberg:1994bz}.
More precisely, we assume the quark masses
in the $R$ symmetry-breaking sector to be
controlled by the VEV of a singlet, which we call $P$.


Inflation is then driven by the SUSY-breaking vacuum energy
density in the IYIT sector, which results in an inflaton
potential equivalent to that of supersymmetric
F-term hybrid inflation (FHI)~\cite{Copeland:1994vg},
including corrections from supergravity~\cite{Panagiotakopoulos:1997qd}
as well as from higher-dimensional terms in the tree-level
K\"ahler potential~\cite{Asaka:1999jb,BasteroGil:2006cm}.
This is reminiscent of the inflation models presented
in~\cite{Izawa:1997jc} and~\cite{Dimopoulos:1997fv}.
The model in~\cite{Dimopoulos:1997fv}, however, corresponds to a reduced
version of the IYIT model with less singlets, which leads to the vanishing
of the inflationary vacuum energy at the end of inflation.
Contrary to our approach, it, thus, does not establish a connection between
inflation and supersymmetry breaking.
Meanwhile, the model in~\cite{Izawa:1997jc} identifies a different singlet field,
other than the Polonyi field, as the inflaton.
This allows the author of~\cite{Izawa:1997jc}
to separate the scales of inflation and supersymmetry breaking
by imposing a hierarchy among the Yukawa couplings of the inflaton
and the SUSY-breaking field.
We, on the other hand, will show how to implement inflation into
the full IYIT model, sticking to the notion that the best motivated inflaton
candidate in the IYIT model is still the Polonyi field itself.
An important consequence of this approach is that, in our case, F-term
hybrid inflation does \textit{not} end in a waterfall transition in the
inflationary sector.
Instead, we simply retain the inflationary vacuum energy density at low energies,
which then continues to act as the vacuum energy density associated with
the spontaneous breaking of supersymmetry.
This also means that our ``waterfall transition-free'' scenario
of F-term hybrid inflation does not suffer from the usual
production of topological defects, such as cosmic strings
in the case of a $U(1)$ waterfall transition, which would otherwise
exert some serious phenomenological pressure on our model
(see, e.g., \cite{Buchmuller:2014epa}).


We assume that the SUSY-breaking sector and the $R$-symmetry breaking
sector only communicate with each other via interactions
in the K\"ahler potential (i.e, not via interactions in the superpotential).
This is sufficient to stabilize the scalar field $P$ during inflation at
$\left<P\right> = 0$ by means of its Hubble-induced mass.
During inflation, the (fermionic) quarks in the $R$-symmetry breaking
sector are, therefore, massless and the discrete $Z_4^R$ symmetry remains unbroken
in this sector.
As anticipated, the superpotential then lacks a constant term during inflation,
which relieves the inflationary dynamics from the most dangerous
gravitational corrections in supergravity.
In particular,
the inflaton potential is free of the notorious ``tadpole term'' linear
in the inflaton field~\cite{Buchmuller:2000zm,Buchmuller:2014epa,Nakayama:2010xf}.
Towards the end of inflation, the Hubble-induced mass of the scalar field
$P$ decreases.
Adding an appropriately chosen superpotential for the field $P$,
we can use this fact to trigger a waterfall transition
\textit{in the $R$ symmetry-breaking sector} at small inflaton field values.
The field $P$ then acquires a large VEV and the quarks in the $R$ symmetry-breaking
sector become very massive.
Consequently, the $R$ symmetry-breaking sector turns into
a pure super-Yang-Mills (SYM) theory and $R$ symmetry becomes spontaneously
broken via gaugino condensation~\cite{Veneziano:1982ah}.
This external waterfall transition is associated with the breaking
of a $Z_2$ parity, which is, however, only an approximate symmetry.
For this reason, we do not have to fear the production of topological
defects (i.e., domain walls) during the waterfall transition in the
$R$ symmetry-breaking sector.


After these introductory remarks, we are now in the position to present our
analysis.
The remainder of this paper is organized as follows.
In the next section, we will show how the IYIT DSB model may
give rise to dynamical F-term hybrid inflation. 
Here, we will first argue why the original Polonyi model~\cite{Polonyi:1977pj}
of supersymmetry breaking is \textit{not} sufficient for a successful realization of
Polonyi inflation, even if we assume zero constant in the superpotential during
inflation.
We, therefore, conclude that we only have a chance of successfully realizing
Polonyi inflation in the presence of radiative corrections---which leads us
to consider the IYIT model as a possible UV completion of the original Polonyi model.
We then derive the scalar potential of F-term hybrid inflation
in the IYIT model and discuss its embedding into supergravity.
As an interesting aside, we demonstrate that Polonyi
inflation is incompatible with the concept of an approximate shift
symmetry in the inflaton direction~\cite{Kawasaki:2000yn}.
Instead, it turns out that Polonyi inflation 
requires a near-canonical K\"ahler potential.
In Sec.~\ref{sec:Rbreaking}, we expand on our mechanism of late-time
$R$ symmetry breaking, showing how a small inflaton field value 
may trigger gaugino condensation in a separate hidden sector.
Related to that, we comment on the backreaction of the
$R$ symmetry-breaking sector on the inflationary dynamics
and discuss how the two sectors of supersymmetry and $R$
symmetry breaking have to conspire to yield a zero CC
in the true vacuum after inflation.
In Sec.~\ref{sec:pheno}, we turn to the phenomenological implications
of our scenario.
Here, we identify the viable region in 
parameter space that leads to agreement with the latest PLANCK
data on the inflationary CMB observables~\cite{Ade:2015xua}.
As we are able to show, a scalar spectral index of $n_s \simeq 0.968$
can be easily achieved for an $\mathcal{O}(1)$ Yukawa coupling, $\lambda \simeq 2$,
and a slightly suppressed coefficient in the noncanonical K\"ahler
potential, $\epsilon \simeq 0.2$.
The amplitude of the scalar power spectrum, $A_s$, fixes the dynamical
scale of the SUSY-breaking sector to a value close to the unification
scale, $\Lambda \simeq 1 \times 10^{16}\,\textrm{GeV}$, suggesting
that our setup may eventually be part of a grand unified theory (GUT).
As a characteristic feature of Polonyi inflation, we highlight the fact
that the relation between the gravitino mass and the inflationary Hubble
rate in Eq.~\eqref{eq:m32Hinf} directly translates into a one-to-one
correspondence between the gravitino mass $m_{3/2}$ and the tensor-to-scalar ratio $r$,
\begin{align}
m_{3/2} \simeq \frac{\pi}{\sqrt{2}}\left(r A_s\right)^{1/2} M_{\rm Pl}
\sim 10^{12}\,\textrm{GeV} \:\left(\frac{r}{10^{-4}}\right)^{1/2} \,.
\label{eq:m32r}
\end{align}
At this point, it is interesting to note that the observed value of the scalar
spectral amplitude, $A_s \sim 10^{-9}$, might be the result of
anthropic selection~\cite{Tegmark:1997in}.
Together with the paradigm of slow-roll inflation (which implies $r \ll 1$),
the anthropic value of $A_s$ could,
therefore, explain why the soft SUSY mass scale is so
much higher than the electroweak or TeV scale.
Moreover, we point out in Sec.~\ref{sec:pheno} that, after inflation,
the Polonyi field mostly decays into gravitinos.
Polonyi inflation is, thus, followed by a phase of gravitino
domination~\cite{Jeong:2012en}, which leads to reheating
around temperatures of $\mathcal{O}\left(10^8\right)\,\textrm{GeV}$.
This paves the way for thermal wino dark matter (DM) as well as
thermal leptogenesis~\cite{Fukugita:1986hr} enhanced
by resonance effects~\cite{Pilaftsis:1997jf}.


Appendix~\ref{app:potential} contains some technical details regarding the
derivation of the effective inflaton potential.
In particular, we show how the one-loop corrections to the effective potential
may be obtained, to leading order, from an effective K\"ahler potential.
In Appendix~\ref{app:viability}, we explain in more detail why Polonyi inflation
in the IYIT model does not work, if the superpotential already contains a constant
term from the very beginning.
This completes our argument that, in the context of our minimal model,
successful Polonyi inflation requires
(i) radiative corrections,
(ii) a near-canonical K\"ahler potential, (iii) as well as late-time $R$ symmetry breaking.
Among all possible choices regarding (i) the type of interactions that the Polonyi
field participates in, (ii) the shape of the K\"ahler potential, and (iii)
the chronology of supersymmetry and $R$ symmetry breaking, this
leave one unique possibility for how to realize Polonyi inflation.


\section{Dynamical inflation in the IYIT supersymmetry breaking model}


\subsection{Inflation in the original Polonyi model and the need for radiative corrections}
\label{subsec:original}


The Polonyi model~\cite{Polonyi:1977pj} is the simplest
O'Raifeartaigh model of supersymmetry breaking via a nonvanishing F-term.
Its superpotential consists of a tadpole term and a constant,
\begin{align}
W = \mu^2\, \Phi + w \,.
\label{eq:WPolonyi}
\end{align}
Here, $\Phi$ denotes the chiral Polonyi field, $\mu$ is the scale of supersymmetry
breaking and $w$ is a constant that breaks $R$ symmetry and which determines the
potential energy density in the ground state.
For a canonical K\"ahler potential and fine-tuning $w$, so that it takes the particular value
$w_0 = \left(2-\sqrt{3}\right)\mu^2 M_{\rm Pl}$, this
model has a Minkowski vacuum at
$\left<\Phi\right> = \left(\sqrt{3}-1\right)M_{\rm Pl}$, in which 
supersymmetry is broken by the Polonyi F-term, $\left|F_\Phi\right| = \mu^2$.
It has been known for a long time, that the scalar potential for the Polonyi field
around this vacuum is unfortunately too steep to support
slow-roll inflation~\cite{Ovrut:1983my}.
In Appendix~\ref{app:viability}, we review and extend this argument,
showing for various choices of the K\"ahler potential, that,
with $w = w_0$ from the very beginning, the Polonyi model does not
give rise to inflation.
We consider, in particular, a canonical K\"ahler potential supplemented
by higher-dimensional corrections as well as K\"ahler potentials featuring an 
approximate shift symmetry either along the real or the imaginary axis in the
complex plane.
In none of the cases under consideration inflation is viable---either
because we fail to satisfy the slow-roll conditions
or because the scalar potential does not exhibit a global Minkowski vacuum in the first place.


This immediately raises the question whether inflation might perhaps become possible
in the Polonyi model, if we impose a discrete $R$ symmetry at high energies,
so that $w = 0$ initially.
Let us address this question for a canonical K\"ahler potential
supplemented by a higher-dimensional correction,%
\footnote{In Sec.~\ref{subsec:SUGRA}, we will discuss the same question for
an approximately shift-symmetric K\"ahler potential.
In this case, inflation turns out be unfeasible because, with the superpotential
being given as $W = \mu^2 \,\Phi$,
the $-3\,\exp\left[K/M_{\rm Pl}^2\right]\left|W\right|^2/M_{\rm Pl}^2$ SUGRA term
in the scalar potential induces a tachyonic mass for the Polonyi field.
This results in a global AdS minimum.
\label{fn:shift}}
\begin{align}
K = \Phi^\dagger \Phi + \frac{\epsilon}{\left(2!\right)^2}
\left(\frac{\Phi^\dagger \Phi}{M_{\rm Pl}}\right)^2 + 
\mathcal{O}\left(\epsilon^2,M_{\rm Pl}^{-4}\right) \,, \quad
\epsilon \lesssim 1 \,.
\label{eq:Kphican}
\end{align}
Here, we assume that the K\"ahler potential is always dominated by the
canonical term, $K \supset \Phi^\dagger \Phi$, also at
field values above the Planck scale.
An exhaustive study of
arbitrary choices for the K\"ahler potential is beyond the scope of this paper.
Under this assumption, the scalar potential
always picks up a SUGRA correction, $V \propto e^{K/M_{\rm Pl}^2}$,
which spoils the flatness of the potential at super-Planckian field values. 
For this reason, we only have a chance of realizing slow-roll inflation
at field values below the Planck scale.
For the K\"ahler potential in Eq.~\eqref{eq:Kphican}, the scalar potential
in supergravity then takes the following form,
\begin{align}
V\left(\varphi\right) = V_0 \left[1 - \frac{\epsilon}{2} \left(\frac{\varphi}{M_{\rm Pl}}\right)^2
+ \frac{1}{8}\left(1-\frac{7\,\epsilon}{2}+\frac{8\,\epsilon^2}{3}\right)
\left(\frac{\varphi}{M_{\rm Pl}}\right)^4 + \mathcal{O}\left(\varphi^6\right) \right]
\,, \quad V_0 = \mu^4 \,,
\label{eq:VPolonyi}
\end{align}
where the real scalar field $\varphi$ denotes the \textit{canonically normalized}
radial component of the complex Polonyi scalar
$\tilde\phi = \tilde\varphi/\sqrt{2}\,e^{i\tilde\theta}$ contained in $\Phi$
(see Sec.~\ref{subsec:SUGRA}) and where we have introduced $V_0$ as
the SUSY-breaking vacuum energy density at $\varphi = 0$.
From the form of the scalar potential in Eq.~\eqref{eq:VPolonyi},
it is evident that, even with $w$ being set to zero,
the Polonyi model fails to yield successful inflation.
For instance, if we choose $\epsilon$ to be negative, the field $\varphi$
is driven towards the origin by a positive mass squared,
similarly as in chaotic inflation~\cite{Linde:1983gd}.
Inflation may then take place at small field values close to the
origin---but not in accord with the observational data.
To see this, consider the slow-roll parameters $\varepsilon$ and $\eta$,
\begin{align}
\varepsilon = \frac{M_{\rm Pl}^2}{2}\left(\frac{V'}{V}\right)^2 \,, \quad
\eta = M_{\rm Pl}^2\, \frac{V''}{V} \,, \quad
V' = \frac{d V}{d\varphi} \,, \quad
V'' = \frac{d^2 V}{d\varphi^2} \,.
\label{eq:SRpara}
\end{align}
Independent of the sign of $\epsilon$, we have $\eta \geq -\epsilon$.
For negative $\epsilon$, the slow-roll parameter $\eta$ is, therefore, bound to be
positive, while the slow-roll parameter $\varepsilon$ turns out to be negligibly
small during inflation, $\varepsilon \ll \left|\eta\right|$.
According to the slow-roll formula for the scalar spectral index,
$n_s = 1 + 2\,\eta_* - 6\,\varepsilon_*$, we will then always obtain a blue-tilted
scalar spectrum ($n_s > 1$).
In view of the latest best-fit value for $n_s$ reported by the PLANCK
collaboration, $n_s^{\rm obs} = 0.9677 \pm 0.0060$~\cite{Ade:2015xua},
such a spectrum is clearly ruled out by the observational data.
On the other hand, if we choose $\epsilon$ to be positive,
the Polonyi field acquires a tachyonic mass around the origin and
inflation proceeds from small to large field values, similarly as
in new inflation~\cite{Linde:1981mu}.
It is then hard to imagine how the generation of the constant
$w$ in the superpotential should be triggered, after a sufficient amount
of inflation, at field values close or even above the Planck scale.
But more than that, even if we assume that this problem could somehow be solved,
the scalar potential in Eq.~\eqref{eq:VPolonyi} still does not
lead to an acceptable phenomenology.
In new inflation, the scalar spectral index turns out to be bounded from above,
$n_s \lesssim 0.95$~\cite{Izawa:1996dv,Nakayama:2012dw},
which deviates from the observed value by at least $3\,\sigma$.
Therefore, also for $\epsilon > 0$, we fail to
reach consistency with the observational data.


In summary, we conclude that the \textit{bare Polonyi model} based on the superpotential
in Eq.~\eqref{eq:WPolonyi}---and for reasonable ans\"atze regarding the shape of
the K\"ahler potential---does not allow for a successful realization of slow-roll inflation.
In respect of this null result, two comments are in order:
(i) We emphasize that our analysis of the Polonyi K\"ahler potential
in this paper does \textit{not} mount up to a general no-go theorem.
In the most general case, the K\"ahler potential
for the Polonyi field is given by an arbitrary function $F$ of the field
$\Phi$ and its conjugate,
$K = M_{\rm Pl}^2\, F\left(\Phi / M_{\rm Pl},\Phi^\dagger / M_{\rm Pl}\right)$.
In absence of any other scale, it is, in particular, clear that $M_{\rm Pl}$
can be the only relevant scale in the K\"ahler potential.
It may then well be that certain fine-tuned functions $F$ do allow for
successful inflation in the Polonyi model, after all
(see, e.g.,~\cite{Izawa:2007qa} for a discussion of fine-tuned
K\"ahler potentials in the context of SUGRA models of inflation).
In the following, we will, however, ignore the possibility of such a
biasedly chosen K\"ahler potential and focus on the usual suspects:
K\"ahler potentials that are either near-canonical or approximately shift-symmetric.
(ii) The fact that we are unable to realize successful inflation in the bare
Polonyi model is not a serious problem, as the Polonyi model is not expected
to be a fundamental description of spontaneous supersymmetry breaking, anyway.
It should rather be seen as the effective theory resulting from some
UV dynamics that provide a dynamical explanation for the origin of
the parameters $\mu$ and $w$ in Eq.~\eqref{eq:WPolonyi}.
From this perspective, it is then more likely than not that the Polonyi field
is not only subject to its gravitational self-interaction,
but that it also participates in Yukawa interactions with heavy
matter fields in the UV theory.
In the corresponding \textit{effective Polonyi model} at low energies, these matter fields
are integrated out, so that they no longer appear in the superpotential.
But their couplings to the Polonyi field still
yield radiative corrections to the scalar potential in Eq.~\eqref{eq:VPolonyi},
which affect the inflationary dynamics.
In such a modified setup, i.e., in the original Polonyi model supplemented by
radiative corrections, successful Polonyi inflation may, therefore, 
very well be an option.
In the following, we will construct a minimal extension
of the Polonyi model where this is indeed the case.
We shall consider a minimal UV completion of
the Polonyi model---consisting of two strongly coupled $SU(2)$ sectors that
account for the dynamical origin of the parameters $\mu$ and $w$, respectively---and
demonstrate that, in the presence of radiative corrections, successful
Polonyi inflation is indeed feasible for a natural choice of parameters values.


\subsection{Dynamical supersymmetry breaking in the low-energy regime of the IYIT model}
\label{subsec:SUSY}


One of the simplest ways to generate the SUSY breaking scale $\mu$ in
Eq.~\eqref{eq:WPolonyi} is to identify the Polonyi field as part of
the IYIT model---the simplest vector-like model of dynamical
supersymmetry breaking~\cite{Izawa:1996pk}.
In its most general formulation, the IYIT model is based on a strongly coupled
$Sp(N)$ gauge theory featuring $N_f = N+1$ pairs of ``quark fields'' $\Psi^i$
that transform in the fundamental representation of $Sp(N)$.
The gauge dynamics of this model are associated with a dynamical
scale $\Lambda$, which denotes the energy scale at which the $Sp(N)$
gauge coupling formally diverges.
The low-energy effective theory below the dynamical scale exhibits
a quantum moduli space of degenerate supersymmetric vacua, which is spanned
by $N_f\left(2 N_f-1\right)$ gauge-invariant composite flat directions
(or ``meson fields'') $M^{ij}$,
\begin{align}
M^{ij} \simeq \frac{1}{\eta\,\Lambda} \left<\Psi^i \Psi^j\right> \,, \quad
i,j = 1,2,\cdots 2\,N_f \,.
\label{eq:Mij}
\end{align}
Here, the parameter $\eta$ is a dimensionless numerical factor, which ensures
the canonical normalization of the meson fields $M^{ij}$ at low energies.
\textit{Naive dimensionful analysis}~\cite{Manohar:1983md} leads us to
expect that $\eta$ should be of $\mathcal{O}\left(4\pi\right)$ and, for definiteness,
we will, therefore, simply set $\eta = 4\pi$ in the following.
The quantum moduli space of the IYIT model is subject to the following
constraint pertaining to the meson VEVs,
\begin{align}
\textrm{Pf}\left(M^{ij}\right) \simeq \left(\frac{\Lambda}{\eta}\right)^2 \,,
\label{eq:PfMij}
\end{align}
which represents the quantum mechanically deformed version
of the classical constraint $\textrm{Pf}\left(M^{ij}\right) = 0$~\cite{Seiberg:1994bz}.
In order to break supersymmetry in the IYIT model, one has to lift the
flat directions $M^{ij}$, so that Eq.~\eqref{eq:PfMij} is longer compatible
with a vanishing vacuum energy density.
This is readily done by coupling the quark pairs $\Psi^i\Psi^j$
to a corresponding number of $Sp(N)$ singlet fields, $Z_{kl}$, in the
tree-level superpotential,
\begin{align}
W_{\rm IYIT}^{\rm tree} = \frac{1}{4} \lambda_{ij}^{kl}\, Z_{kl}\, \Psi^i \Psi^j \,, \quad
Z_{kl} = - Z_{lk} \,, \quad 
\lambda_{ij}^{kl} = - \lambda_{ji}^{kl} = \lambda_{ji}^{lk} \,, \quad
i,j,k,l = 1,2,\cdots 2\,N_f \,,
\label{eq:WIYITtree}
\end{align}
where $\lambda_{ij}^{kl}$ denotes a matrix of Yukawa couplings with at most
$N_f\left(2N_f-1\right)$ independent eigenvalues.
These Yukawa couplings induce Dirac mass terms for the meson and singlet fields
at low energies,
\begin{align}
W_{\rm IYIT}^{\rm eff} \simeq \frac{1}{4} \lambda_{ij}^{kl}\, \frac{\Lambda}{\eta}
\, Z_{kl}\, M^{ij} \,,
\label{eq:WIYITeff}
\end{align}
so that the singlet F-term conditions, $\lambda_{ij}^{kl}  M^{ij} = 0$, 
are incompatible with the deformed moduli constraint, $\textrm{Pf}\left(M^{ij}\right)\neq 0$.
Supersymmetry is then broken \`a la O'Raifeartaigh via nonvanishing F-terms.


In global supersymmetry and for all Yukawa couplings in Eq.~\eqref{eq:WIYITtree} being equal,
$\lambda_{ij}^{kl} = \lambda\, \delta_i^k \delta_j^l$,
the anomaly-free global flavor symmetry $G_F$ of the IYIT model is given as follows,
\begin{align}
G_F = SU(2N_f)\times Z_{2N_f}\times U(1)_R  \,.
\label{eq:GF}
\end{align}
Here, the discrete $Z_{2N_f}$ symmetry is the anomaly-free subgroup of the anomalous
$U(1)$ that is contained in the full flavor symmetry at the classical level,
$U(2N_f) \times U(1)_R \cong SU(2N_f) \times U(1)\times U(1)_R$.
Under the $Z_{2N_f}$ symmetry, all quarks carry charge $1$, while 
the singlet fields carry charge $-2$.
Meanwhile, the presence of the global continuous $R$ symmetry 
is characteristic for a large class of DSB models~\cite{Nelson:1993nf}.
Under $U(1)_R$, the quark and singlet fields carry charges $0$
and $2$, respectively.
In the SUSY-breaking vacuum of the IYIT model, where $\left<M\right> \neq 0$
and $\left<Z\right> = 0$, $R$ symmetry, therefore, remains unbroken.
As a global symmetry, the continuous $R$ symmetry of the IYIT model
is, of course, only an approximate symmetry, which we expect to be broken
by quantum gravitational effects~\cite{Banks:2010zn}.
On the other hand, recall that we assume the discrete subgroup
$Z_4^R \subset U(1)_R$ to be gauged (see our discussion in Sec.~\ref{subsec:ingredients}).
This protects the quality of the $U(1)_R$ symmetry; and
it is reasonable to assume that all gravity-induced $U(1)_R$-breaking
effects in the IYIT sector are suppressed.
In the following, we will, therefore, stick to the effective superpotential
in Eq.~\eqref{eq:WIYITeff} and neglect the possibility of small $R$ symmetry-breaking corrections.


From now on, let us restrict ourselves to the simplest version of
the IYIT model: a strongly coupled $Sp(1) \cong SU(2)$ gauge theory featuring
four matter fields $\Psi^i$ and six singlet fields $Z_{kl}$.
The non-Abelian flavor symmetry $SU(2N_f)$ then corresponds to a global $SU(4)$, 
under which $M^{ij}$ and $Z_{kl}$ transform as six-dimensional
antisymmetric rank-2 tensor representations.
Here, note that $SU(4)$ is the double cover of $SO(6)$.
This allows us to rewrite the meson
and singlet fields as vector representations of $SO(6)$,
\begin{align}
\begin{pmatrix}
X^0 \\ 
X^1 \\
X^2 \\
X^3 \\
X^4 \\
X^5
\end{pmatrix} = \frac{1}{\sqrt{2}}
\begin{pmatrix}
+1 \left(M^{12} + M^{34}\right) \\
-1 \left(M^{13} - M^{24}\right) \\
+1 \left(M^{14} + M^{23}\right) \\
-i \left(M^{14} - M^{23}\right) \\
-i \left(M^{13} + M^{24}\right) \\
+i \left(M^{12} - M^{34}\right) 
\end{pmatrix} \,, \qquad 
\begin{pmatrix}
S_0 \\ 
S_1 \\
S_2 \\
S_3 \\
S_4 \\
S_5
\end{pmatrix} = \frac{1}{\sqrt{2}}
\begin{pmatrix}
+1 \left(Z_{12} + Z_{34}\right) \\
-1 \left(Z_{13} - Z_{24}\right) \\
+1 \left(Z_{14} + Z_{23}\right) \\
+i \left(Z_{14} - Z_{23}\right) \\
+i \left(Z_{13} + Z_{24}\right) \\
-i \left(Z_{12} - Z_{34}\right) 
\end{pmatrix} \,.
\label{eq:XS}
\end{align}
As we will see shortly, it will turn out to be convenient to study the SUSY-breaking
dynamics of the IYIT sector in terms of this ``$SO(6)$ language'' rather than in
terms of the original ``$SU(4)$ language''.


Trading the mesons $M^{ij}$ for the new fields
$X^a$, the Pfaffian constraint in Eq.~\eqref{eq:PfMij} can be written as,
\begin{align}
\textrm{Pf}\left(M^{ij}\right) \equiv \frac{1}{2}\,\left(X\cdot X\right) \equiv
\frac{1}{2}\sum_{a=0}^5 X^a X^a \equiv \frac{1}{2} \left(X^a\right)^2 \simeq
\left(\frac{\Lambda}{\eta}\right)^2 \,,
\label{eq:PfXX}
\end{align}
which defines a sphere in the six-dimensional space spanned by the six
meson coordinates $X^a$.
An elegant way to enforce this constraint is to directly incorporate it into
the effective superpotential~\cite{Seiberg:1994bz,Izawa:1996pk},
\begin{align}
W_{\rm eff}^{\rm dyn} \simeq \frac{\kappa}{\eta}\,\Lambda^2\, T\,C\left(x^a\right) \,, \quad
C\left(x^a\right) = \frac{1}{2} \left(x^a\right)^2  - 1 \,, \quad
x^a = \frac{X^a}{\Lambda/\eta} \,.
\label{eq:Weffdyn}
\end{align}
Here, the field $T$ denotes a Lagrange multiplier, the corresponding
F-term condition of which, $C\left(x^a\right) = 0$, is nothing but a
reformulation of the moduli constraint in Eq.~\eqref{eq:PfXX}.
The physical status of the field $T$ is unfortunately rather unclear and depends on
whether (uncalculable) strong-coupling effects induce a kinetic term
for $T$ or not.
If the field $T$ should be physical, it might represent a dynamical
glueball field, $T \sim \left<gg\right>$, and the dimensionless coupling
constant $\kappa$ in Eq.~\eqref{eq:Weffdyn} would be expected
to take some value of $\mathcal{O}(1)$.
In that case, the Pfaffian constraint in Eq.~\eqref{eq:PfXX} would be satisfied
only approximately, depending on the competition between the different
F-term conditions that enter into the determination of the true ground state.
If, on the other hand, $T$ should be unphysical, we would have to treat
it as a mere auxiliary field. 
In that case, the Pfaffian constraint should be satisfied exactly,
which would require us to eventually take the limit
$\kappa \rightarrow \infty$ in our analysis.
In the following, we will suppose that the Lagrange multiplier field $T$ is,
indeed, a physical (glueball) field and set $\kappa = 1$ for definiteness.%
\footnote{For an extended
discussion of this point, see also \cite{Harigaya:2015soa,Domcke:2014zqa}.}


In terms of the fields $X^a$ and $S_a$, the effective tree-level superpotential takes
the following form,
\begin{align}
W_{\rm eff}^{\rm tree} \simeq \frac{\lambda_a}{\eta}\, \Lambda\, S_a X^a \,,
\label{eq:Wefftree}
\end{align}
where we assume, w.l.o.g., the Yukawa couplings $\lambda_a$ to be ordered by size,
$\lambda_b \leq \lambda_{b+1}$ for all $b = 0,\cdots4$.
Note that we will refer to the smallest Yukawa coupling, $\lambda_0$,
simply as $\lambda$ in the following, $\lambda \equiv \lambda_0$.
For generic values of the six Yukawa couplings $\lambda_a$,
the non-Abelian flavor symmetry is completely broken,
\begin{align}
\lambda_a \textrm{ all different} \quad\Rightarrow\quad
G_F = SO(6) \times Z_4  \times U(1)_R \rightarrow Z_4  \times U(1)_R \,.
\label{eq:GFbreaking}
\end{align}
The total effective superpotential is
given by the sum of $W_{\rm eff}^{\rm dyn}$ in Eq.~\eqref{eq:Weffdyn}
and $W_{\rm eff}^{\rm tree}$ in Eq.~\eqref{eq:Wefftree},
\begin{align}
W_{\rm eff} \simeq \frac{\lambda_a}{\eta}\, \Lambda\, S_a X^a + 
\frac{\kappa}{\eta}\,\Lambda^2\, T\,C\left(x^a\right) \,.
\end{align}
We mention once more that, as a consequence of the nonrenormalization
theorem, the effective superpotential does not receive radiative corrections
in perturbation theory~\cite{Grisaru:1979wc}.
Given this form of the total superpotential and assuming a quadratic
K\"ahler potential for all meson and singlet fields,
the global minimum of the resulting F-term scalar potential is located at
\begin{align}
\left<X^0\right> = \pm\,\sqrt{2}\left(1-\zeta\right)^{1/2} \frac{\Lambda}{\eta} \,, \quad 
\left<S_0\right> = \mp\,\sqrt{2}\left(\frac{1-\zeta}{\zeta}\right)^{1/2} \left<T\right> \,, \quad
\left<X^n\right> = \left<S_n\right> = 0 \,, \quad n = 1,\cdots5 \,,
\label{eq:XS0vac}
\end{align}
where $\left<T\right>$ is undetermined at tree level.
The sign ambiguity is a consequence of the $Z_4$ flavor symmetry
(see Eq.~\eqref{eq:GFbreaking}).
Moreover, the parameter $\zeta$ measures how well the deformed moduli
constraint is satisfied,
\begin{align}
\zeta = \left<\left|C\left(x^a\right)\right|\right> =
\left(\frac{\lambda}{\kappa\eta}\right)^2 = \frac{\lambda^2}{16\pi^2} \,.
\label{eq:zeta}
\end{align}
For perturbative values of the Yukawa coupling $\lambda$, i.e., for $\lambda \sim 1$
or smaller, we have $\zeta \ll 1$, which tells us that the deformed moduli
constraint is fulfilled almost exactly.
On the other hand, for nonperturbative values of $\lambda$, i.e., for $\lambda$ as large
as $\lambda \sim 4\pi$, the parameter $\zeta$ becomes of $\mathcal{O}\left(1\right)$,
indicating that the constraint function $C\left(x^a\right)$ significantly
deviates from zero.
To put this result into perspective, we must remember that,
for nonperturbative values of the Yukawa coupling, uncalculable
corrections to the effective K\"ahler potential due to strong-coupling effects
become important.
In fact, as pointed out by Chacko et al., these nonperturbative corrections
are only negligible as long as $\lambda \ll 4\pi$~\cite{Chacko:1998si}.
We can, therefore, trust our above analysis, based on a canonical
K\"ahler potential, only as long as $\lambda$ remains in the perturbative regime.
For this reason, we will, from now on, only consider $\lambda$
values at most as large as $\lambda_{\rm max} \simeq 4$,
so that we may always maintain a hierarchy among $\lambda$ and $\eta$
(i.e., so that $\lambda /\eta \lesssim 10^{-0.5}$).
For the parameter $\zeta$, this then means that it can take
values at most as large as $\zeta_{\rm max} \simeq 0.1$.
This translates into the statement that the moduli constraint is always satisfied
in our analysis---up to a deviation of at most $10\,\%$,
$\left<\left|C\left(x^a\right)\right|\right> \lesssim 0.1$.


In passing, we also mention that the scalar potential exhibits a saddle
point at the origin in field space as well as two saddle points along
each direction $X^n$ in moduli space.
Here, the loci of the saddle points away from the origin have the same
functional form as $\left<X^0\right>$ and $\left<S^0\right>$ in Eq.~\eqref{eq:XS0vac},
the only difference being that $\lambda$ in Eq.~\eqref{eq:zeta} needs to
be exchanged with the respective Yukawa coupling $\lambda_n$.
The low-energy vacuum along the $X^0$ axis, therefore, not only
marks the global (and only local) minimum of the scalar potential, it is
also the stationary point at which the deformed moduli
constraint is fulfilled best.


In the true vacuum, supersymmetry is broken
by the nonvanishing F-terms of the fields $S_0$ and $T$,
\begin{align}
\left<\left|F_{S_0}\right|\right> = \sqrt{2}\left(1-\zeta\right)^{1/2}
\lambda \left(\frac{\Lambda}{\eta}\right)^2 \,, \quad
\left<\left|F_T\right|\right> = \zeta^{1/2} \lambda \left(\frac{\Lambda}{\eta}\right)^2 \,.
\end{align}
The F-term of the meson field $X^0$, on the other hand, (which seems to be nonzero
at first sight) cancels,
\begin{align}
\left<\left|F_{X^0}\right|\right> = \frac{\lambda}{\eta}\, \Lambda \left<S_0\right> +
\kappa\eta \left<T\right>\langle X^0\rangle =
\left(\frac{\lambda}{\eta}-\kappa\,\zeta^{1/2}\right) \Lambda \left<S_0\right> = 0 \,.
\end{align}
In order to identify the mass eigenstates around the true vacuum,
we now shift the field $X^0$ by its VEV,
\begin{align}
X^0 = \left<X^0\right> + \Xi^0 \,, 
\end{align}
and rotate the SUSY-breaking fields $S_0$ and $T$ by their mixing angle $\beta$,
\begin{align}
\begin{pmatrix}
\Phi \\
\Sigma 
\end{pmatrix} = 
\begin{pmatrix}
\cos\beta & -\sin\beta \\
\sin\beta &  \cos\beta
\end{pmatrix}
\begin{pmatrix}
S_0 \\
T
\end{pmatrix} \,, \qquad
\tan\beta = \frac{\left<\left|T\right|\right>}{\left<\left|S_0\right|\right>} = 
\frac{1}{\sqrt{2}}\left(\frac{\zeta}{1-\zeta}\right)^{1/2} \,.
\label{eq:PhiSigma}
\end{align}
After these field transformations, the effective superpotential for
$\Phi$, $\Sigma$, $\Xi^0$, $S_n$, and $X^n$ reads as follows,
\begin{align}
W_{\rm eff} \simeq \mu^2\,\Phi + m_0\, \Sigma\, \Xi^0 + m_n\, S_n X^n
-\left(\kappa_\Phi\,\Phi - \kappa_\Sigma\,\Sigma\right)
\left[\frac{1}{2}\left(\Xi^0\right)^2 +\frac{1}{2}\left(X^n\right)^2\right] \,,
\label{eq:Weff}
\end{align}
which is (apart from the missing constant term $w$)
exactly of the form anticipated at the end of Sec.~\ref{subsec:original}.


First of all, note that the first term on the right-hand side
of Eq.~\eqref{eq:Weff}, $\mu^2\, \Phi$, is nothing but the
dynamical realization of the SUSY-breaking tadpole term
in Eq.~\eqref{eq:WPolonyi} within the IYIT model.
The field $\Phi$ is, thus, to be identified as the chiral Polonyi
field which breaks supersymmetry via its nonzero F-term,
\begin{align}
\left<\left|F_\Phi\right|\right> = \mu^2 \,, \quad
\mu = \left(2-\zeta\right)^{1/4} \lambda^{1/2}\, \frac{\Lambda}{\eta} \,,
\label{eq:mu}
\end{align}
where the parameter $\mu$ denotes again the SUSY breaking scale.
In this sense, Eqs.~\eqref{eq:Weff} and \eqref{eq:mu} show
that the IYIT model serves, indeed, as a viable UV completion
for at least half the Polonyi model:
The SUSY-breaking dynamics of the IYIT model manage to provide
a dynamical explanation for the SUSY breaking scale $\mu$.
However, as the IYIT model preserves $R$ symmetry in its ground state,
it is not capable of accounting for the origin of the $R$
symmetry-breaking constant $w$ in the Polonyi superpotential.


Second of all, the effective superpotential in Eq.~\eqref{eq:Weff} contains
(also as envisaged at the end of Sec.~\ref{subsec:original})
Yukawa couplings between the Polonyi field $\Phi$ and a number of
massive matter fields, $\Xi^0$ and $X^n$.
Here, the meson masses, $m_0$ and $m_n$, follow from Dirac 
mass terms together with the singlet fields $\Sigma$ and $S_n$,
\begin{align}
m_0 = \frac{m}{r_0} \,, \quad 
m_n = \frac{m}{r_n} \,, \quad
m = \kappa_\Phi^{1/2} \mu = \lambda\,\frac{\Lambda}{\eta} \,, \quad
r_0 = \left(\frac{\zeta}{2-\zeta}\right)^{1/2} = \sin\beta \,, \quad
r_n = \frac{\lambda}{\lambda_n} \,,
\label{eq:mara}
\end{align}
where we have introduced the flavor-independent mass scale $m$
as well as the respective ratios $r_0$ and $r_n$ between this
scale $m$ and the masses $m_0$ and $m_n$.
From the fact that $m = \kappa_\Phi^{1/2} \mu$ (see Eq.~\eqref{eq:kappa} below),
it immediately follows that the scale $m$ represents the amount of SUSY-breaking
mass splitting within the respective meson and singlet multiplets that is
induced by the tadpole term in Eq.~\eqref{eq:Weff}, see
Appendix~\ref{subsec:massspectrum} for details.
Given the definition of $\zeta$ in Eq.~\eqref{eq:zeta}
and recalling that we assume $\lambda$ to be the smallest
among all Yukawa couplings, $\lambda \leq \lambda_n$,
we also find that the ratios $r_0$ and $r_n$ are bounded from above,
\begin{align}
r_0 = \frac{m}{m_0} \leq 1 \,, \quad
r_n = \frac{m}{m_n} \leq 1 \,,
\label{eq:rbound}
\end{align}
so that the SUSY-breaking mass splitting $m$ never exceeds the supersymmetric
Dirac masses $m_a$.
Moreover, for the parameter range of interest, $\lambda \lesssim 4$ and
$\lambda_n \lesssim 4\pi$, the ratio $r_0$ always turns out to be the smallest,
$r_0 < r_n$, which leads to the interesting (and to some extent counter-intuitive)
result that the zeroth flavor, i.e., the flavor with the smallest Yukawa coupling,
ends up being stabilized the most.
In addition to that, the $\Xi^0$ flavor is also singled out by the
fact that its Dirac mass partner is none of the original singlet fields $S_a$,
but the linear combination $\Sigma$ that we introduced in Eq.~\eqref{eq:PhiSigma}
and which, for small values of $\zeta$, mostly consists of the Lagrange
multiplier field $T$.
This also explains why, for large $\kappa$ (and, hence, small
$\zeta$), the $\Xi^0$ mass $m_0$ diverges.
In this limit, the Pfaffian constraint is fulfilled exactly,
which results in the decoupling of $T$ and removes one meson
multiplet (i.e., $\Xi^0$) from the spectrum.


Next to the Polonyi field $\Phi$, also the ``stabilizer field'' $\Sigma$
couples to the meson fields $\Xi^0$ and $X^n$.
Here, the strengths of the respective Yukawa couplings, $\kappa_\Phi$ and $\kappa_\Sigma$,
are given by $\kappa\eta$ and the mixing angle $\beta$,
\begin{align}
\kappa_\Phi = \sin\beta\, \kappa\eta =
\frac{\lambda}{\left(2-\zeta\right)^{1/2}} \,, \quad
\kappa_\Sigma = \cos\beta\, \kappa\eta =
\left(\frac{2}{\zeta}\right)^{1/2} \left(\frac{1-\zeta}{2-\zeta}\right)^{1/2} \lambda \,.
\label{eq:kappa}
\end{align}
Just like the mass $m_0$, the Yukawa coupling $\kappa_\Sigma$ diverges in the limit
$\kappa\rightarrow\infty$.
This is a trivial consequence of its proportionality to $\kappa\eta$.
As for the Polonyi coupling $\kappa_\Phi$, this divergence is, however, canceled out
by the $\sin\beta$ factor.
In contrast to $\kappa_\Sigma$, the coupling $\kappa_\Phi$, therefore, always
remains finite.
In the limit $\kappa\rightarrow\infty$, it reproduces, in particular,
the Yukawa coupling of the singlet $S_0$ at energies above the dynamical
scale,
\begin{align}
W_{\rm IYIT}^{\rm tree} \supset \frac{\lambda}{\sqrt{2}} \,S_0
\left(\Psi^1\Psi^2 + \Psi^3\Psi^4\right) \,.
\end{align}
Finally, we mention that the four new parameters $\mu$, $m_0$,
$\kappa_\Phi$, and $\kappa_\Sigma$ introduced in Eq.~\eqref{eq:Weff}
are not linearly independent.
In fact, they must be dependent, as they can all be expressed in
terms of the three old parameters $\lambda$, $\Lambda$, and $\kappa$.
By making use of Eqs.~\eqref{eq:mu}, \eqref{eq:mara}, and \eqref{eq:kappa},
one easily convinces oneself that
\begin{align}
\mu^2 = \frac{\kappa_\Phi}{\kappa_\Phi^2 + \kappa_\Sigma^2}\,m_0^2 \,.
\end{align}


In order to see how the superpotential in Eq.~\eqref{eq:Weff} may give
rise to Polonyi inflation, it is instructive to forget about
the Yukawa couplings of the field $\Sigma$ for a moment and to rewrite 
Eq.~\eqref{eq:Weff} as follows,
\begin{align}
W_{\rm eff} \simeq \kappa_\Phi\,\Phi \left[v^2 - \frac{1}{2}\left(\Xi^0\right)^2 +
\frac{1}{2}\left(X^n\right)^2\right] + m_0\, \Sigma\, \Xi^0 + m_n\, S_n X^n + \cdots \,,
\label{eq:Weffv}
\end{align}
where the ellipsis stands for the Yukawa couplings involving
the stabilizer field $\Sigma$ and where we have introduced the mass scale $v$,
\begin{align}
v = \frac{m}{\kappa_\Phi}
= \left(2- \zeta\right)^{1/2} \,\frac{\Lambda}{\eta} \,.
\label{eq:v}
\end{align}
Remarkably enough, the first part of the superpotential in
Eq.~\eqref{eq:Weffv} has the same form as the superpotential of
supersymmetric F-term hybrid inflation~\cite{Copeland:1994vg} based
on $SO(6)$.
In the context of this interpretation, the Polonyi field plays the role
of the chiral inflaton singlet, while the meson fields $\Xi^0$ and $X^n$
act as a multiplet of FHI waterfall fields that transform in
the vector representation of $SO(6)$.
The mass scale $v$ is then to be identified as the energy scale
of the waterfall transition at the end of inflation,
while the mass splitting $m$ should be understood as
the tachyonic mass of the FHI waterfall fields at $\Phi = v$.
We note that it is this picture that
the authors of~\cite{Dimopoulos:1997fv} arrive at.
In their model, no other singlets except for the Polonyi field $\Phi$
are introduced.
The meson fields $\Xi^0$ and $X^n$, therefore, lack
their Dirac mass partners, so that the resulting effective
superpotential is exactly identical to the one of F-term
hybrid inflation.
This allows the authors of~\cite{Dimopoulos:1997fv}
to realize F-term hybrid inflation at large inflaton field
values, $\left|\Phi\right| > v$, where the flatness of the
inflaton potential is lifted by logarithmic loop corrections.


In our case, the situation at large field values is quite similar,
as we will discuss shortly; but at small field values, it is drastically different.
The second part of the superpotential in
Eq.~\eqref{eq:Weffv} introduces explicit Dirac mass terms for
the ``would-be waterfall fields'' $\Xi^0$ and $X^n$ that are
absent in standard F-term hybrid inflation
(as well as in the model in~\cite{Dimopoulos:1997fv}).
Accounting for the presence of these Dirac masses
in the low-energy effective theory, the tachyonic waterfall mass $m$ is
always compensated (see Eq.~\eqref{eq:rbound}),
so that none of the meson fields ever becomes destabilized.
Because of that, the total superpotential in Eq.~\eqref{eq:Weffv} fails
to give rise to a waterfall transition and, even in
the low-energy vacuum, we retain the vacuum energy density
resulting from the nonzero Polonyi F-term, $V_0 = \left<\left|F_\Phi\right|\right>^2 = \mu^4$.
As anticipated in Sec.~\ref{subsec:setup}, this is a characteristic
feature of our construction, in which we intend to use one and the
same vacuum energy density for driving inflation and breaking supersymmetry.
Moreover, independent of the symmetry group under which the meson fields transform,
the absence of the waterfall transition automatically implies
that the end of inflation is not accompanied by the production
of topological defects.
This may be regarded as a significant phenomenological advantage
of our scenario over standard F-term hybrid inflation.


\subsection{Pseudomodulus potential in global supersymmetry}
\label{subsec:potential}


To the best of our knowledge, the superpotential in Eq.~\eqref{eq:Weffv}
has not been considered as the dynamical origin of inflation, so far.
Here, part of the reason certainly is that successful inflation based on
Eq.~\eqref{eq:Weffv} is bound to require a rather high SUSY breaking scale $\mu$.
As explained in the introduction, in supergravity, this necessitates a
large constant in the superpotential to cancel the CC in the true vacuum,
which then spoils slow-roll inflation (see Sec.~\ref{subsec:origin}).
In the rest of this paper, we will, however, show that 
the superpotential in Eq.~\eqref{eq:Weffv} can yield successful
Polonyi inflation, after all, if we generate the constant in
the superpotential only towards the end of inflation.
To this end, we shall now examine the effective one-loop potential
for the complex Polonyi scalar $\phi = \varphi/\sqrt{2}\,e^{i\theta}$
in global supersymmetry more closely.
In the next sections, we will then turn to the embedding
of the IYIT model into supergravity (see Sec.~\ref{subsec:SUGRA})
as well as to the generation of the constant term
in a separate hidden sector (see Sec.~\ref{sec:Rbreaking}).


The Yukawa interactions between the Polonyi field $\Phi$ and 
the meson fields $\Xi^0$ and $X^n$ in Eq.~\eqref{eq:Weff}
lead to radiative corrections to the Polonyi potential, $V_{\rm 1-loop}\left(\varphi\right)$,
that may be calculated according to the Coleman-Weinberg (CW) formula for
the effective one-loop potential~\cite{Coleman:1973jx}.
The details of our calculation may be found in Appendix~\ref{app:potential};
in the following, we will merely summarize our results.
Generally speaking, the effective potential may be divided into
two regimes:
(i) At large field values, all of the meson fields
acquire a large inflaton-dependent Majorana mass
$M\left(\varphi\right) =  \kappa_\Phi \left|\phi\right|$.
For $M\left(\varphi\right) \gg m_a$, the supersymmetric Dirac
masses $m_a$ in Eq.~\eqref{eq:Weff} are, hence, negligible and the effective potential
takes the usual logarithmic form as in standard F-term hybrid inflation,
$V_{\rm 1-loop}\left(\varphi\right) \propto m^4 \,\ln M\left(\varphi\right)$.
(ii) On the other hand, at small Polonyi field values, such that
$M\left(\varphi\right) \ll m_a$, the Dirac masses $m_a$ become more relevant.
Integrating out the ``heavy fields'' then leads
to a quadratic Polonyi potential around the origin,
$V_{\rm 1-loop}\left(\varphi\right) \propto m^2 M^2\left(\varphi\right)$.
We note that, as has been shown for the first time in~\cite{Chacko:1998si},
the effective potential around the origin has \textit{positive} curvature.
The low-energy vacuum at $\left<\Phi\right> = 0$ is, therefore, indeed stable.
Moreover, we find that, at large as well as at small field values, the effective
potential scales with the soft SUSY-breaking mass scale $m$, i.e., the smallest
mass scale in our model (see Eq.~\eqref{eq:rbound}).
This illustrates how supersymmetry succeeds
in protecting the inflaton potential from picking up too large
radiative corrections~\cite{Ellis:1982ed}.


To quantify the above statements, it is convenient to introduce the
following mass ratios,
\begin{align}
R_a\left(\varphi\right) = \frac{M\left(\varphi\right)}{m_a} \,, \quad
M\left(\varphi\right) = \kappa_\Phi \left|\phi\right| = 
\frac{\lambda}{\left(2-\zeta\right)^{1/2}}\frac{\varphi}{\sqrt{2}} \,.
\label{eq:Ra}
\end{align}
For large values of $R_a\left(\varphi\right)$, we then obtain a logarithmic one-loop
potential, while for small values of
$R_a\left(\varphi\right)$, the one-loop corrections take a quadratic form.
This behavior can be captured by studying the effective potential
as a function of a single order parameter $x\left(\varphi\right)$,
the geometric mean of all ratios $R_a\left(\varphi\right)$,
\begin{align}
x\left(\varphi\right) = \Big(\prod_a R_a\left(\varphi\right)\Big)^{1/N_X}
= \frac{M\left(\varphi\right)}{\overline{m}} \,, \quad 
\overline{m} = \Big(\prod_a m_a\Big)^{1/N_X} \,, \quad
N_X = 6 \,.
\label{eq:x}
\end{align}
Here, $N_X = 6$ counts the number of meson fields in the IYIT sector,
while $\overline{m}$ stands for the geometric mean of all explicit
mass parameters in Eq.~\eqref{eq:Weff}. 
In this sense, $\overline{m}$ denotes the ``supersymmetric mass
scale'' of the IYIT sector, i.e., a characteristic value for
the nonperturbatively generated Dirac masses in the low-energy effective theory.
In the following, we will set $\overline{m}$ to the dynamical scale $\Lambda$, for definiteness,
\begin{align}
\overline{m} = \Lambda \,.
\label{eq:mbar}
\end{align}
This mainly serves the purpose to account, in an effective way, for heavy
composite states with masses around the dynamical scale $\Lambda$
that we expect to be present, but which we can unfortunately not explicitly
describe in terms of our perturbative language at low energies.
\textit{Formally}, we can always set $\overline{m}$ to $\Lambda$
in our calculation by choosing an appropriate value for the effective heavy-flavor
Yukawa coupling $\tilde{\lambda}$,
\begin{align}
\overline{m} = m_0^{1/6} \,\widetilde{m}_{\phantom{0}}^{5/6} \,, \quad
\widetilde{m} = \tilde{\lambda}\,\frac{\Lambda}{\eta}  \,, \quad
\tilde{\lambda} = \left(\lambda_1\, \lambda_2\, \lambda_3\, \lambda_4\, \lambda_5\right)^{1/5} \,.
\end{align}
By fixing $\tilde{\lambda}$ at a nonperturbative
value, we are, therefore, able to enforce our designated value
for $\overline{m}$,
\begin{align}
\tilde{\lambda} = \left(\frac{r_0}{\lambda}\right)^{1/5} \eta^{6/5} \simeq \eta \,.
\end{align}

\begin{figure}
\centering
\includegraphics[width=0.65\textwidth]{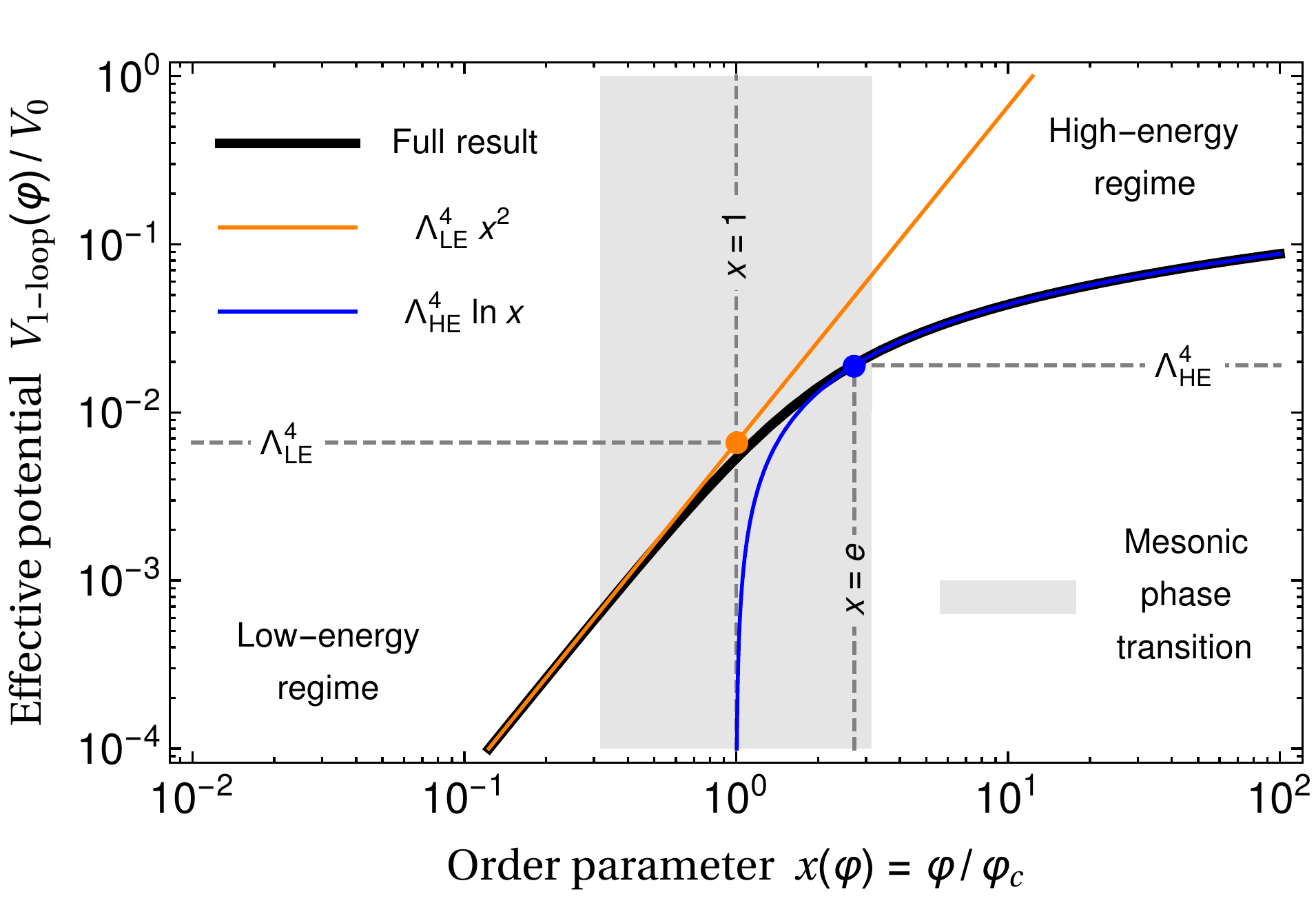}
\caption{Effective CW potential for the Polonyi field, $V_{\rm 1-loop}$,
as a function of  the order parameter $x$ for $\lambda = 1$,
see Eq.~\eqref{eq:V1loop}.
The potential energy scales $\Lambda_{\rm LE}$ and $\Lambda_{\rm HE}$
are given in Eq.~\eqref{eq:LambdaLEHE}.
For values $x\sim1$, we can only have limited trust in our perturbative result
because of potentially important strong-coupling effects at energies close
to the dynamical scale $\Lambda$.}
\label{fig:effpot}
\end{figure}

The value of $x\left(\varphi\right)$ indicates
whether the Dirac masses $m_a$ are negligible or not and, thus,
decides whether we are in the logarithmic or the quadratic part
of the effective potential.
The transition between both regimes takes place
at field values close to what we shall refer to as the critical
field value $\varphi_c$,
\begin{align}
x\left(\varphi_c\right) = 1  \quad\Leftrightarrow\quad
M\left(\varphi_c\right) = \overline{m} \quad\Rightarrow\quad
\varphi_c = \sqrt{2}\, \frac{\overline{m}}{\kappa_\Phi} \,.
\label{eq:phic}
\end{align}
This implies that the order parameter $x\left(\varphi\right)$
can also be regarded as the ratio of the actual and the critical
field value, $x\left(\varphi\right) = \varphi/\varphi_c$.
Far away from the critical field value, i.e., at
$x\left(\varphi\right) \ll 1$ and $x\left(\varphi\right) \gg 1$,
we now find the following expressions for the effective potential
(see Appendix~\ref{app:potential} for details),
\begin{align}
x\left(\varphi\right) \ll 1 \quad\Rightarrow\quad
V_{\rm 1-loop}^{\rm LE}\left(\varphi\right) & =
\frac{1}{2}\, m_{\rm eff}^2 \,\varphi^2 + \mathcal{O}\left(x^4\right) \,, 
\label{eq:V1loopLEHE}\\ \nonumber
x\left(\varphi\right) \gg 1 \quad\Rightarrow\quad
V_{\rm 1-loop}^{\rm HE}\left(\varphi\right) & =
\frac{N_X}{16\pi^2}\, m^4 \ln x\left(\varphi\right)
+ \mathcal{O}\left(x^{-4}\right) \,,
\end{align}
with $m_{\rm eff}$ denoting the effective one-loop mass
of the Polonyi field around the origin,%
\footnote{A similar expression has been derived for the first
time in~\cite{Chacko:1998si}.
Our result differs from the one in~\cite{Chacko:1998si} to the extent
that we allow for nonzero $\zeta$ (and, hence, nonzero $r_0$), which means
that we do not necessarily enforce the moduli constraint exactly.
The calculation in~\cite{Chacko:1998si}, on the other hand, is based on the
assumption that the moduli constraint \textit{is} fulfilled exactly,
so that $r_0 \equiv 0$.
For a recent derivation and discussion of the effective Polonyi
mass in $SU(4)$ language, see \cite{Domcke:2014zqa,Harigaya:2015soa}.
\label{fn:meff}}
\begin{align}
m_{\rm eff}^2 = \left(2\ln2-1\right) N_X^{\rm eff}\left(r_a\right)
\frac{\kappa_\Phi^2}{16\pi^2}\, m^2 \,, \quad 
N_X^{\rm eff}\left(r_a\right) =  \sum_a \omega\left(r_a\right) \,, \quad
\omega\left(r_a\right) \approx r_a^2 \,.
\label{eq:meff2}
\end{align}
Here, $N_X^{\rm eff}\left(r_a\right)$ counts the effective number of mesons
that contribute to the effective Polonyi mass.
The full functional form of $N_X^{\rm eff}\left(r_a\right)$ is a sum of
complicated loop factors $\omega\left(r_a\right)$.
To good approximation, these loop functions, however, happen to coincide
with the mass ratios $r_a$ squared, $\omega\left(r_a\right) \approx r_a^2$.
We can write the result in Eq.~\eqref{eq:V1loopLEHE} more compactly,
if we make use of the following two potential energy scales,
\begin{align}
\Lambda_{\rm LE}^4  = 
\sum_a \frac{m^4}{16\pi^2}\left(2\ln2-1\right)\left(\frac{\overline{m}}{m_a}\right)^2 =
\frac{1}{2}\, m_{\rm eff}^2\, \varphi_c^2 \,, \quad
\Lambda_{\rm HE}^4 = \sum_a \frac{m^4}{16\pi^2} = N_X\frac{m^4}{16\pi^2} \,.
\label{eq:LambdaLEHE}
\end{align}
The effective potential far away from the critical field value $\varphi_c$
then takes the following form (see Fig.~\ref{fig:effpot}),
\begin{align}
V_{\rm 1-loop}\left(\varphi\right) \approx 
\begin{cases}
\Lambda_{\rm LE}^4 \,\,x^2\left(\varphi\right) & ; \: x \ll 1 \\ 
\Lambda_{\rm HE}^4 \,\ln x\left(\varphi\right) & ; \: x \gg 1 
\end{cases} \,, \quad
x\left(\varphi\right) = \frac{\varphi}{\varphi_c} =
\frac{\left|\phi\right|}{\Lambda/\kappa_\Phi} \,.
\label{eq:V1loop}
\end{align}


The crucial question which we need to answer in the following is:
Can we use either the low-energy or the high-energy part of this effective
potential to realize successful Polonyi inflation?
Let us first investigate whether inflation might occur in the quadratic
part close to the origin.
As we know from standard chaotic inflation~\cite{Linde:1983gd},
the effective inflaton mass then needs to take a value of
$\mathcal{O}\left(10^{13}\right)\,\textrm{GeV}$ to ensure
the correct normalization of the scalar power spectrum.
This requires the coupling $\lambda$ to take
a value at least as large $\lambda \simeq 0.2$, since
otherwise the dynamical scale $\Lambda$ would have to be
super-Planckian,
\begin{align}
m_{\rm eff} \simeq 10^{13}\,\textrm{GeV} \left(\frac{\lambda}{0.2}\right)^3
\left(\frac{\Lambda}{M_{\rm Pl}}\right) + \mathcal{O}\left(\lambda^5\right) \,.
\label{eq:meff}
\end{align}
At the same time, we know that chaotic inflation requires
a large super-Planckian field excursion to yield
a sufficient number of $e$-folds.
The scalar perturbations probed in CMB observations, e.g.,
cross outside the Hubble horizon at a field value
$\varphi_* \sim 15\, M_{\rm Pl}$.
However, in the context of our SUSY breaking model, this large field range
does not ``fit'' into the low-energy part of the effective potential.
This follows from the fact that, for $\lambda \simeq 0.2$ and $\Lambda \simeq M_{\rm Pl}$,
the critical field value $\varphi_c$ only becomes as large as $\varphi_c \simeq 10\,M_{\rm Pl}$,
\begin{align}
\varphi_c \simeq 10 \, M_{\rm Pl} \left(\frac{0.2}{\lambda}\right)
\left(\frac{\Lambda}{M_{\rm Pl}}\right)
+ \mathcal{O}\left(\lambda\right) \,.
\end{align}
Therefore, to raise $\varphi_c$, so as to make the field
range required for chaotic inflation fit into the quadratic part of
the effective potential, $\varphi_c \gtrsim \varphi_*$,
we would have to go to smaller values of $\lambda$.
But then, we are either forced to push $\Lambda$ beyond the Planck
scale or we fail to reproduce the correct scalar spectral amplitude.
This eliminates the possibility of Polonyi inflation in the
low-energy part of the effective potential, which is why we will
focus on inflation in the logarithmic part of the effective potential from now on.


Before continuing, we, however, point out that inflation close to
the origin might become possible, after all, if we relax our assumptions.
That is, if we allowed for values of the dynamical
scale as large as, say, $\Lambda \sim \left(8\pi\right)^{1/2} M_{\rm Pl}$,
we would, in fact, be able to raise $\varphi_c$ above $\varphi_*$.
If we then trusted the full effective potential also at
field values close to $\varphi_c$ (see Fig.~\ref{fig:effpot}),
inflation in the transitioning regime between the quadratic
and the logarithmic part of the effective potential might become
feasible.
Such a scenario would promise to interpolate between the predictions
of chaotic inflation and F-term hybrid inflation, so that we would
expect it to result in interesting predictions for the tensor-to-scalar
ratio, $r \sim 0.1$.
Because of the uncertainties involved in such a scenario,
we, however, do not pursue this idea any further in this paper
and leave a more detailed study for future work.
In closing, we remark that a similar model of \textit{subcritical hybrid inflation},
based on a dynamically generated D-term~\cite{Domcke:2014zqa},
may be found in~\cite{Buchmuller:2014rfa}.
This model illustrates how to realize chaotic inflation after the
waterfall transition of \textit{D-term} hybrid inflation.


Let us now turn to the possibility of inflation in the logarithmic
part of the effective potential. 
As we will show, in this part of the potential,
successful Polonyi inflation is indeed feasible.
To be on the safe side, we will limit our analysis in the following
to field values that are larger than the critical field value by at least
half an order of magnitude, $\varphi \gtrsim 10^{0.5}\varphi_c$.
We do so because the effective Polonyi potential may receive
nonperturbative corrections around $\varphi\sim\varphi_c$ that
we do not have under control.
In fact, around the critical field value, the inflaton-dependent mass
$M\left(\varphi\right)$ drops below the dynamical scale $\Lambda$
(see Eq.~\eqref{eq:phic}).
This triggers the IYIT sector to transition
from the high-energy quark-gluon regime into the low-energy meson regime.
During this \textit{mesonic phase transition}, the IYIT quarks become confined
in the composite mesons and the strongly coupled
$SU(2)$ gauge group becomes completely broken by the nonzero
squark VEVs $\left<\Psi^1\Psi^2\right>$ and $\left<\Psi^3\Psi^4\right>$
that contribute to the meson VEV $\left<X^0\right>$
(see Eqs.~\eqref{eq:XS} and \eqref{eq:XS0vac}),%
\footnote{Here, $\left<\Psi^1\Psi^2\right> = 
\left<\Psi^3\Psi^4\right>$ follows from the fact
that the linear combination $X^5$ vanishes in the
vacuum, $\left<X^5\right> = 0$.\smallskip}
\begin{align}
\left<\Psi^1\Psi^2\right> = \left<\Psi^3\Psi^4\right> \neq 0 
\quad\Rightarrow\quad SU(2)\rightarrow\mathbbm{1} \,.
\end{align}
Thanks to the fact that the $SU(2)$ symmetry is spontaneously broken down to ``nothing'',
no topological defects are formed during the confining phase 
transition~\cite{Vilenkin:1984ib}.
We emphasize that this is an important phenomenological feature
of the IYIT model based on $SU(2)$, as it allows for the particular
breaking pattern $ SU(2)\rightarrow\mathbbm{1}$.
In summary, our scenario of Polonyi inflation, therefore,
crucially differs from ordinary F-term hybrid inflation in the following
respect:
While ordinary F-term hybrid inflation ends in a waterfall
transition---in the course of which the inflationary vacuum energy density
is ``eaten up'' by the FHI waterfall fields and which potentially leads to
the production of troublesome topological defects---our scenario of Polonyi
inflation undergoes a confining quark-meson phase transition that
\textit{conserves} the inflationary vacuum energy density and that does \textit{not}
lead to the production of topological defects.


\subsection{Embedding into supergravity and choice of the K\"ahler potential}
\label{subsec:SUGRA}


In supergravity, the flatness of the tree-level Polonyi potential in global supersymmetry,
$V\left(\varphi\right) = V_0 = \mu^4$, is not only lifted
by the radiative corrections in Eq.~\eqref{eq:V1loop}, but also by gravitational
corrections (see our discussion at the end of Sec.~\ref{subsec:ingredients}).
In our case, these SUGRA corrections turn out to be rather mild for basically
two reasons:
(i) Since the superpotential in Eq.~\eqref{eq:Weff} only contains terms
linear in $\Phi$, the tree-level SUGRA mass of the Polonyi field accidentally
cancels, as long as we assume a canonical K\"ahler potential.
As far as the embedding into supergravity is concerned, this represents an important
advantage of F-term hybrid inflation (and of our model) over alternative
models that do feature higher powers of the inflaton field in the superpotential.
(ii) Since we intend to realize Polonyi inflation in the logarithmic
part of the effective potential, we will consistently work with sub-Planckian
field values.
Our scenario of Polonyi inflation will, hence, turn out
to be a \textit{small-field} model of inflation.
Accordingly, the SUGRA corrections in our model are
bound to be less significant than in alternative \textit{large-field} models of inflation.


The total scalar potential for the complex
Polonyi field $\tilde\phi \subset \Phi$ in supergravity now reads,%
\footnote{The field $\tilde\phi$ is not necessarily canonically normalized,
which we indicate by placing a tilde on top of the symbol $\phi$.}
\begin{align}
V\big(\tilde\phi\big) = V_F\big(\tilde\phi\big)
+ V_{\rm 1-loop}\big(\big|\tilde\phi\big|\big) \,,
\end{align}
with $V_{\rm 1-loop}$ being given in Eq.~\eqref{eq:V1loop} and
where $V_F$ denotes the tree-level F-term potential in supergravity,
\begin{align}
V_F = \left|F\right|^2
- 3\, \exp\left[\frac{K}{M_{\rm Pl}^2}\right] \frac{\left|W\right|^2}{M_{\rm Pl}^2} \,.
\label{eq:VF}
\end{align}
Here, $\left|F\right|$ denotes the norm of the
generalized F-term vector $\mathcal{F}^i$ induced by the K\"ahler metric $K_i^{\:\bar{\jmath}}$,
\begin{align}
\left|F\right| = \left(\mathcal{F}\cdot\mathcal{F}^*\right)^{1/2} \,,\quad
\mathcal{F}\cdot\mathcal{F}^* = 
\mathcal{F}^i K_i^{\:\bar{\jmath}} \mathcal{F}_{\bar{\jmath}}^* \,, \quad
K_i^{\:\bar{\jmath}} =
\frac{\partial^2 K}{\partial \phi^i\partial \phi_{\bar{\jmath}}^*}  \,.
\end{align}
The individual components of the the F-term vector
$\mathcal{F}^i$ are proportional to the conjugate the of
K\"ahler-covariant derivatives of the superpotential, $\left(D_j W\right)^*$,
multiplied by the inverse of the K\"ahler metric, $K_{\:\bar{\jmath}}^i$,
\begin{align}
\mathcal{F}^i = -K_{\:\bar{\jmath}}^i \,
\exp\left[\frac{K}{2 M_{\rm Pl}^2}\right] \left(D_j W\right)^* \,, \quad
K_{\:\bar{\jmath}}^i = \left(K^{-1}\right)_{\:\bar{\jmath}}^i\,, \quad
D_i W = \frac{\partial W}{\partial \phi^i} +
\frac{W}{M_{\rm Pl}^2}\frac{\partial K}{\partial \phi^i} \,.
\label{eq:Fivector}
\end{align}
After integrating out the heavy fields $\Xi^0$, $X^n$, $\Sigma$, and $S_n$
in Eq.~\eqref{eq:Weff}, the effective superpotential of the IYIT sector
reduces to the SUSY-breaking tadpole term for the Polonyi field
(see Eq.~\eqref{eq:WPolonyi}),
\begin{align}
W_{\rm eff} \simeq \mu^2\, \Phi \,.
\label{eq:Wtadpole}
\end{align}


Meanwhile, the tree-level K\"ahler potential $K$ is
not unambiguously defined.
At first sight, there are several well motivated choices for $K$
that come into question. 
For instance, we might think that an approximately shift-symmetric
K\"ahler potential could help in keeping the SUGRA corrections small,%
\footnote{A shift symmetry in the direction of the Polonyi field can never
be an exact symmetry, as it is always
explicitly broken by the superpotential, $W_{\rm eff} \simeq \mu^2 \,\Phi$,
and the radiative corrections in the scalar potential
(see Eq.~\eqref{eq:V1loop}).
For this reason, we expect that, also in the tree-level K\"ahler potential, the shift
symmetry should be explicitly broken.}
\begin{align}
K = \pm \frac{1}{2} \left(\Phi \pm \Phi^\dagger\right)^2 \mp
\frac{\epsilon}{2} \left(\Phi \mp \Phi^\dagger\right)^2 +
\mathcal{O}\left(\epsilon^2,M_{\rm Pl}^{-2}\right) \,, \quad \epsilon \ll 1 \,.
\end{align}
Depending on the sign choice, the real-valued (not canonically normalized)
inflaton field $\tilde\varphi$ is then identified either as
the real ($-$) or the imaginary ($+$)
part of the complex scalar $\tilde\phi \subset \Phi$.
Likewise, the real-valued  (not canonically normalized) scalar
inflaton partner $\tilde\sigma$ (i.e., the ``sinflaton'')
corresponds to the imaginary ($-$) or the real ($+$) part
of the complex scalar $\tilde\phi\subset \Phi$.
As long as we do not assume an exact shift symmetry in the K\"ahler potential
(i.e., $\epsilon \equiv 0$), $\tilde\varphi$ and $\tilde\sigma$
need to be canonically normalized,
\begin{align}
\varphi = \left(1+\epsilon\right)^{1/2} \tilde\varphi \,, \quad
\sigma = \left(1+\epsilon\right)^{1/2} \tilde\sigma \,.
\end{align}
Unlike in the case of a canonical K\"ahler potential,
the SUGRA F-term potential $V_F$ now contains nonzero tree-level
masses  for the inflaton and its scalar partner.
At the origin in field space, we find
\begin{align}
m_\varphi^2 = - \frac{1-\epsilon}{\left(1+\epsilon\right)^2}\frac{3\,\mu^4}{M_{\rm Pl}^2} \,, \quad
m_\sigma^2 = + \frac{1-\epsilon}{\left(1+\epsilon\right)^2}\frac{3\,\mu^4}{M_{\rm Pl}^2} \,.
\label{eq:mphisigma}
\end{align}
For $\epsilon \ll 1$, the inflaton direction is, hence, tachyonic!
If we forget about the radiative corrections $V_{\rm 1-loop}$ for a moment, this
immediately implies that Polonyi inflation based on an approximately
shift-symmetric K\"ahler potential is impossible---which
completes our argument in Sec.~\ref{subsec:original} (see Footnote~\ref{fn:shift}).
On the other hand, taking the radiative corrections into account,
we may hope that the effective inflaton mass $m_{\rm eff}$
around the origin could possibly compensate the tachyonic
tree-level mass $m_{\varphi}$
and, thus, stabilize the vacuum at $\varphi = 0$.
It is clear that this will not be easy, since
$m_{\varphi}$ in Eq.~\eqref{eq:mphisigma} is typically very large,
\begin{align}
\left|m_\varphi\right| \simeq 8 \times 10^{15}\,\textrm{GeV} \left(\frac{\lambda}{0.2}\right)
\left(\frac{\Lambda}{M_{\rm Pl}}\right)^2
+ \mathcal{O}\left(\epsilon,\lambda^3\right) \,.
\end{align}
In order to sufficiently suppress it,
we need to go to smaller values of the dynamical scale $\Lambda$.
Requiring that the total inflaton mass at the origin be non-tachyonic,
$m_\varphi^2 + m_{\rm eff}^2 > 0$, we find (see Eq.~\eqref{eq:meff}),
\begin{align}
\Lambda \lesssim 3 \times 10^{15} \,\textrm{GeV} 
\left(\frac{\lambda}{0.2}\right)^2 \,. 
\end{align}
For such small values of the dynamical scale, the critical field value
$\varphi_c$ is clearly sub-Planckian (see Eq.~\eqref{eq:phic}).
As discussed in the previous section, this eliminates the possibility
of realizing inflaton in the quadratic part of the potential close to the origin.
Our only remaining hope, therefore, is to realize inflation at field
values $\varphi \gtrsim \varphi_c$.
To this end, it is important that the total scalar potential does not
exhibit a local minimum above the critical field value---otherwise
the inflaton will never reach the origin and inflation would never end.
But as it turns out, exactly such a local minimum is always present.


The point is that the quadratic part of the effective potential
can only compensate the tachyonic tree-level mass up to field values
of $\mathcal{O}\left(\varphi_c\right)$.
At the same time, the gradient of the tree-level potential only 
turns positive at very large field values, $\varphi \geq \varphi_{\rm min} \propto M_{\rm Pl}/\epsilon$,
where $\varphi_{\rm min}$ denotes the position of a global AdS vacuum.
As long as $\varphi_c \lesssim \varphi_{\rm min}$, the gradient
of the total inflaton potential is, therefore, guaranteed to turn
negative in between $\varphi_c$ and $\varphi_{\rm min}$---which
indicates the presence of a local minimum.
To evaluate the relation between $\varphi_c$ and $\varphi_{\rm min}$
more precisely, consider the tree-level inflaton potential for $\sigma = 0$,
\begin{align}
V_F\left(\varphi\right) = \exp\left[\frac{\epsilon}{1+\epsilon}\frac{\varphi^2}{M_{\rm Pl}^2}\right]
\left(v_0 + \frac{1}{2}\, M_\varphi^2\, \varphi^2
+ \frac{\lambda_\varphi}{4!}\, \varphi^4\right) \,,
\end{align}
with the three parameters $v_0$, $M_\varphi$, and $\lambda_\varphi$ being given as follows,
\begin{align}
v_0 = \frac{\mu^4}{1+\epsilon} \,, \quad
M_\varphi^2 = - \frac{3-\epsilon}{1+\epsilon}\frac{v_0}{M_{\rm Pl}^2} \,, \quad
\lambda_\varphi = 24 \left(\frac{\epsilon}{1+\epsilon}\right)^2 \frac{v_0}{M_{\rm Pl}^4} \,.
\end{align}
The local minimum in the SUGRA tree-level potential, $\varphi_{\rm min}$,
is then located at
\begin{align}
\varphi_{\rm min} = \left(\frac{1+\epsilon}{\epsilon}\right)^{1/2}
\left[\left(1+\textrm{sgn}\left(\epsilon\right)\right)a - 1\right]^{1/2} M_{\rm Pl} \,, \quad
a = -\frac{\epsilon}{1+\epsilon}\frac{6}{\lambda_\varphi}\frac{M_\varphi^2}{M_{\rm Pl}^2} \,,
\end{align}
so that the AdS vacuum energy density takes the following value,
\begin{align}
V_{\rm min} = - \exp\left[\frac{\epsilon}{1+\epsilon}
\frac{\varphi_{\rm min}^2}{M_{\rm Pl}^2}\right]
\left(\frac{3-\epsilon}{2\left|\epsilon\right|}-2\right) v_0 \,.
\end{align}
Given the above expression for $\varphi_{\rm min}$, we find that, in the entire
parameter space of interest ($\lambda/\eta \lesssim 10^{-0.5}$ and
$\left|\epsilon\right| \lesssim 10^{-0.5}$), $\varphi_c$
and $\varphi_{\rm min}$ are always separated by at least one order of magnitude,
$\varphi_c / \varphi_{\rm min} \lesssim 0.1$.
In the shift-symmetric limit, $\epsilon\rightarrow 0$, the AdS vacuum
disappears, in particular, altogether, $\varphi_{\rm min} \rightarrow \infty$,
so that the total inflaton potential becomes unbounded from below.
In summary, this shows that the effective potential
manages to compensate the tachyonic tree-level mass only for too small
a field range.
It only ``bends around'' the tachyonic part of the tree-level potential
up to $\varphi \sim \varphi_c$, while it
should actually do this up to field values that are at least
ten times as large, $\varphi \sim \varphi_{\rm min} \gtrsim 10\, \varphi_c$.
This finally completes our argument that Polonyi inflation is incompatible with
an approximate shift symmetry in the K\"ahler potential,
independent of whether we account for the presence of radiative corrections or not.%
\footnote{Note that this conclusion does not hold for ordinary F-term hybrid
inflation.
As has been shown in~\cite{Pallis:2014xva}, F-term hybrid inflation
can, in fact, be successfully embedded into supergravity based on a shift-symmetric
K\"ahler potential---provided that supersymmetry is broken at a high scale and
\textit{broken in a different sector}.
That is, once the inflaton is \textit{not} identified as the SUSY-breaking
Polonyi field, the inflaton may receive a large soft mass that compensates
its tachyonic tree-level mass.}


This result leaves a near-canonical tree-level K\"ahler potential as the
best possible choice,
\begin{align}
K = \Phi^\dagger \Phi + \frac{\epsilon}{\left(2!\right)^2}
\left(\frac{\Phi^\dagger \Phi}{M_{\rm Pl}}\right)^2 + 
\mathcal{O}\left(\epsilon^2,M_{\rm Pl}^{-4}\right) \,.
\label{eq:Kphi}
\end{align}
Of course, simply assuming a canonical K\"ahler potential, $K=\Phi^\dagger\Phi$,
would represent an even more minimal choice.
In that case, the Polonyi field would not ``equip itself''
with a Hubble-induced mass (see our discussion at the beginning of this section)
and we would not have to fear any dangerous SUGRA corrections.
But such a scenario is rather unlikely.
Terms featuring higher powers of $\Phi^\dagger\Phi$
are not forbidden by any symmetry and, hence, we cannot exclude
their presence in the K\"ahler potential.
Quite the contrary, we rather expect that they are the unavoidable
consequence of radiative corrections in quantum gravity that manifest themselves
as Planck-suppressed operators in the low-energy effective theory.
The only open question regarding the higher-dimensional terms
in the K\"ahler potential pertains to the size of their coefficients.
The leading correction to the canonical K\"ahler potential, e.g.,
results in an inflaton mass squared of
$m_\varphi^2 = -3\,\epsilon\, H_{\rm inf}^2$,
which shifts the slow-roll parameter $\eta$ by $\Delta\eta = -\epsilon$.
Therefore, if $\epsilon$ is of $\mathcal{O}(1)$, as one might naively 
expect, we would encounter the $\eta$ problem~\cite{Dine:1983ys}.
This tells us that the coefficient $\epsilon$ needs to be
suppressed, $\epsilon \lesssim 0.1$, which introduces
a parameter fine-tuning at the level of $1\cdots10\,\%$.
In the following, we will not speculate what the physical reason for this
fine-tuning might be and simply assume that $\epsilon$ is sufficiently small.
We merely remark that there is a number of models
in the literature that attempt to explain why $\epsilon$
ends up being suppressed.
For instance, a small coefficient $\epsilon$ may be the result of a sub-Planckian
cut-off scale, $M_* \lesssim 0.1\, M_{\rm Pl}$, in the sector giving rise to the
higher-dimensional term in Eq.~\eqref{eq:Kphi}~\cite{Watari:2000jh}
(see~\cite{Watari:2004xh} for a realization of this idea in string theory); 
or $\epsilon$ may be small because it is, for one reason or another,
suppressed by NDA factors of $\mathcal{O}\left(4\pi\right)$~\cite{Craig:2008tv}.


Having defined the K\"ahler potential, we now need to canonically normalize
the complex Polonyi field $\tilde\phi = \tilde\varphi/\sqrt{2}\,e^{i\tilde\theta}$.
Here, we will restrict ourselves to normalizing the radial component
$\tilde\varphi$ only, as the complex phase $\tilde\theta$ will be irrelevant
during inflation, anyway.
The canonically normalized field $\varphi$ is then given as
\begin{align}
\varphi\left(\tilde\varphi\right) =
\int d\tilde\varphi \left(\frac{\partial^2 K}{\partial\tilde\phi\,\partial\tilde\phi^*}\right)^{1/2}
= \tilde\varphi \left[1 + \frac{\epsilon}{12}\left(\frac{\tilde\varphi}{M_{\rm Pl}}\right)^2 +
\mathcal{O}\left(\epsilon^2\right)\right] \,.
\end{align}
This relation is readily inverted, which provides us with an expression for
$\tilde\varphi$ in terms of $\varphi$,
\begin{align}
\tilde\varphi\left(\varphi\right) =
\varphi \left[1 - \frac{\epsilon}{12}\left(\frac{\varphi}{M_{\rm Pl}}\right)^2 +
\mathcal{O}\left(\epsilon^2\right)\right] \,.
\end{align}
We are now in the position to
write down the full tree-level SUGRA potential $V_F$ for $\varphi$ and $\tilde\theta$.
In doing so, we shall first use the full Polonyi superpotential in Eq.~\eqref{eq:WPolonyi}
for illustrative purposes (i.e., we will add a constant $w$ to the
effective superpotential, although $w = 0$ in the IYIT model).
Eq.~\eqref{eq:VF} then yields
\begin{align}
V_F\big(\varphi,\tilde\theta\big) = c_0
+ c_1 \,\varphi\cos\tilde\theta
+ \frac{c_2}{2}\,\varphi^2
+ \frac{c_3}{3!}\,\varphi^3 \cos\tilde\theta
+ \frac{c_4}{4!}\,\varphi^4
+ \mathcal{O}\left(\varphi^5\right) \,,
\label{eq:VFw}
\end{align}
where the coefficients $c_0$, $c_1$, $c_2$, $c_3$, and $c_4$ are given as follows,%
\footnote{A dimension-6 operator in the K\"ahler
potential, $K \supset \epsilon'/\left(3!\right)^2\left(\Phi^\dagger\Phi\right)^3/M_{\rm Pl}^4$,
only contributes to the coefficient of the quartic inflaton term,
$c_4 \rightarrow c_4 - 3/2\, \epsilon' V_0 / M_{\rm Pl}^4$.
This effect is negligibly small, which is why we will set $\epsilon'= 0$ in the following.}
\begin{align}
c_0 & = V_0 - 3\,\frac{w^2}{M_{\rm Pl}^2} \,, \quad
c_1 = - 2\sqrt{2}\, \frac{w}{M_{\rm Pl}^2} V_0^{1/2}  \,, \quad
c_2 = - \epsilon\,\frac{V_0}{M_{\rm Pl}^2} - 2\,\frac{w^2}{M_{\rm Pl}^4} \,,
\label{eq:c01234}\\ \nonumber
c_3 & = - 3\sqrt{2} \left(1+\frac{\epsilon}{6}\right)\frac{w}{M_{\rm Pl}^4} V_0^{1/2} \,, \quad
c_4 = 3 \left(1 - \frac{7\,\epsilon}{2} + \frac{8\,\epsilon^2}{3}\right) \frac{V_0}{M_{\rm Pl}^4}
-3\left(1+\frac{\epsilon}{6}\right)\frac{w^2}{M_{\rm Pl}^6} \,.
\end{align}
This form of the scalar potential illustrates the impact of 
the constant term in the superpotential, $W\supset w$.
Not only does $w$ induce a dependence on 
the phase $\tilde\theta$ of the complex inflaton field
(thereby introducing odd powers of the radial component $\varphi$ in
the scalar potential), it also results in a large tachyonic tree-level mass.
In Appendix~\ref{subsec:canonical}, we show that this potential
is always too steep for slow-roll inflation.
Moreover, we note that the potential in Eq.~\eqref{eq:VFw} is
similar to the tree-level potential of F-term hybrid inflation.
Eq.~\eqref{eq:VFw} yields the  SUGRA FHI tree-level potential
up to $\mathcal{O}\left(\varphi^4\right)$,
if we make the following replacements,
\begin{align}
c_0 \rightarrow c_0 + 3\,\frac{w^2}{M_{\rm Pl}^2} \,, \quad
c_1 \rightarrow c_1                                 \,, \quad
c_2 \rightarrow c_2 + 3\,\frac{w^2}{M_{\rm Pl}^4}   \,, \quad
c_3 \rightarrow c_3                                 \,, \quad 
c_4 \rightarrow c_4 + 9\left(1-\frac{\epsilon}{6}\right)\frac{w^2}{M_{\rm Pl}^6} \,,
\end{align}
where the constant $w$ is to be understood as a measure for the
gravitino mass, $w = m_{3/2}\,M_{\rm Pl}^2$.
In F-term hybrid inflation, the functional dependence of the scalar potential
on the phase of the complex inflaton field is, therefore, exactly the same
as in the full Polonyi model (see \cite{Buchmuller:2014epa} for the first
analysis that properly treats F-term hybrid inflation as a two-field model in the complex plane).
By contrast, the soft inflaton mass is no longer tachyonic in F-term hybrid inflation,
but identical to the gravitino mass.
Meanwhile, the change in the coefficient $c_4$ is suppressed by
$m_{3/2}^2/M_{\rm Pl}^2$ and, thus, irrelevant.
Furthermore, it is worthwhile pointing out that
F-term hybrid inflation is \textit{not} compatible with the idea of a gravitino
mass as large as the inflationary Hubble rate.
Instead, $m_{3/2}$ and $H_{\rm inf}$ must always maintain a hierarchy
of at least three orders of magnitude, $m_{3/2} \lesssim 10^{-3} \,H_{\rm inf}$,
since otherwise slow-roll inflation becomes spoiled
by too large SUGRA corrections~\cite{Buchmuller:2014epa}.
This is in contrast to our scenario of Polonyi inflation,
where $m_{3/2}$ and $H_{\rm inf}$ are related to each other at
a fundamental level and, in fact, equal to each other (see Eq.~\eqref{eq:m32Hinf}).


Let us now turn to the scalar potential of $\textit{our}$ inflationary scenario.
For our purposes, the constant $w$ needs to be zero during inflation.
The total inflaton potential for Polonyi inflation, therefore, follows from the sum 
of the logarithmic one-loop potential in Eq.~\eqref{eq:V1loop} and
$V_F$ in Eq.~\eqref{eq:VFw} evaluated for $w = 0$,
\begin{align}
V\left(\varphi\right) = V_0 - \frac{\epsilon}{2} \frac{V_0}{M_{\rm Pl}^2} \varphi^2
+ \frac{1}{8}\left(1 - \frac{7\epsilon}{2} + \frac{8\,\epsilon^2}{3}\right)
\frac{V_0}{M_{\rm Pl}^4} \varphi^4
+ \Lambda_{\rm HE}^4 \,\ln\left(\frac{\varphi}{\varphi_c}\right)
+ \mathcal{O}\left(x^{-4},\varphi^6\right) \,.
\label{eq:infpot}
\end{align}
This inflaton potential is identical to the potential of F-term hybrid inflation
(including corrections from supergravity and a noncanonical K\"ahler potential)
in the limit of a vanishing gravitino mass~\cite{BasteroGil:2006cm}.


\section{Spontaneous \boldmath{$R$} symmetry breaking after the end of inflation}
\label{sec:Rbreaking}


\subsection[Implications of broken $R$ symmetry for $R$ parity and the gravitino mass]
{Implications of broken \boldmath{$R$} symmetry for \boldmath{$R$} parity and the gravitino mass}


In the previous section, we have seen what it takes
to identify the SUSY-breaking Polonyi field $\Phi$ as the chiral inflaton field.
We considered three different types of radiative corrections to the scalar
Polonyi potential (none, quadratic corrections, logarithmic corrections), two different
shapes of the tree-level K\"ahler potential (approximately shift-symmetric
and near-canonical), and two different choices for the constant in the superpotential
($w = 0$ and $w\neq0$)---and we found only one possible combination of ``model ingredients''
that could not be ruled out immediately: the Polonyi model supplemented by \textit{logarithmic}
radiative corrections in combination with a \textit{near-canonical}
K\"ahler potential and $\mathit{w = 0}$ during inflation.


In the following, we will focus on this scenario and and present a dynamical model that
illustrates how the constant in the superpotential, $W \supset w$,
may be generated at the end of inflation.
As we will see, the constant $w$ then represents (not the only one, but) the
dominant contribution to the VEV of the superpotential in the
low-energy vacuum, $\left<W\right> \simeq w$.
Consequently, $w$ turns out to be the main source of spontaneous $R$ symmetry breaking
at low energies.
This is the reason why, throughout the entire paper, we refer to \textit{the generation
of the constant term in the superpotential} after inflation also as the
\textit{spontaneous breaking of $R$ symmetry at late times}.
Technically speaking, this is not quite correct, as, globally, $R$ symmetry
is already spontaneously broken during inflation by the VEV of the
Polonyi field,
\begin{align}
\left<\Phi\right> \neq 0 \quad\Rightarrow\quad Z_4^R \rightarrow Z_2^R \,.
\end{align}
This is because $\Phi$ carries $R$ charge $2$.
What we actually mean by speaking of \textit{late-time $R$ symmetry breaking}
is, therefore, not the spontaneous breaking of $R$ symmetry from a global perspective,
but only the spontaneous breaking of $R$ symmetry in the sector responsible for
the generation of the constant term $w$.
From the perspective of the low-energy effective theory, the process of
late-time $R$ symmetry breaking then amounts
to the (explicit) breaking of $R$ symmetry at the level of the Lagrangian
by means of a constant in the superpotential,
as opposed to the (spontaneous) breaking of $R$ symmetry
at the level of a scalar field VEV during inflation.
Here, it is important to note that the latter kind of $R$ symmetry breaking mostly
vanishes in the true vacuum after inflation, whereas the
former kind of $R$ symmetry also remains at low energies.
Likewise, we shall refer to the sector responsible for
the generation of the constant $w$ (i.e., the sector responsible for
$R$ symmetry breaking at low energies) as the \textit{$R$ symmetry-breaking sector}
for short.
We also point out that the temporary breaking of $R$ symmetry by
the VEV of the inflaton field is, in fact, mandatory
from a phenomenological point of view.
Otherwise, the spontaneous breaking of $R$ symmetry during the generation of
the constant $w$ would result in the production of domain walls at the end
of inflation, which would render our scenario phenomenologically
unviable~\cite{Dine:2010eb}.%
\footnote{If the inflationary dynamics did not break $R$ symmetry
(in the context of some \textit{other} model), $R$ symmetry would need to
be broken in a separate sector way before  the end of inflation,
to make sure that all domain walls are sufficiently diluted.
Such an alternative scenario provides a different explanation for why supersymmetry
is necessarily broken at a high scale~\cite{Harigaya:2015yla}.}


Similarly to the Polonyi VEV, also the constant $w$
breaks the $Z_4^R$ symmetry down to a $Z_2^R$ parity,
\begin{align}
w\neq 0 \quad\Rightarrow\quad Z_4^R \rightarrow Z_2^R \,,
\end{align}
so that, during inflation as well as in the low-energy vacuum after inflation,
the only discrete symmetry that remains globally
unbroken corresponds to an exact $Z_2^R$.
Interestingly enough, this parity is suited to be identified as the
$R$ parity of the MSSM, $P_R$~\cite{Buchmuller:1982ye}.
Here, note that $R$ parity is not a proper $R$ symmetry in the actual sense,
as it is equivalent to matter parity, $P_M$, which is a non-$R$ symmetry~\cite{Dine:2009swa}.
Our model, therefore, automatically offers a possible explanation for the origin of
$R$ parity in the MSSM.
In contrast to other models, it does not depend on extra continuous
symmetries, such as a global or local $U(1)_{B-L}$, that would leave behind
$P_M \cong P_R$ as a discrete remnant subgroup after spontaneous symmetry
breaking.


Given the above sources of $R$ symmetry breaking during and
after inflation, the physical gravitino mass $m_{3/2}$
varies as a function of time.
In general, $m_{3/2}$ is controlled by the VEV of the superpotential,
\begin{align}
m_{3/2} = \exp\left[\frac{\left<K\right>}{2M_{\rm Pl}^2}\right]
\frac{\left<W\right>}{M_{\rm Pl}^2}
\,, \quad W = W_{\rm eff} + w \,.
\label{eq:m32}
\end{align}
During inflation, the constant $w$ vanishes and, thus,
$m_{3/2}$ turns out to be roughly proportional to the Polonyi field value.
Meanwhile, after inflation, $m_{3/2}$ is dominated
by the contribution from the constant $w$.
In order to tune the CC to zero, we have to require that
$m_{3/2}$ eventually takes the following value,
\begin{align}
m_{3/2} = \frac{\Lambda_{\rm SUSY}^2}{\sqrt{3}\, M_{\rm Pl}} \,, \quad
\Lambda_{\rm SUSY} = \left<\left|F\right|\right>^{1/2} \,.
\end{align}
This condition can always be satisfied by fine-tuning the constant $w$
to the particular value $w_0$,
\begin{align}
w_0 = \exp\left[-\frac{\left<K\right>}{2M_{\rm Pl}^2}\right]
\frac{M_{\rm Pl} }{\sqrt{3}}\left<\left|F\right|\right>_{w=w_0}-\left<W_{\rm eff}\right> \,.
\label{eq:w0}
\end{align}
Here, it is important to note that the SUSY-breaking F-term $\left|F\right|$ also depends
on the constant $w$ via the covariant derivative of the superpotential
(see Eq.~\eqref{eq:Fivector}).
For this reason, the above relation is, in fact, only an implicit definition of $w_0$,
which, \textit{in principle}, still needs to be solved for $w_0$.
\textit{In practice}, the dependence of $\left|F\right|$ on $w$ is,
however, negligible in most cases, so that the right-hand
side of Eq.~\eqref{eq:w0} readily yields the required value for $w$.
In the following, we shall now show how the constant $w$ may be generated
after inflation (see Secs.~\ref{subsec:Rbreakingsector} and \ref{subsec:trigger})
and discuss under which conditions it may be successfully matched
with the desired value $w_0$, so that the CC vanishes at low energies
(see Sec.~\ref{subsec:backreaction}).


\subsection{Gaugino condensation in a mass-deformed strongly coupled hidden sector}
\label{subsec:Rbreakingsector}


The simplest way to break $R$ symmetry via strong dynamics is to make use of
dynamical gaugino condensation in a pure SYM theory~\cite{Veneziano:1982ah}.
For instance, in a strongly coupled pure SYM theory based on $SU(N_c)$
and associated with a dynamical scale $\Lambda'$, gaugino
condensation results in an $R$ symmetry-breaking constant
$w$ of $\mathcal{O}\big(\Lambda^{\prime\,3}\big)$.
Let us now illustrate how we can use this property for our purposes.


Our starting point is supersymmetric quantum chromodynamics (SQCD): a
strongly coupled $SU(N_c)$ gauge theory with $N_c$ colors and $N_f<3 N_c$ flavors,
where each flavor consists of a quark/antiquark pair $\left\{Q^i,\bar{Q}^i\right\}$.
We will constrain the viable values for $N_c$ and $N_f$ shortly.
For the moment, however, let us stay as general as possible
and leave the concrete values of $N_c$ and $N_f$ unspecified.
We assume that all quark/antiquark pairs are \textit{a priori}
massless, i.e., we do \textit{not} introduce any bare mass terms
of the form $m_i\, Q^i\bar{Q}^i$ in the tree-level superpotential.
Instead, we suppose that all of the $N_f$ flavors share a Yukawa
coupling with some $SU(N_c)$ singlet field $P$, which, thus,
results in \textit{field-dependent} masses for all flavors,
\begin{align}
W^R =  c_i\, P \, Q^i \bar{Q}^i \,.
\label{eq:WR}
\end{align}
Here, the coefficients $c_i$ denote dimensionless Yukawa couplings
of $\mathcal{O}\left(1\right)$.
By varying the VEV of the singlet field $P$, we are then able to control
the quark mass matrix, $M_Q$, in our SQCD theory,
\begin{align}
M_Q = \textrm{diag} \left\{M_{Q^i}\right\} =
\textrm{diag} \left\{c_i\left<P\right>\right\} \,.
\end{align}


As long as $P$ is stabilized at the origin, $\left<P\right> = 0$,
all quarks are massless and the SQCD sector remains what it is:
an $SU(N_c)$ gauge theory with $N_f$ dynamical flavors, described
by a quantum moduli space at low energies~\cite{Seiberg:1994bz}.%
\footnote{During inflation, the scalar quark/antiquark
fields acquire Hubble-induced masses.
In the special case of $N_f < N_c$, these stabilize the runaway
directions on the quantum moduli space induced by the
nonperturbative ADS superpotential~\cite{Affleck:1983rr,Affleck:1983mk}.}
On the other hand, once $P$ acquires a large VEV,
all flavors become heavy and can be integrated out.
This mass deformation transforms the SQCD sector into
a pure SYM theory, in which $R$ symmetry is spontaneously
broken:
First, $R$ symmetry is broken down to $Z_{2 N_c}^R$ due to $SU(N_c)$ instantons.
Then, the discrete $Z_{2 N_c}^R$ symmetry is broken further down to
$Z_2^R$ via gaugino condensation.
This results in the eagerly anticipated constant term $w$ in the
effective superpotential,
\begin{align}
W_{\rm eff}^R  = w = \frac{1}{\tilde{\eta}^2}\,N_c\,\tilde{\Lambda}_{\rm eff}^3 \,.
\label{eq:WeffR}
\end{align}
Here, $\tilde\eta$ is a dimensionless ``fudge factor'' that encompasses all numerical factors
entering into the expression for $w$ except for $N_c$ and $\tilde{\Lambda}_{\rm eff}^3$.
Based on NDA, we again expect $\tilde\eta$ to be of $\mathcal{O}\left(4\pi\right)$,
which is why we will set $\tilde\eta = 4\pi$ in the following, for definiteness.
Meanwhile, $\tilde{\Lambda}_{\rm eff}$ denotes the effective dynamical scale
of the SYM theory after integrating out all heavy flavors.
We obtain an expression for $\tilde{\Lambda}_{\rm eff}$ by matching the
running of the $SU(N_c)$ gauge coupling constant $\tilde{g}$
at the respective heavy-quark mass thresholds,
\begin{align}
\tilde{\Lambda}_{\rm eff}^{3N_c} = \overline{M}_Q^{N_f} \,\tilde\Lambda^{3N_c-N_f} \,,
\label{eq:GCscale}
\end{align}
where $\tilde\Lambda$ denotes the dynamical scale of the original
high-energy theory and where $\overline{M}_Q$ represents the effective quark mass
scale, i.e., the geometric mean of all quark mass eigenvalues $M_{Q^i}$,
\begin{align}
\overline{M}_Q = \Big(\prod_i M_{Q^i}\Big)^{1/N_f} = \left(\det M_Q\right)^{1/N_f}
= \bar{c} \, \left<P\right> \,, \quad 
\bar{c} = \Big(\prod_i c_i\Big)^{1/N_f} \,.
\end{align}
In the following, we will set $\bar{c}$ (the geometric mean of all Yukawa
couplings $c_i$) to $\bar{c} = 1$, for simplicity.
For given values of $N_c$ and $N_f$, the constant $w$ in
Eq.~\eqref{eq:WeffR} then ends up being a function of
$\left<P\right>$ and $\tilde\Lambda$,
\begin{align}
w = \frac{N_c}{16\pi^2} \left<P\right>^{N_f/N_c} \tilde\Lambda^{3-N_f/N_c} \,.
\label{eq:w}
\end{align}
Here, $\left<P\right> = \overline{M}_Q$ must be larger than
$\tilde\Lambda$ to ensure that the heavy quark flavors can be
integrated out perturbatively in the high-energy theory.
Our result for $w$ in Eq.~\eqref{eq:w} is, therefore, bounded from above,
\begin{align}
\tilde\Lambda \lesssim \left<P\right> \quad\Rightarrow\quad
w \lesssim w_{\rm max} = \frac{N_c}{16\pi^2} \left<P\right>^3 \,.
\label{eq:wmax}
\end{align}
The dynamical scale of the high-energy theory, $\tilde\Lambda$, is generated
via dimensional transmutation and, hence, solely depends on the value of the $SU(N_c)$
gauge coupling constant $\tilde{g}$ at the Planck scale,
\begin{align}
\tilde{\Lambda} = M_{\rm Pl} \exp\left[-\frac{8\pi^2}{b}
\frac{1}{\tilde{g}^2\left(M_{\rm Pl}\right)}\right] \,, \quad
b = 3\, N_c - N_f \,.
\label{eq:Lambdatilde}
\end{align}
We assume the $SU(N_c)$ beta function coefficient $b$ to be positive,
which means that $\tilde\Lambda$ can basically take any desired
value below the Planck scale.
For a given VEV $\left<P\right>$, the constant $w$
can then be tuned simply by varying the value of the gauge coupling
constant $\tilde{g}$ at the UV cutoff scale.
From this perspective, the fine-tuning of the CC in the 
true vacuum turns into an issue that pertains to
the UV boundary conditions of the low-energy effective
theory. 
It is no longer a problem within the low-energy effective
theory itself.


How can we now apply these results in the context of Polonyi inflation?
The crucial idea is to relate the value of the VEV $\left<P\right>$
to the inflationary dynamics.
We must make sure that the singlet $P$ is stabilized at $\left<P\right> = 0$
during inflation and that it becomes destabilized only towards the end of inflation.
In other words, we need to achieve a situation in which small inflaton
field values trigger a mass deformation of the SQCD sector,
$\overline{M}_Q = 0 \rightarrow \overline{M}_Q \neq 0$, so
that this sector turns into a pure SYM theory and spontaneously
breaks $R$ symmetry via gaugino condensation.
This approach shares some similarities with the models presented
in~\cite{Savoy:2007jb}, in which small inflaton field values trigger
the mass generation in a separate ISS sector~\cite{Intriligator:2006dd},
which then spontaneously breaks supersymmetry.
Moreover, our scenario of $R$ symmetry breaking should be compared with
the approaches in~\cite{Izawa:1996dv,Harigaya:2013pla} and~\cite{Kawasaki:2013re}.
In~\cite{Izawa:1996dv,Harigaya:2013pla}, inflation ends in a supersymmetric, but $R$
symmetry-breaking ground state, while in~\cite{Kawasaki:2013re},
$R$ symmetry is broken by a ``field-dependent constant'' term in
the superpotential that only becomes large at the end of inflation.


\subsection{Triggering a late-time mass deformation by a small inflaton field value}
\label{subsec:trigger}


We can realize the scenario described in the previous section by introducing
the following superpotential,
\begin{align}
W^P = \alpha\, Y \left(v_P^2-\frac{1}{2}\,P^2\right) + \frac{\beta}{3!}\, Y^3  \,.
\label{eq:WP}
\end{align}
Here, $Y$ is a singlet field that carries $R$ charge $2$; $\alpha$
and $\beta$ are dimensionless coupling constant; and $v_P$ is a
mass scale.
It is tempting to speculate that also the scale $v_P$ might be
dynamically generated in a strongly coupled hidden sector,
similar to the scale $v$ in Eq.~\eqref{eq:v}.
But for our purposes, it will not be necessary to specify its
origin.
Instead, we will treat it as a hard dimensionful input scale,
the only one in our model.
Before we explain in more detail how $W^P$
allows us to give a VEV to the field $P$ at the end
of inflation, let us examine for which values of $N_c$ and
$N_f$ the superpotential in Eq.~\eqref{eq:WP} is consistent
with the (approximate) global symmetries of the theory
in the first place.
Promoting these global symmetries to gauge symmetries
(and/or assuming that they are at least sufficiently protected by other gauge symmetries),
we will then be able to narrow down the viable choices for $N_c$ and $N_f$.


The superpotential in Eq.~\eqref{eq:WR} is invariant under
an anomaly-free global $U(1)_B \times Z_{2N_f} \times U(1)_R$ flavor
symmetry, under which the fields $Q^i$, $\bar{Q}^i$, $P$, and $Y$ 
are charged as follows,

\begin{center}
\begin{tabular}{c|cccc}
           & $Q^i$       & $\bar{Q}^i$ & $P$           & $Y$ \\ \hline
$U(1)_R$   & $1-N_c/N_f$ & $1-N_c/N_f$ & $2 N_c / N_f$ & $2$ \\
$U(1)_B$   & $1/3$       & $-1/3$      & $0$           & $0$ \\
$Z_{2N_f}$ & $1$         & $1$         & $-2$          & $0$
\end{tabular}
\end{center}

\noindent As mentioned earlier, in the pure SYM theory after integrating
out the heavy quarks, the continuous $U(1)_R$ is broken to a
discrete $Z_{2N_c}^R$ symmetry by $SU(N_c)$ instantons. 
This immediately suggests to set the number of colors $N_c$ to $2$,
so that the  global $Z_{2N_c}^R$ symmetry of the pure SYM theory
can be identified with our gauged $Z_4^R$ symmetry.%
\footnote{Here, we suppose some kind of K\"ahler interaction
between the two hidden sectors,
such as, e.g., $K \supset \epsilon' \Phi P^\dagger + \textrm{h.c.}$, that
relates $R$ symmetry transformations in the one sector to
$R$ symmetry transformations in the other sector.
Besides that, we assume the coefficients
of these mixing operators to be negligibly small for all practical purposes, $\epsilon'\lll1$.
This might, e.g., be achieved if the cross terms in the K\"ahler potential
are suppressed by (several powers of) a large cut-off scale.}
According to the above table,
the operator $YP^2$ in Eq.~\eqref{eq:WP} then only complies with the
exact $Z_{2N_c}^R \equiv Z_4^R$ symmetry,
if we choose the number of flavors $N_f$ to be either $1$ or $2$,
\begin{align}
\left[YP^2\right]_R = 2 + \frac{4N_c}{N_f} =  2 \textrm{ mod } 2N_c
\quad \Rightarrow \quad N_f = 1,2 \,.
\end{align}
On second sight, $N_f = 1$ is, however, not a viable option.
In that case, the $R$ charge of the field $P$ vanishes, $\left[P\right]_R = 0$,
so that we cannot prevent the unwanted operator $YP$ from popping up in
the superpotential.
Also the approximate discrete $Z_{2N_f} \equiv Z_2$ flavor symmetry cannot
help us in this situation, since the singlet field $P$
is required to transform even,
$\left[P\right]_{Z_2} = 0$, under this parity.
At the same time, the singlet field $Y$ must also transform even,
$\left[Y\right]_{Z_2} = 0$, to allow for the presence of the $Y$ tadpole 
term in Eq.~\eqref{eq:WP}.
The discrete $Z_2$ flavor symmetry, therefore, does not allow
us to eliminate the operator $YP$, as long as we want to keep the $Y$ tadpole term.
This leaves us with the following unique choice for $N_c$ and $N_f$,
\begin{align}
N_ c = 2 \,, \quad N_f = 2 \,,
\end{align}
which results in exactly the same matter
and gauge field content as in the IYIT sector.


For $N_c = N_f = 2$, the singlet field $P$ carries $R$ charge $2$.
From the perspective of $R$ symmetry, the fields $P$ and $Y$
are, hence, indistinguishable from each other, so that, next to the three wanted operators
$Y$, $YP^2$, and $Y^3$, also the three unwanted operators $P$, $PY^2$, and $P^3$
are allowed to appear in the superpotential.
We are, therefore, led to impose the $Z_{2N_f} \equiv Z_4$ flavor symmetry
as an approximate symmetry---which we assume to be protected by some exact
gauge symmetry in the UV.%
\footnote{If we did \textit{not} identify the $Z_{2N_c}^R$ of the pure SYM
theory with our anomaly-free $Z_4^R$, i.e., if we allowed
for $N_c > 2$ (while keeping $N_f = 2$ fixed), we would not be forced
to impose any \textit{additional} approximate flavor symmetry.
In that case, the $Z_{2N_c}^R$ alone would suffice to restrict the
set of allowed operators to $Y$ and $YP^2$ (meaning that the $Y^3$ term
would be suppressed).
Such a scenario would, however, not present an advantage over the $N_c = 2$
case. 
Either way, we have to assume a sufficiently intact global symmetry.
While for $N_c = 2$ the $Z_4$ needs to be imposed as
an approximate symmetry, the $Z_{2N_c}^R$ itself would need to be play the
role of an approximate symmetry for $N_c > 2$.
We will, therefore, ignore this possibility in the following.}
This approximate $Z_4$ acts on the fields $P$ and $Y$ again as a $Z_2$ parity---the
crucial difference to the $N_f = 1$ case being that, now, $P$
transforms odd, $\left[P\right]_{Z_2} = 1$,
while $Y$ still transforms even, $\left[Y\right]_{Z_2} = 0$.
By virtue of this $Z_2$ parity, the coefficients of the unwanted
operators $P$, $PY^2$, and $P^3$ then end up being suppressed.
Moreover, we emphasize the importance of the fact that
the $Z_4$ symmetry is only an approximate symmetry. 
If it was an exact symmetry, its spontaneous breaking at the end
of inflation would result in the formation of dangerous domain walls.
As the $Z_4$ symmetry is, however, bound to be explicitly broken (e.g., by the suppressed
$Z_2$-odd operators $P$, $PY^2$, and $P^3$), we are safe from running into
this problem.


Given the superpotential in Eq.~\eqref{eq:WP}, let us now demonstrate
how it enables us to trigger a mass deformation of the $R$ symmetry-breaking
sector towards the end of inflation.
The first thing to note is that the superpotential in Eq.~\eqref{eq:WP} 
results in two tachyonic mass eigenstates in global supersymmetry,
\begin{align}
\textrm{Global SUSY:} \qquad
m_{p^\pm}^2 = \pm\, \alpha^2 v_P^2 \,, \quad
m_{y^\pm}^2 = \pm\, \alpha \beta\, v_P^2 \,,
\end{align}
where $p^\pm$ and $y^\pm$ denote the real scalar degrees of freedom (DOFs)
contained in $P$ and $Y$, respectively.
These masses, however, receive corrections in supergravity that may
render them non-tachyonic,
\begin{align}
m_{p^\pm}^2\left(\varphi\right) & = \pm\, \alpha^2 v_P^2 \:\,+
\left(\frac{V\left(\varphi\right)}{M_{\rm Pl}^2} +
\frac{\Delta V_P}{M_{\rm Pl}^2} \right) \quad\:\,\,\,+
\Delta m_{p^\pm}^2\left(\varphi\right) \,, \quad
\Delta V_P = \alpha^2 v_P^4 \,, \label{eq:m2pySUGRA} \\ \nonumber
m_{y^\pm}^2\left(\varphi\right) & = \pm\, \alpha \beta\, v_P^2 +
\left(\frac{V\left(\varphi\right)}{M_{\rm Pl}^2} +
\mathcal{O}\left(M_{\rm Pl}^{-4}\right)\right) +
\Delta m_{y^\pm}^2\left(\varphi\right) \,.
\end{align}
Here, the corrections proportional to $V\left(\varphi\right)$
are nothing but the usual Hubble-induced
masses that scalar fields typically acquire during inflation.
Apart from that, we collect all further corrections that explicitly
depend on $\varphi$ in the field-dependent mass contributions
$\Delta m_{p^\pm}^2\left(\varphi\right)$
and $\Delta m_{y^\pm}^2\left(\varphi\right)$.
Just like the Hubble-induced masses, these corrections result 
from the F-term tree-level potential in supergravity.


As an important aside, we point out that the Hubble-induced masses
in Eq.~\eqref{eq:m2pySUGRA} are, in fact, sensitive to higher-dimensional
operators in the K\"ahler potential between the fields $P$
and $Y$ and the Polonyi field,
\begin{align}
K_{\rm mix} = F_P \big(\Phi/M_*,\Phi^\dagger/M_*\big)\left|P\right|^2 +
F_Y \big(\Phi/M_*',\Phi^\dagger/M_*'\big)\left|Y\right|^2 \,,
\label{eq:Kmix}
\end{align}
where $F_P$ and $F_Y$ denote two unknown functions of $\Phi$ and $\Phi^\dagger$
and where $M_*$ and $M_*'$ represent appropriate cut-off scales.
Note that such couplings between $P$, $Y$, and $\Phi$ in the
K\"ahler potential are the only relevant means of communication between the SUSY-breaking
sector and the $R$ symmetry-breaking sector in our model.
For instance, we could imagine that both sectors only communicate with each other via
gravitational interactions.
In that case, the scales $M_*$ and $M_*'$ should be identified as
the Planck scale. 
Alternatively, we may assume that the $R$ symmetry-breaking sector couples
to the SUSY-breaking sector via the exchange of heavy messenger particles
of mass $M_{\rm mess}$.
Then, $M_*^{(\prime)}$ would have to be identified as a typical mass scale of the
messenger sector, $M_*^{(\prime)} \sim M_{\rm mess}$.
To illustrate the effect of the K\"ahler potential in Eq.~\eqref{eq:Kmix},
let us suppose that the leading-order contributions to $F_P$ and $F_Y$ take
the following form,
\begin{align}
K_{\rm mix} = \epsilon_P\, \frac{\left|\Phi\right|^2\left|P\right|^2}{M_*^2} +
\epsilon_Y\, \frac{\left|\Phi\right|^2\left|Y\right|^2}{M_*^2} + \cdots \,.
\label{eq:Kmixeps}
\end{align}
Such ``mixing terms'' shift the the Hubble-induced masses
in Eq.~\eqref{eq:m2pySUGRA}.
Consider, e.g., the mass of $p^-$,
\begin{align}
m_{p^-}^2\left(\varphi\right) \subset
\frac{V\left(\varphi\right)}{M_{\rm Pl}^2} \rightarrow
\left(1-\epsilon_P\,\frac{M_{\rm Pl}^2}{M_*^2} +
\frac{\epsilon_P^2}{2}\frac{M_{\rm Pl}^4}{M_*^4}\frac{\varphi^2}{M_{\rm Pl}^2} + \cdots\right)
\frac{V\left(\varphi\right)}{M_{\rm Pl}^2} \,,
\label{eq:epsPHubble}
\end{align}
where the ellipsis stands for further, subdominant corrections.
$m_{p^+}^2$ and $m_{y^\pm}^2$ are modified in a similar way.
For negative coefficients $\epsilon_P$ and $\epsilon_Y$, these corrections may help
establish a strong hierarchy between the masses $m_{p^\pm}$ and $m_{y^\pm}$
on the one hand and the inflationary  Hubble rate $H_{\rm inf}$ on the other hand,
\begin{align}
- \epsilon_{P,Y}\,\frac{M_{\rm Pl}^2}{M_*^2} \gg 1 \quad\Rightarrow\quad
\left|m_{p^\pm}\right| \,, \left|m_{y^\pm}\right| \gg H_{\rm inf} \,,
\label{eq:mHhierarchy}
\end{align}
so as to make sure that, during inflation, all of the scalar fields
$p^\pm$ and $y^\pm$ are sufficiently stabilized.
At the same time, it is clear that the K\"ahler potential
in Eq.~\eqref{eq:Kmix} significantly complicates the analysis of the
$R$ symmetry-breaking sector.
For presentational purposes, we will, therefore, neglect the effect of
$K_{\rm mix}$ in the following.
We have checked explicitly that, accounting for the exemplary K\"ahler potential
in Eq.~\eqref{eq:Kmixeps}, all results remain quantitatively
the same---the only difference being the hierarchy in Eq.~\eqref{eq:mHhierarchy}.
In this sense, all results in this and the following section should be regarded
as approximate results that convey the general flavor of our mechanism
of late-time $R$ symmetry breaking.
An exact numerical study including a more systematic treatment of the functions
$F_P$ and $F_Y$ in Eq.~\eqref{eq:Kmix} is left for future work.
For our purposes, it is merely important to remember that the masses
of the fields $p^\pm$ and $y^\pm$ during inflation may be further increased
by adding higher-dimensional terms to the K\"ahler potential.


The idea now is to choose the coupling constants $\alpha$ and $\beta$
such that the scalar DOF $p^-$ is stabilized during inflation,
turning tachyonically unstable only at a certain (small)
inflaton field value $\varphi_0$,
\begin{align}
\varphi \geq \varphi_0 & \quad\Rightarrow\quad
m_{p^+}^2\left(\varphi\right) > 0 \,, \quad
m_{p^-}^2\left(\varphi\right) \geq 0 \,, \quad
m_{y^\pm}^2\left(\varphi\right) > 0 \,, \\ \nonumber
\varphi < \varphi_0 & \quad\Rightarrow\quad
m_{p^+}^2\left(\varphi\right) > 0 \,, \quad
m_{p^-}^2\left(\varphi\right) < 0 \,, \quad
m_{y^\pm}^2\left(\varphi\right) > 0 \,.
\end{align}
To determine the required values for $\alpha$ and $\beta$ in order to
implement such a scenario, we first note that the field-dependent mass
$m_{p^-}^2$ in Eq.~\eqref{eq:m2pySUGRA} vanishes
at $\varphi = 0$ for the following choice of $\alpha$,
\begin{align}
m_{p^-}^2\left(0\right) = -\alpha_0^2 v_P^2 + \frac{V_0}{M_{\rm Pl}^2} + 
\frac{\alpha_0^2 v_P^4}{M_{\rm Pl}^2} = 0 \quad\Rightarrow\quad
\alpha_0 = \frac{\mu^2}{v_P\left(M_{\rm Pl}^2 - v_P^2\right)^{1/2}} \,.
\label{eq:alpha0}
\end{align}
In other words, for $\alpha = \alpha_0$, the Hubble-induced
mass compensates the tachyonic mass, $-\alpha^2v_P^2$, for all field values
during inflation, i.e., as long as
$\varphi > 0$, and only at the origin the scalar $p^-$ becomes massless.
At the same time, we need to make sure that the scalar DOFs contained in $Y$
are stabilized at all times. 
This is achieved by requiring that their respective masses in global supersymmetry,
$\pm \alpha\beta\,v_P^2$, are always outweighed by the Hubble-induced mass, even
at the origin, where the Hubble-induced mass is smallest,
\begin{align}
\left|\alpha\beta\,v_P^2\right| < 
\frac{V_0}{M_{\rm Pl}^2}
\quad\Rightarrow\quad
\left|\beta\right| < \beta_0 =
\frac{V_0}{\alpha\, v_P^2 M_{\rm Pl}^2} \,.
\label{eq:beta0}
\end{align}
As long as this condition is satisfied, the field $Y$ is
stabilized at a nonzero VEV,%
\footnote{In global supersymmetry, the VEV $\left<Y\right>$ vanishes during and 
after inflation.
In supergravity, the scalar potential, however, contains a tadpole term linear in
$y^+$ (due to the nonzero F-term of the field $Y$) that displaces
$\left<Y\right>$ from the origin.}
\begin{align}
\left<Y\right> = \frac{\alpha\,v_P^2 \,\mu^2}{m_{y^+}^2 M_{\rm Pl}^2}
\frac{\varphi}{\sqrt{2}} + \mathcal{O}\left(\varphi^5\right) \,,
\end{align}
which induces a positive Majorana mass for the field $P$
via the waterfall superpotential in Eq.~\eqref{eq:WP},
\begin{align}
m_{p^{\pm}}\left(\varphi\right) \supset \alpha \left<Y\right> \,.
\label{eq:aY}
\end{align}
This field-dependent mass is contained in $\Delta m_{p^\pm}^2\left(\varphi\right)$
in Eq.~\eqref{eq:m2pySUGRA} and, similarly as in ordinary F-term hybrid inflation,
it helps to stabilize the field $P$ during inflation.
Only at the end of inflation, $\Delta m_{p^\pm}^2\left(\varphi\right)$ and, in particular,
the mass term in Eq.~\eqref{eq:aY} vanish, so that $P$ has a chance of becoming unstable.


Let us now study the inflaton-dependent mass corrections in Eq.~\eqref{eq:m2pySUGRA}
more closely.
In doing so, it turns out to be convenient to use $\alpha_0$ and $\beta_0$ as
reference values for $\alpha$ and $\beta$.
To this end, we define
\begin{align}
a = \left(\frac{\alpha}{\alpha_0}\right)^2
= \frac{\alpha^2v_P^2-\alpha^2v_P^4/M_{\rm Pl}^2}{V_0/M_{\rm Pl}^2} \,, \quad
b = \frac{\beta}{\beta_0} = 
\frac{\alpha\beta\,v_P^2}{V_0/M_{\rm Pl}^2} \,.
\label{eq:ab}
\end{align}
The field-dependent shift $\Delta m_{p^-}^2\left(\varphi\right)$ in the mass
of the scalar DOF $p^-$ can then be written as follows,
\begin{align}
\Delta m_{p^-}^2\left(\varphi\right) = \left[1 - \frac{2\,a}
{\left(1+b\right)^2} \left(1-\frac{a}{2} +\frac{7\,b}{4}+\frac{b^2}{2}\right)+
\mathcal{O}\left(M_{\rm Pl}^{-2}\right)\right]
\frac{V_0}{M_{\rm Pl}^2}
\frac{\varphi^2}{2M_{\rm Pl}^2} + \mathcal{O}\left(\varphi^4\right) \,.
\end{align}
The parameter $a$ is supposed to be chosen, such that the total mass
$m_{p^-}^2$ vanishes at $\varphi = \varphi_0$,
\begin{align}
m_{p^-}^2\left(\varphi_0\right) = \left(1-a\left(\varphi_0\right)\right) \frac{V_0}{M_{\rm Pl}^2}
+ \frac{1}{2}\,m_{\rm eff}^2 \left(\frac{\varphi_0}{M_{\rm Pl}}\right)^2
+ \Delta m_{p^-}^2\left(\varphi_0\right) = 0 \,.
\end{align}
For $\varphi_0 \ll M_{\rm Pl}$, this condition yields
a quadratic equation for $a\left(\varphi_0\right)$, which has two independent solutions,
\begin{align}
a_1\left(\varphi_0\right) & = 1 + \frac{1}{2}\frac{m_{\rm eff}^2\,\varphi_0^2}{V_0}
-\frac{3\,b}{4\left(1+b\right)^2}
\left(\frac{\varphi_0}{M_{\rm Pl}}\right)^2 + \mathcal{O}\left(\varphi_0^4\right) \,,
\label{eq:a12} \\ \nonumber
a_2\left(\varphi_0\right) & = 2\left(1+b\right)^2 \left(\frac{M_{\rm Pl}}{\varphi_0}\right)^2
+ \mathcal{O}\left(\varphi_0^0\right)\,.
\end{align}
Here, the first solution is close to unity, $a_1\left(\varphi_0\right) \approx 1$,
so that $\alpha\left(\varphi_0\right) = a^{1/2}\left(\varphi_0\right) \alpha_0 \approx \alpha_0$.
For $a = a_1\left(\varphi_0\right)$, the Hubble-induced mass $V/M_{\rm Pl}^2$
cancels the tachyonic mass, $-\alpha^2v_P^2$, to good precision and
the field-dependent correction $\Delta m_{p^-}^2\left(\varphi_0\right)$
nearly vanishes on its own.
The second solution is, by contrast, much larger,
$a_2\left(\varphi_0\right) \gg 1$.
In that case, the individual contributions to $m_{p^-}^2\left(\varphi_0\right)$
do not vanish independently.
Instead, all contributions are large, have different signs, and simply happen to
cancel each other at $\varphi = \varphi_0$.
As we will see further below, the solution $a_2$
will turn out to be unviable for phenomenological reasons.
For now, we will, however, remain impartial and treat both
solutions on the same footing.


Combining our results in Eqs.~\eqref{eq:alpha0}, \eqref{eq:beta0}, \eqref{eq:ab},
and \eqref{eq:a12}, we eventually obtain our final expressions for the parameter $\alpha$.
Again, we find two solutions,
\begin{align}
\alpha_1\left(\varphi_0\right) & = \frac{1}{v_P}\frac{V_{\rm eff}^{1/2}\left(\varphi_0\right)}
{\left(M_{\rm Pl}^2-v_P^2\right)^{1/2}} -
\frac{3\,\beta\,\varphi_0^2}{8\left(M_{\rm Pl}^2-v_P^2\right)} +
\mathcal{O}\left(\beta^2,\varphi_0^4\right)
\,, \quad
V_{\rm eff}\left(\varphi_0\right) = V_0 + \frac{1}{2}\,m_{\rm eff}^2\,\varphi_0^2 \,,
\label{eq:alpha12} \\ \nonumber
a_2\left(\varphi_0\right) & = \frac{2\,V_0}{\sqrt{2}\,v_P\mu^2\left(M_{\rm Pl}^2-v_P^2\right)^{1/2}
\varphi_0/M_{\rm Pl}-2\,\beta\,v_P^2 M_{\rm Pl}^2}
+ \mathcal{O}\left(\varphi_0^0\right)\,.
\end{align}
These expressions mark one of the main results in this paper:
Provided that the parameter $\alpha$ in the superpotential $W^P$ in Eq.~\eqref{eq:WP}
takes either of these two values, the scalar field $p^-$ becomes unstable,
once the inflaton field $\varphi$ reaches the value $\varphi_0$ at the end
of inflation or after inflation.
This demonstrates how varying the value of the parameter $\alpha$ puts us
in the position to control the time at which $R$ symmetry is spontaneously broken
via gaugino condensation at the end of inflation.
For small values of $\beta$ and $v_P$ (such that $\beta \ll \beta_0$ and
$v_P \ll M_{\rm Pl}$), the solutions for $\alpha$ in Eq.~\eqref{eq:alpha12}
reduce to more compact expressions,
\begin{align}
\alpha_1\left(\varphi_0\right) \approx
\frac{V_{\rm eff}^{1/2}\left(\varphi_0\right)}{v_P\,M_{\rm Pl}}
\,, \quad
a_2\left(\varphi_0\right) \approx \frac{\sqrt{2}\,\mu^2}{v_P\,\varphi_0} \,.
\end{align}
As evident from Eq.~\eqref{eq:alpha12}, the first solution,
$\alpha = \alpha_1\left(\varphi_0\right)$, depends
only weakly on the field value $\varphi_0$.
In fact, for given values of all other free parameters ($\mu$, $\lambda$, $v_P$, and $\beta$),
matching $\alpha$ with $\alpha_1\left(\varphi_0\right)$ for a reasonable value of $\varphi_0$
requires some amount of fine-tuning.
To first approximation, all values $\alpha_1\left(\varphi_0\right)$ correspond
to the reference value $\alpha_0$, their variation with $\varphi_0$
being a subdominant effect,
\begin{align}
\alpha_1\left(\varphi_0\right) \approx \alpha_0 \simeq
\frac{\mu^2}{v_P M_{\rm Pl}}
\simeq 4 \times 10^{-6}\: \bigg(\frac{\mu}{10^{15}\,\textrm{GeV}}\bigg)^2
\left(\frac{10^{17}\,\textrm{GeV}}{v_P}\right) \,.
\end{align}
In the context of our model, this is, however, not a problem.
Recall that the ultimate purpose of the superpotential $W^P$
in Eq.~\eqref{eq:WP} is to help us in addressing the CC problem.
From that point of view, it is expected that some (if not all) of the parameters
involved in determining the final value of the CC end up being
subject to some amount of fine-tuning.
In addition, we emphasize that the value of $\alpha$ may have an important
impact on the time when inflation ends.
In regions of space (or the ``string landscape'') where $\alpha$ takes too small
a value, gaugino condensation never occurs, i.e., inflation never ends.
On the other hand, in regions of space where $\alpha$ takes too large
a value, the constant in the superpotential, $W \supset w$, is already
generated at early times.
Inflation then only lasts for a few $e$-folds or does not take place at all.
Universes in which $\alpha$ matches $\alpha_1\left(\varphi_0\right)$
for a reasonable value of $\varphi_0$, i.e., universes in which inflation lasts
for a long but finite time interval, therefore, eventually occupy
the largest spatial volume among all post-inflationary universes.
This might explain why, in habitable universes, $\alpha$ appears
to be fine-tuned.


Finally, we evaluate the total $p^-$ mass $m_{p^-}^2$
in Eq.~\eqref{eq:m2pySUGRA} for the two solutions for $\alpha$ in Eq.~\eqref{eq:alpha12},
\begin{align}
\alpha \rightarrow \alpha_1 \quad\Rightarrow\quad
m_{p^-}^2\left(\varphi\right) & =\left[
\frac{1}{2}\, m_{\rm eff}^2 -
\frac{3\,b}{4\left(1+b\right)^2}\frac{V_0}{M_{\rm Pl}^2}
+ \mathcal{O}\left(M_{\rm Pl}^{-4}\right)\right]
\frac{\varphi^2-\varphi_0^2}{M_{\rm Pl}^2} + \mathcal{O}\left(\varphi^4\right)
\,, \label{eq:mpm2}\\ \nonumber
\alpha \rightarrow \alpha_2 \quad\Rightarrow\quad
m_{p^-}^2\left(\varphi\right) & =
\left[2\left(1+b\right)^2\vphantom{\frac{3(-b)}{4\left(1+b\right)^2}}
\frac{V_0}{\varphi_0^2}
+ \mathcal{O}\left(M_{\rm Pl}^{-2}\right)\right]
\frac{\varphi^2-\varphi_0^2}{\varphi_0^2}
+ \mathcal{O}\left(\varphi^4\right) \,,
\end{align}
which nicely illustrates how the scalar DOF $p^-$ turns tachyonic,
once the inflaton field $\varphi$ reaches the value $\varphi_0$.
In this sense, Polonyi inflation does end in a ``waterfall transition'', after all.
But, instead of taking place in the inflaton sector itself, this transition
occurs in separate hidden sector.
The inflaton and waterfall sectors are, therefore, separated in our model,
which allows us to bring together two phenomena that would otherwise
mutually exclude each other:
(i) Inflation in a scalar potential equivalent to that of ordinary F-term hybrid
inflation and (ii) supersymmetry breaking at a very high scale.
For a discussion of the tension between these two phenomena
and a possible resolution in the context of string theory,
see~\cite{Higaki:2012iq}.
We note that, in contrast to~\cite{Higaki:2012iq}, our scenario represents a fully
field-theoretic construction.


In closing, we also mention that, adding the particular K\"ahler potential
in Eq.~\eqref{eq:Kmixeps}, the expressions for $m_{p^-}$ in
Eq.~\eqref{eq:mpm2} remain quantitatively the same.
The only effect of $K_{\rm mix}$ is that it increases the overall magnitude
of the terms in square brackets in Eq.~\eqref{eq:mpm2}. 
We have explicitly checked numerically that, if we choose
$\epsilon_{P,Y} \sim -1$,  $M_* \sim 10^{17}\,\textrm{GeV}$,
and $M_*' \sim M_{\rm Pl}$ in Eq.~\eqref{eq:Kmixeps}, our first solution
for $m_{p^-}$ exceeds the inflationary
Hubble rate for all times during inflation, $m_{p^-} \gg H_{\rm inf}$.
At the same time, for field values slightly below $\varphi \simeq \varphi_0$, the
absolute value of  $m_{p^-}$ is again much larger than the Hubble rate,
$m_{p^-} \ll - H_{\rm inf}$.
This ensures that the scalar field $p^-$
quickly reaches the minimum of the scalar potential.
Meanwhile, all other scalars always have masses that are large compared
to $H_{\rm inf}$, so that they are safely stabilized at the origin.
A possible interpretation of these numerical findings is to suppose
the existence of an other strongly coupled hidden gauge sector around the
scale $M_* \sim 10^{17}\,\textrm{GeV}$.
If the singlet $P$ is the only field that is strongly coupled to this
new dynamical sector (meaning that $Y$ and $\Phi$ are only weakly
coupled), simple arguments from NDA
automatically result in a K\"ahler potential of just the desired form.%
\footnote{For $\epsilon_P<0$, the kinetic
term for the field $P$ becomes singular at $\varphi \simeq M_*$,
as long as we only assume the K\"ahler potential in Eq.~\eqref{eq:Kmixeps}.
For this reason, we expect additional higher-dimensional operators in $F_P$
to become important around $\varphi \simeq M_*$, which regulate the kinetic term for
$P$ (see Eq.~\eqref{eq:Kmix}).
Here, one simple solution is to assume that $F_P$ as a function of
$\varphi$ simply saturates at a maximal
negative value, $F_P \rightarrow -F_0 > - 1$, around field values of
$\mathcal{O}\left(M_*\right)$, so that the prefactor of the kinetic term for $P$
always remains positive, $K \supset \left(1+F_P\right)P^\dagger P
\rightarrow\left(1-F_0\right)P^\dagger P$.
At large field values, $\varphi \gtrsim M_*$, the field $P$ is then stabilized
by the field-dependent mass in Eq.~\eqref{eq:aY}, while for
$\varphi_0 \lesssim \varphi \lesssim M_*$, it is stabilized
by the terms in Eq.~\eqref{eq:epsPHubble}.}
In particular, such new strong
dynamics do not induce a dangerous $\left|\Phi\right|^4/M_*^2$ term
in the K\"ahler potential (which would otherwise spoil slow-roll inflation),
as long as the Polonyi field is only weakly coupled to the new sector.
However, we emphasize that this picture of a new gauge sector around
$M_* \sim 10^{17}\,\textrm{GeV}$ is just a simple example to illustrate what additional
dynamics could possibly lead to a cut-off scale of $\mathcal{O}\left(10^{17}\right)\,\textrm{GeV}$
in the $\left|\Phi\right|^2\left|P\right|^2$ operator in the noncanonical
K\"ahler potential $K_{\rm mix}$.
From the perspective of our model, the main consequence of $K_{\rm mix}$
in Eq.~\eqref{eq:Kmixeps} is that it
provides us with the ability to increase the absolute value as
well as the gradient of $m_{p^-}$ as a function of $\varphi$.
Apart from that, our simplified analysis based on $K_{\rm mix} = 0$ 
captures all relevant aspects of our mechanism of $R$ symmetry breaking.
For this reason, we leave a more detailed study of $K_{\rm mix}$ and its UV origin
for future work.


\subsection{Backreaction on inflation and low-energy ground state after inflation}
\label{subsec:backreaction}


The ``waterfall superpotential'' $W^P$ in Eq.~\eqref{eq:WP} not only triggers a mass deformation
in the $R$ symmetry-breaking sector, it also results in Planck-suppressed
corrections to the inflaton potential in Eq.~\eqref{eq:infpot}.
During inflation, i.e., as long as $w = 0$, the scalar Polonyi potential now reads
\begin{align}
V\left(\varphi\right) = C_0
+ \frac{C_2}{2}\,\varphi^2
+ \frac{C_4}{4!}\,\varphi^4
+ \mathcal{O}\left(\varphi^{-6}\right) \,,
\end{align}
where the coefficients $C_0$, $C_2$, and $C_4$ correspond to the coefficients $c_0$,
$c_2$, and $c_4$ in Eq.~\eqref{eq:c01234} evaluated at $w=0$ and multiplied by
correction factors that stem from the $R$ symmetry-breaking sector,
\begin{align}
C_0 & = \left(1+\nu\right) \left.c_0\right|_{w=0}  \,, \quad 
C_2 = \left(1-\frac{b}{1+b}\frac{\nu}{\epsilon}\right) \left.c_2\right|_{w=0} \,, 
\label{eq:C024}\\ \nonumber
C_4 & = \left[1 +
\left(3+\frac{32\,b}{3}+\frac{79\,b^2}{6}+6\,b^3+b^4 +
\mathcal{O}\left(\epsilon,\nu\right)\right)\frac{\nu}
{\left(1+b\right)^4}\right] \left.c_4\right|_{w=0} \,.
\end{align}
Here, we have introduced the parameter $\nu$ to measure the
relative importance of these new corrections,
\begin{align}
\nu = \frac{\Delta V_P}{V_0} = \frac{\alpha^2 v_P^4}{\mu^4} \,,
\label{eq:nu}
\end{align}
which is nothing but the ratio of the $Y$ and $\Phi$ F-terms during inflation,
$\nu = \left|F_Y/F_\Phi\right|$.
To make sure that the $R$ symmetry-breaking sector does not
disturb the inflationary dynamics, we have to require that $\nu$
is sufficiently small.
In view of the expression for $C_2$ in Eq.~\eqref{eq:C024},
we have to demand in particular that
\begin{align}
\nu \ll \epsilon \,,
\label{eq:nueps}
\end{align}
since otherwise the inflaton field will receive too large a Hubble-induced
mass, $m_\varphi^2 \sim 3\,\nu H_{\rm inf}^2$.
This constraint then translates into upper bounds on the mass
scale $v_P$ in the $R$ symmetry-breaking sector,
\begin{align}
\alpha = \alpha_1\left(\varphi_0\right) \:\:\Rightarrow\:\:
v_P \ll v_P^{\rm max,1} = \epsilon^{1/2} M_{\rm Pl} \,, \qquad
\alpha = \alpha_2\left(\varphi_0\right) \:\:\Rightarrow\:\:
v_P \ll v_P^{\rm max,2} = \frac{\epsilon^{1/2}\,\varphi_0}{\sqrt{2}\left(1+b\right)} \,.
\label{eq:vmax12}
\end{align}
The scale $v_P$ is a free parameter, which suggests that we should
actually always be able to satisfy these bounds.
At the end of this section, we will, however, see that the upper
bound $v_P^{\rm max,2}$ turns out to be too restrictive, so that only
our first solution, $\alpha = \alpha_1\left(\varphi_0\right)$,
remains phenomenologically viable.


To get there, we first need to examine the true vacuum after
the end of inflation as well as the tuning of the CC in more detail.
In global supersymmetry, the VEVs of the fields $\Phi$, $Y$, and $P$
are found to be
\begin{align}
\textrm{Global SUSY:} \qquad
\left<\Phi\right> = 0 \,, \quad
\left<Y\right> = 0 \,, \quad
\left<P\right> = \sqrt{2}\, v_P \,.
\label{eq:VEVsglobal}
\end{align}
The fluctuations around the true vacuum are, therefore, described by the following
superpotential,
\begin{align}
W = \mu^2\, \Phi - \sqrt{2}\, \alpha\, v_P\, Y P' - \frac{\alpha}{2}\, Y P^{\prime\,2}
+ \frac{\beta}{3!}\, Y^3 + w \,, \label{eq:W0}
\end{align}
where $P'$ denotes the singlet field $P$ shifted by its VEV in global supersymmetry,
$P \rightarrow \sqrt{2}\,v_P + P'$.
Eq.~\eqref{eq:W0} tells us that, from the perspective of global supersymmetry, the
F-term of the field $Y$ completely vanishes in the true vacuum, so that the Polonyi
field $\Phi$ remains as the only SUSY-breaking field.
In addition, we find that the fields $Y$ and $P$ now share a common
supersymmetric Dirac mass.
In global supersymmetry, the scalar mass eigenvalues at low energies
are consequently given as follows,
\begin{align}
\textrm{Global SUSY:} \qquad
m_\phi = m_{\rm eff} \,, \quad m_y = m_p = \sqrt{2}\, \alpha\, v_P \,,
\label{eq:massesglobal}
\end{align}
where $\phi$, $y$, and $p$ stand for the complex scalars contained
in the chiral fields $\Phi$, $Y$, and $P'$, respectively.


In order to find the SUGRA corrections to the VEVs in Eq.~\eqref{eq:VEVsglobal}
and the masses in Eq.~\eqref{eq:massesglobal}, we expand the full SUGRA
scalar potential up to second order in the fluctuations $\phi$, $y$, and $p$,
\begin{align}
V\left(\phi,y,p\right) & = c_0
+ c_\phi \left(\phi + \phi^*\right)
+ c_y \left(y + y^*\right)
+ c_p \left(p + p^*\right)
+ m_\phi^2 \left|\phi\right|^2 
+ m_y^2 \left|y\right|^2 
+ m_p^2 \left|p\right|^2 \label{eq:Vphiyp} \\ \nonumber
& + m_{\phi y}^2 \left(\phi y^* + \phi^* y\right)
+ m_{\phi p}^2 \left(\phi p^* + \phi^* p\right)
+ m_{yp}^2 \left(y p^* + y^*p\right) + \cdots \,.
\end{align}
Here, the coefficients of the linear terms ($c_\phi$, $c_y$, $c_p$),
the masses around the true vacuum ($m_\phi$, $m_y$, $m_p$)
as well as the mass mixing parameters ($m_{\phi y}$, $m_{\phi p}$, $m_{yp}$)
take, to leading order, the following form,
\begin{align}
\label{eq:cphiyp}
c_\phi & \simeq - 2\,\frac{\mu^2\,w}{M_{\rm Pl}^2} \,, &
c_y    & \simeq - 2\,\frac{\alpha\, v_P^2\,w}{M_{\rm Pl}^2} \,, &
c_p    & \simeq \sqrt{2}\,\frac{v_P\,V_0}{M_{\rm Pl}^2}
                - 2\sqrt{2}\,\frac{v_P\, w^2}{M_{\rm Pl}^4}\,, \\ \nonumber
m_\phi^2 & \simeq m_{\rm eff}^2-\epsilon\,\frac{V_0}{M_{\rm Pl}^2} - 2\,\frac{w^2}{M_{\rm Pl}^4} \,, &
m_y^2 & \simeq 2\, \alpha^2 v_P^2 + \frac{V_0}{M_{\rm Pl}^2} - 2\,\frac{w^2}{M_{\rm Pl}^4} \,, &
m_p^2 & \simeq 2\, \alpha^2 v_P^2 + \frac{V_0}{M_{\rm Pl}^2} - 2\,\frac{w^2}{M_{\rm Pl}^4} \,, \\ \nonumber
m_{\phi y}^2 & \simeq -2\,\frac{\alpha\,v_P^2\,\mu^2}{M_{\rm Pl}^2}     \,,  &
m_{\phi p}^2 & \simeq -2\sqrt{2}\,\frac{v_P\,\mu^2\,w}{M_{\rm Pl}^4}    \,, &
m_{yp}^2     & \simeq -6\sqrt{2}\,\frac{\alpha\,v_P^3\,w}{M_{\rm Pl}^4} \,.
\end{align}
Given that $\nu \ll1$ (see Eqs.~\eqref{eq:nu} and \eqref{eq:nueps}),
the mass mixing among $\phi$, $y$, and $p$ turns out be negligible.
Setting the off-diagonal masses $m_{\phi y}$, $m_{\phi p}$, and $m_{yp}$ in
Eq.~\eqref{eq:Vphiyp} to zero, we then arrive at the following simple expressions
for the VEVs of the fields $\Phi$, $Y$, and $P$ in supergravity,
\begin{align}
\left<\Phi\right> \simeq - \frac{c_\phi}{m_\phi^2} \,, \quad
\left<Y\right> \simeq - \frac{c_y}{m_y^2} \,, \quad
\left<P\right> \simeq \sqrt{2}\, v_P - \frac{c_p}{m_p^2} \,,
\label{eq:VEVphiyp}
\end{align}
which results in the following vacuum energy density,
\begin{align}
\left<V\right> \simeq c_0 + c_\phi \left<\Phi\right> +
c_y \left<Y\right> + c_p\, \big(\negthinspace\left<P\right>-\sqrt{2}\,v_P\big)
\,, \quad c_0 = V_0 - 3\,\frac{w^2}{M_{\rm Pl}^2} \,,
\end{align}
where the inflaton term, $c_\phi \left<\Phi\right>$, clearly constitutes the largest
correction to the leading $c_0$ term.
In order to make the CC $\left<V\right>$ vanish, we have to fine-tune the constant $w$,
so that it takes the following value,
\begin{align}
w_0 \simeq \frac{1}{\sqrt{3}}\left(1-\gamma\right)\mu^2 M_{\rm Pl} \,, \quad
\gamma = \frac{2\,m_{3/2}^{\prime\,2}}
{m_{\rm eff}^2- \epsilon\,V_0/M_{\rm Pl}^2 + 2\,m_{3/2}^{\prime\,2}} \,, \quad
m_{3/2}' = \frac{\mu^2}{\sqrt{3}M_{\rm Pl}^2} \,. \label{eq:w0gamma}
\end{align}
Here, $m_{3/2}'$ denotes the ``asymptotic gravitino mass''
in the limit $\left<\Phi\right> \rightarrow 0$.
The physical gravitino mass, on the other hand, follows
from plugging our results for $w_0$ as well as for the field VEVs into Eq.~\eqref{eq:m32},
\begin{align}
m_{3/2} \simeq \exp\left[\frac{\left<\Phi\right>^2}{2M_{\rm Pl}^2}\right]
\left(\frac{1-\gamma}{\sqrt{3}}+\frac{\left<\Phi\right>}{M_{\rm Pl}}\right)
\frac{\mu^2}{M_{\rm Pl}} \,, \quad
\left<\Phi\right> \simeq \frac{\sqrt{3}\,\gamma}{1-\gamma+\gamma^2}\, M_{\rm Pl}
\approx \frac{4}{3}\sqrt{3}\,\gamma M_{\rm Pl} \,. \label{eq:m32Phi}
\end{align}


As evident from Eq.~\eqref{eq:m32Phi}, the parameter $\gamma$ controls the size
of the Polonyi VEV in the true vacuum.
In fact, it is a convenient measure for the relation between the individual
contributions to the ``asymptotic inflaton mass'' $m_\phi'$ in the
limit $\left<\Phi\right> \rightarrow 0$.
In that limit, we have $m_{3/2} \rightarrow m_{3/2}'$, $\gamma\rightarrow0$,
and $m_\phi \rightarrow m_\phi'$,
with the asymptotic inflaton mass $m_\phi'$ being given as follows,
\begin{align}
m_\phi^{\prime\,2} = m_{\phi,0}^2 + m_{\phi,w}^2 \,, \quad
m_{\phi,0}^2 = m_{\rm eff}^2- \epsilon\,\frac{V_0}{M_{\rm Pl}^2} \,, \quad
m_{\phi,w}^2 = - 2\,m_{3/2}^{\prime\,2} \,.
\end{align}
Here, $m_{\phi,0}$ denotes the effective inflaton mass close to the origin for $w=0$,
whereas $m_{\phi,w}$ represents the correction to $m_\phi'$
in the true vacuum, i.e., the correction appearing after the generation of
the constant $w$.
Making use of these definitions, we recognize that $\gamma$
parametrizes nothing else than the following ratio,
\begin{align}
\gamma = \frac{\left|m_{\phi,w}\right|^2}{\left|m_{\phi,0}\right|^2
+ \left|m_{\phi,w}\right|^2} \,.
\end{align}
In order to stabilize the Polonyi field at a sub-Planckian field value
after the end of inflation, $\left<\Phi\right> \ll M_{\rm Pl}$,
we have to require that the additional SUGRA correction
generated in the course of late-time $R$ symmetry breaking, $m_{\phi,w}$,
always remains smaller than the effective one-loop mass $m_{\rm eff}$,
\begin{align}
\left<\Phi\right> \ll M_{\rm Pl} \quad\Leftrightarrow\quad \gamma \ll 1
\quad\Leftrightarrow\quad m_{3/2}' \ll m_{\rm eff} \,.
\label{eq:gamma}
\end{align}
We remark that this requirement is a consequence of the fact that $m_{\rm eff}$ and $m_{3/2}'$
are controlled by the same dynamics, i.e., by
the dynamical scale $\Lambda$ and the Yukawa coupling $\lambda$.
These two masses are, therefore, potentially of the same order
of magnitude, in which case the SUGRA correction $m_{\phi,w}$ threatens to
destabilize the minimum of the effective potential in global supersymmetry.
For this reason, we have to explicitly impose the requirement
$m_{3/2}' \ll m_{\rm eff}$ as an extra condition.
Once this condition is satisfied, our results in Eqs.~\eqref{eq:cphiyp},
\eqref{eq:w0gamma}, and \eqref{eq:m32Phi} simplify considerably.
For small values of $\gamma$, we find
\begin{align}
w_0     \simeq \frac{\mu^2 M_{\rm Pl}}{\sqrt{3}}\,, \quad
m_{3/2} \simeq m_{3/2}'  \,, \quad
m_\phi^2 \simeq m_{\rm eff}^2 - 2\,m_{3/2}^2 \,, \quad 
m_y^2    \simeq 2\,\alpha^2 v_P^2 + m_{3/2}^2 \,, \quad m_p^2 \simeq m_y^2 \,.
\label{eq:wm32mphiyp}
\end{align}
In combination with Eqs.~\eqref{eq:cphiyp} and \eqref{eq:VEVphiyp},
these results lead to simple expressions for $\left<\Phi\right>$,
$\left<Y\right>$, and $\left<P\right>$ that are valid for $\gamma \ll 1$.
Neglecting all effects of $\mathcal{O}\left(\gamma\right)$
and $\mathcal{O}\left(\epsilon\right)$ wherever possible, we obtain
\begin{align}
\left<\Phi\right> \simeq \frac{2\sqrt{3}\, m_{3/2}^2}{m_{\rm eff}^2 - 2\,m_{3/2}^2} \,M_{\rm Pl} \,, \quad
\left<Y\right> \simeq \frac{2\,\alpha\,v_P\, m_{3/2}}{2\,\alpha^2 v_P^2 + m_{3/2}^2} \, v_P \,, \quad
\left<P\right> \simeq
\bigg(1- \frac{m_{3/2}^2}{2\,\alpha^2 v_P^2 + m_{3/2}^2}\bigg)\sqrt{2}\, v_P \,.
\label{eq:VEVSUGRA}
\end{align}
These expressions are the SUGRA counterpart to the VEVs in Eqs.~\eqref{eq:VEVsglobal}
and represent our final results for $\left<\Phi\right>$,
$\left<Y\right>$, and $\left<P\right>$.
In this vacuum, also the fields $Y$ and $P$ have nonzero F-terms.
These F-terms are suppressed compared to the Polonyi F-term by
a factor of $\mathcal{O}\left(v_P/M_{\rm Pl}\right)$ and are, hence, negligible.


The masses $m_{3/2}'$ and $m_{\rm eff}$ scale as follows with
the Yukawa coupling $\lambda$ and the dynamical scale $\Lambda$,
\begin{align}
m_{\rm eff} \simeq 4\times 10^{10}\,\textrm{GeV}\:\left(\frac{\lambda}{0.2}\right)^3
\left(\frac{\Lambda}{10^{16}\,\textrm{GeV}}\right) \,, \quad
m_{3/2}' \simeq 4\times 10^{10}\,\textrm{GeV}\:\left(\frac{\lambda}{0.2}\right)
\left(\frac{\Lambda}{10^{16}\,\textrm{GeV}}\right)^2 \,.
\end{align}
For a fixed value of $\Lambda$, the requirement that
$m_{\rm eff}$ must exceed $m_{3/2}'$ then results in a lower bound on $\lambda$,
\begin{align}
m_{3/2}' \lesssim m_{\rm eff} \quad\Rightarrow\quad
\lambda \gtrsim 0.2 \left(\frac{\Lambda}{10^{16}\,\textrm{GeV}}\right)^{1/2} \,.
\end{align}
But this is not the end of the story.
We must also make sure that the final Polonyi VEV, $\left<\varphi\right>$,
lies below the critical field value $\varphi_c$, so as to stay in the
quadratic part of the effective potential close to the origin,
\begin{align}
\left<\varphi\right> \lesssim \varphi_c \,.
\end{align}
Otherwise, the mass term in the effective potential disappears altogether and the Polonyi field
rolls back to field values of the order of the Planck scale.
This requirement yields an even stronger bound on $\lambda$,
\begin{align}
\left<\varphi\right> \lesssim \varphi_c \quad\Rightarrow\quad
\lambda \gtrsim 1.0 \left(\frac{\Lambda}{10^{16}\,\textrm{GeV}}\right)^{1/3} \,.
\label{eq:lambdamin}
\end{align}
Together with the requirement of perturbativity,
this limits the range of viable $\lambda$ values to $1 \lesssim \lambda \lesssim 4$.


Having determined the position of the true vacuum after inflation in field
space, we are now finally in the position to link the $R$ symmetry-breaking
sector to the IYIT sector.
In order to tune the CC in the true vacuum to zero, the constant $w$ generated in the $R$
symmetry-breaking sector (see Eq.~\eqref{eq:w}) needs to be matched with
the value $w_0$ dictated by the IYIT sector (see Eq.~\eqref{eq:wm32mphiyp}).
For $N_c = N_f = 2$, we have
\begin{align}
w \simeq  m_{3/2}\, M_{\rm Pl}^2 \simeq \frac{1}{8\pi^2} \left<P\right> \tilde{\Lambda}^2 \,, \quad
w_0 \simeq \frac{\mu^2M_{\rm Pl}}{\sqrt{3}} \,,
\end{align}
which results in a condition on the dynamical scale $\tilde{\Lambda}$ in the $R$
symmetry breaking sector,
\begin{align}
w = w_0 \quad\Rightarrow\quad
\tilde{\Lambda} \simeq 
\left(\frac{8\pi^2}{\sqrt{3}}\frac{\mu^2M_{\rm Pl}}{\left<P\right>}\right)^{1/2} \,.
\end{align}
This solution for $\tilde\Lambda$ is only consistent as long as
it is smaller than the heavy-quark mass scale $\overline{M}_Q = \left<P\right>$,
i.e., as long as the gaugino condensation scale is smaller than heavy-quark
mass scale, $w \lesssim w_{\rm max}$ (see Eq.~\eqref{eq:wmax}).
This requirement can be reformulated as an upper bound on the SUSY-breaking scale $\mu$,
\begin{align}
\tilde\Lambda \lesssim \left<P\right> \quad\Rightarrow\quad
\mu \lesssim \left(\frac{\sqrt{3}}{8\pi^2} \frac{\left<P\right>^3}{M_{\rm Pl}}\right)^{1/2} \,.
\label{eq:mubound}
\end{align}
In supergravity, $\left<P\right>$ follows from Eq.~\eqref{eq:VEVSUGRA}.
Depending on our choice for $\alpha$, we find two solutions,
\begin{align}
\left<P\right> \simeq \left(1- \frac{1}{1+6\,a}\right) \sqrt{2}\,v_P
\simeq \sqrt{2}\,v_P \times
\begin{cases}
6/7 &  ; \quad a = a_1\left(\varphi_0\right) \simeq 1 \\
1   &  ; \quad a = a_2\left(\varphi_0\right) \gg 1
\end{cases} \,.
\end{align}
Together with the limits on the scale $v_P$ in Eq.~\eqref{eq:vmax12},
these relations imply upper bounds on $\Lambda$,
\begin{align}
\alpha = \alpha_1\left(\varphi_0\right) \quad\Rightarrow\quad 
\Lambda & \lesssim 2\times 10^{18}\,\textrm{GeV}\:\bigg(\frac{\epsilon}{0.2}\bigg)^{3/4}
\left(\frac{1}{\lambda}\right)^{1/2} \bigg(\frac{v_P}{v_P^{\rm max,1}}\bigg)^{3/2}
\,, \\ \nonumber
\alpha = \alpha_2\left(\varphi_0\right) \quad\Rightarrow\quad 
\Lambda & \lesssim 3\times 10^{14}\,\textrm{GeV}\:\bigg(\frac{\epsilon}{0.2}\bigg)^{3/4}
\left(\frac{1}{\lambda}\right)^{1/2}\bigg(\frac{v_P}{v_P^{\rm max,2}}\bigg)^{3/2}
\left(\frac{1}{1+b}\right)^{3/2} \bigg(\frac{\varphi_0}{10^{16}\,\textrm{GeV}}\bigg)^{3/2} \,.
\end{align}
Once this condition is satisfied, there always exists a sufficiently small
value of $\tilde{\Lambda}$ that allows us to tune the CC to zero
(so that $w=w_0$, while at the same time $\tilde\Lambda \lesssim \left<P\right>$).
We, however, note that the upper bound on $\Lambda$ in the case of the second solution
is too severe.
For such a small dynamical scale, the vacuum energy density during inflation 
does not suffice to account for the scalar spectral amplitude $A_s$.
The requirement of successful Polonyi inflation, therefore, singles out the
fine-tuned first solution $\alpha = \alpha_1\left(\varphi_0\right) \approx \alpha_0$.


\section{Phenomenological implications}
\label{sec:pheno}


\subsection{Properties of the scalar potential driving inflation}


We now have everything at our disposal that we need to study
the phenomenological implications of our model.
As argued in Sec.~\ref{subsec:SUGRA}, the scalar potential driving
inflation takes the following form (see Eq.~\eqref{eq:infpot}),
\begin{align}
V\left(\varphi\right) = V_0 + \frac{1}{2}\,m_\varphi^2\, \varphi^2 +
\frac{\lambda_\varphi}{4!}\,\varphi^4 + \Lambda_{\rm HE}^4\,\ln\frac{\varphi}{\varphi_c} \,,
\label{eq:Vinf}
\end{align}
where $V_0$ and $\Lambda_{\rm HE}^4$ are given in Eqs.~\eqref{eq:VPolonyi}
and \eqref{eq:LambdaLEHE}, respectively, and where
$m_\varphi^2$ and $\lambda_\varphi$ follow from Eq.~\eqref{eq:c01234},
\begin{align}
V_0 = \mu^4 \,, \quad
m_\varphi^2 = -\epsilon\,\frac{V_0}{M_{\rm Pl}^2} = - 3\,\epsilon\,H_{\rm inf} \,, \quad
\lambda_\varphi = 3\left(1-\frac{7\,\epsilon}{2}
+ \frac{8\,\epsilon}{3}\right)\frac{V_0}{M_{\rm Pl}^4} \,, \quad
\Lambda_{\rm HE}^4 = N_X\frac{m^4}{16\pi^2} \,.
\end{align}
This scalar potential is identical to the inflaton potential of F-term hybrid inflation
(including corrections from supergravity and a noncanonical K\"ahler potential)
in the limit of a vanishing gravitino mass.
Its implications for the inflationary CMB observables have been studied for the first time
in~\cite{BasteroGil:2006cm}.
In the following, we shall review the aspects of this
inflationary scenario that are most relevant for our purposes
and summarize the predictions for the inflationary CMB observables
in our notation.


First of all, we note that the potential in Eq.~\eqref{eq:Vinf}
has an inflection point at the following field value, 
\begin{align}
\varphi_{\rm flex} = \frac{1}{\lambda_\varphi^{1/2}}
\left[-m_\varphi^2+\left(m_\varphi^4 + 2\,\lambda_\varphi\,\Lambda_{\rm HE}^4\right)^{1/2}
\right]^{1/2} \,, \quad V''\left(\varphi_{\rm flex}\right) = 0 \,.
\label{eq:phiflex}
\end{align}
\textit{A priori}, the sign of the potential gradient at $\varphi = \varphi_{\rm flex}$
is undetermined.
In particular, the inflection point  at $\varphi = \varphi_{\rm flex}$ turns
into a saddle point, $V'\left(\varphi_{\rm flex}\right) = 0$, once
the following relation is satisfied,
\begin{align}
3\,m_\varphi^4 - 2\,\lambda_\varphi\,\Lambda_{\rm HE}^4 = 0 \,.
\label{eq:saddle}
\end{align}
This condition can be fulfilled by setting the coefficient $\epsilon$
to a particular critical value $\epsilon_0$,
\begin{align}
\epsilon_0 = \frac{12\,\rho}{21\,\rho+\left(57\,\rho^2 +72\,\rho\right)^{1/2}}
= \frac{12\,\lambda}{21\,\lambda+\left(45\,\lambda^2+384\pi^2\right)^{1/2}} \,, \quad
\rho = \frac{\Lambda_{\rm HE}^4}{V_0} \,,
\label{eq:epsilon0}
\end{align}
which solely depends on the size of the Yukawa coupling $\lambda$.
For $\lambda$ values of $\mathcal{O}(1)$, the critical coefficient $\epsilon_0$
takes values of $\mathcal{O}(0.1)$. 
Once we have $\epsilon>\epsilon_0$, the potential gradient at
$\varphi = \varphi_{\rm flex}$ is negative, $V'\left(\varphi_{\rm flex}\right) < 0$,
whereas, for $\epsilon<\epsilon_0$, the potential gradient at
$\varphi = \varphi_{\rm flex}$ is positive, $V'\left(\varphi_{\rm flex}\right) > 0$.
For $\epsilon>\epsilon_0$, the scalar potential, therefore, exhibits
a local minimum ($+$) as well as a local maximum ($-$)
in the vicinity of $\varphi_{\rm flex}$,
\begin{align}
\epsilon \geq \epsilon_0 \quad\Rightarrow\quad
\varphi_{\rm min, max} = \frac{\sqrt{3}}{\lambda_\varphi^{1/2}}
\left[-m_\varphi^2\pm \frac{1}{\sqrt{3}}
\left(3\,m_\varphi^4 - 2\,\lambda_\varphi\,\Lambda_{\rm HE}^4\right)^{1/2} \right]^{1/2} \,,
\quad \varphi_{\rm max} \leq \varphi_{\rm flex} \leq \varphi_{\rm min} \,.
\end{align}
Note that these two field values become identical,
once the condition in Eq.~\eqref{eq:saddle} is satisfied,
\begin{align}
\epsilon = \epsilon_0 \quad\Rightarrow\quad
\varphi_{\rm min} = \varphi_{\rm flex} = \varphi_{\rm max}
= \bigg(\frac{-3\,m_\varphi^2}{\lambda_\varphi}\bigg)^{1/2}
\simeq \epsilon_0^{1/2} M_{\rm Pl} \,.
\end{align}

\begin{figure}
\centering
\includegraphics[width=0.67\textwidth]{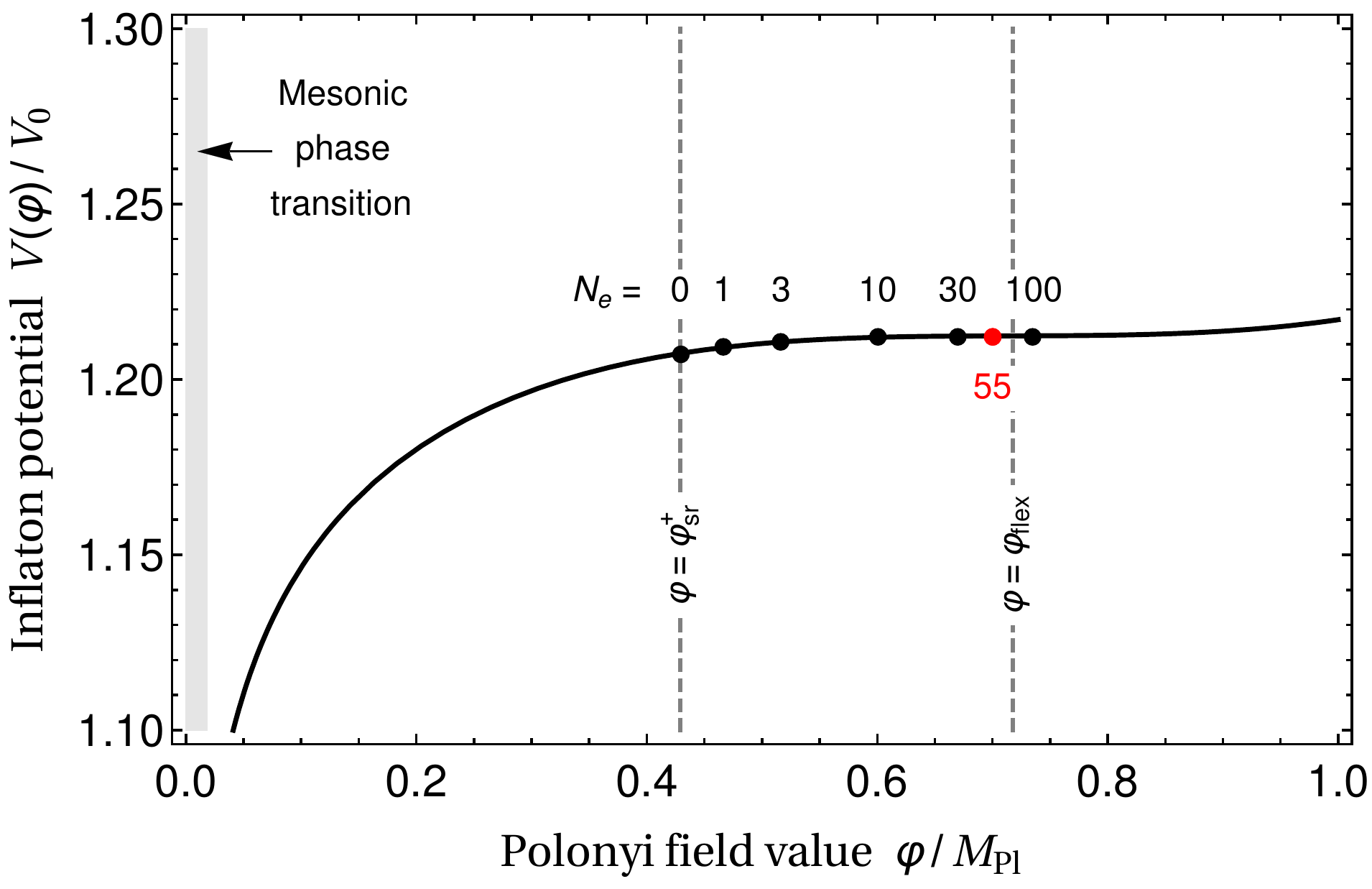}
\caption{Inflaton potential in the inflection-point regime ($\epsilon < \epsilon_0$).
Here, the values for the free parameters in the inflaton sector
have been chosen as in the benchmark scenario discussed in Sec.~\ref{subsec:bench}:
$\Lambda \simeq 1.26 \times 10^{16}\,\textrm{GeV}$,
$\lambda \simeq 1.66$, and $\epsilon \simeq 0.204$.}
\label{fig:inflation}
\end{figure}

The presence of two local extrema in the scalar potential
may be regarded as a disadvantage, as it requires the initial inflaton field
value, $\varphi_{\rm ini}$, to be smaller than $\varphi_{\rm max}$.
For one thing, this requirement may necessitate some amount of fine-tuning,
$\varphi_{\rm ini} \simeq \varphi_{\rm max}$,
so as to achieve a sufficient number of $e$-folds of inflation. 
For another thing, the constraint $\varphi_{\rm ini} \leq \varphi_{\rm max}$
is not compatible with the notion that 
inflation is expected to descend from a ``Planck epoch'', during which the inflaton
field, the potential energy density and all other relevant quantities
take values of the order of the Planck scale.
For these reasons, we consider the option $\epsilon > \epsilon_0$
less likely than the alternative possibility $\epsilon < \epsilon_0$.
We can  take this argument even one step further by making the explicit
assumption that, for one reason or another, $\varphi_{\rm ini}$ is necessarily
larger than the Planck scale, $\varphi_{\rm ini} \gtrsim M_{\rm Pl}$.
Under this assumption, $\epsilon$ \textit{must} be smaller than $\epsilon_0$,
as the inflaton would otherwise get trapped in the false vacuum at
$\varphi = \varphi_{\rm min}$.
From this perspective, the critical coefficient $\epsilon_0$ plays
the role of an upper bound on $\epsilon$, which decides whether the inflaton
field has a chance of reaching the true vacuum or not.
In Fig.~\ref{fig:inflation}, we plot the scalar potential in Eq.~\eqref{eq:Vinf}
for an $\epsilon$ value that is slightly smaller than the critical value $\epsilon_0$.
In this case, the scalar potential features a flat plateau around
$\varphi \sim \varphi_{\rm flex}$, which gives rise to successful inflation
in accord with the observational data.


Last but not least, we note that the scalar potential in Eq.~\eqref{eq:Vinf}
only allows for slow-roll inflation within a limited range of field values.
The slow-roll conditions, $\varepsilon \ll 1$ and $\eta \ll 1$
(see Eq.~\eqref{eq:SRpara}),
are only satisfied in the interval 
$\varphi_{\rm sr}^+ \leq \varphi \leq\varphi_{\rm sr}^-$, where
the two boundary values $\varphi_{\rm sr}^\pm$ are given as follows,%
\footnote{In principle, Eq.~\eqref{eq:phiSR} represents
an implicit definition of $\varphi_{\rm sr}^\pm$, as $M_\pm$
depends on $V\left(\varphi_{\rm sr}^{\pm}\right)$.
In practice, this dependence is, however, very weak,
$V\left(\varphi_{\rm sr}^{\pm}\right) \approx V_0$, so that
the right-hand side of Eq.~\eqref{eq:phiSR} readily yields the desired values
of $\varphi_{\rm sr}^\pm$.}
\begin{align}
\varphi_{\rm sr}^{\pm} = \frac{1}{\lambda_\varphi^{1/2}}
\left[-M_\pm^2+\left(M_\pm^4 + 2\,\lambda_\varphi\,\Lambda_{\rm HE}^4\right)^{1/2}
\right]^{1/2} \,, \quad M_\pm = \left(m_\varphi^2 \pm
\left|\eta_{\rm max}\right|\frac{V\left(\varphi_{\rm sr}^{\pm}\right)}
{M_{\rm Pl}^2}\right)^{1/2} \,.
\label{eq:phiSR}
\end{align}
Here, $\left|\eta_{\rm max}\right|$ denotes the maximally allowed
absolute value of the slow-roll parameter $\eta$.
The slow-roll parameter $\varepsilon$ is, by comparison, always subdominant
during inflation, $\varepsilon \ll \left|\eta\right|$.
For definiteness, we will use $\left|\eta_{\rm max}\right| = 10^{-0.5}$
in the following.
Finally, we also mention that, if we formally take the limit
$\left|\eta_{\rm max}\right| \rightarrow 0$ in Eq.~\eqref{eq:phiSR},
the expression for $\varphi_{\rm sr}^{\pm}$ in Eq.~\eqref{eq:phiSR}
reduces to our result for $\varphi_{\rm flex}$ in Eq.~\eqref{eq:phiflex}.


For the parameter region of interest, $1 \lesssim \lambda \lesssim 4$
(see Eq.~\eqref{eq:lambdamin}), the field value $\varphi_{\rm sr}^+$ is 
always larger than the critical value field $\varphi_c$ associated
with the quark-meson phase transition in the IYIT model.
Inflation, therefore, always ends, once $\varphi$ reaches $\varphi_{\rm sr}^+$.
That is, inflation ends because the slow-roll conditions
become violated, and not because of a sudden phase transition (such as the one
in the $R$ symmetry-breaking sector) triggered by a small inflaton field value.
Correspondingly, the number of $e$-folds of slow-roll inflation, $N_e$,
in dependence of the inflaton field value is always given by the following integral,
\begin{align}
N_e \left(\varphi\right) = \int_{\varphi_{\rm sr}^+}^\varphi
\frac{d \varphi'}{M_{\rm Pl}} \frac{1}{\sqrt{2\,\varepsilon\left(\varphi'\right)}} \,,
\label{eq:Ne}
\end{align}
where $N_e$ is defined such that it counts the remaining number of $e$-folds
until the end of inflation.
In the following, we will work in the approximation that the number of $e$-folds
in between the end of slow-roll inflation and the onset of reheating is negligible.
This is to say that we assume all scalar fields to settle at their respective VEVs
sufficiently fast, once slow-roll inflation has ended.
Solving the integral in Eq.~\eqref{eq:Ne} and inverting the relation
$N_e = N_e \left(\varphi\right)$ then
provides us with the inflaton field value as a function of the number
of $e$-folds, $\varphi = \varphi\left(N_e\right)$.
For the purposes of this paper, we perform these steps numerically.


\subsection{Inflationary CMB observables}


The function $\varphi\left(N_e\right)$ is exactly what we need
to determine our predictions for the CMB observables (the scalar spectral amplitude
$A_s$, the scalar spectral index $n_s$ as well as the tensor-to-scalar ratio $r$),
\begin{align}
A_s = \frac{1}{24\pi^2\,\varepsilon_*}
\frac{V_*}{M_{\rm Pl}^4} \,, \quad
n_s = 1 + 2\,\eta_* - 6\,\varepsilon_* \,, \quad
r = 16\,\varepsilon_* \,.
\label{eq:CMBobs}
\end{align}
Here, the asterisk indicates that all quantities are to be evaluated
at $\varphi_* = \varphi\left(N_e^*\right)$, where $N_e^* \simeq 55$
denotes the required number of $e$-folds during slow-roll inflation.
According to the latest results of the PLANCK collaboration, the $95\,\%$
confidence intervals for these observables are given as follows~\cite{Ade:2015xua},
\begin{align}
A_s^{\rm obs} = e^{3.062\pm0.029} \times 10^{-10} \simeq 2 \times 10^{-9} \,, \quad
n_s^{\rm obs} = 0.9677 \pm 0.0060 \,, \quad r < 0.11 \,.
\end{align}
The scalar potential in Eq.~\eqref{eq:Vinf} is basically controlled by
three free parameters: the dynamical scale $\Lambda$, the Yukawa coupling $\lambda$,
as well as the coefficient in the higher-dimensional K\"ahler potential, $\epsilon$.
This parametric freedom always allows us to choose $\Lambda$, such that we succeed in
reproducing the correct normalization of the scalar power spectrum, $A_s \simeq A_s^{\rm obs}$.
After fixing $\Lambda$ in this way, we then have to deal
with two free parameters---$\lambda$ and $\epsilon$ (see also the discussion
at the end of Sec.~\ref{subsec:ingredients})---which leaves us with the task
of studying the predictions for the inflationary CMB observables as functions
of $\lambda$ and $\epsilon$.


\begin{figure}
\centering
\includegraphics[width=0.48\textwidth]{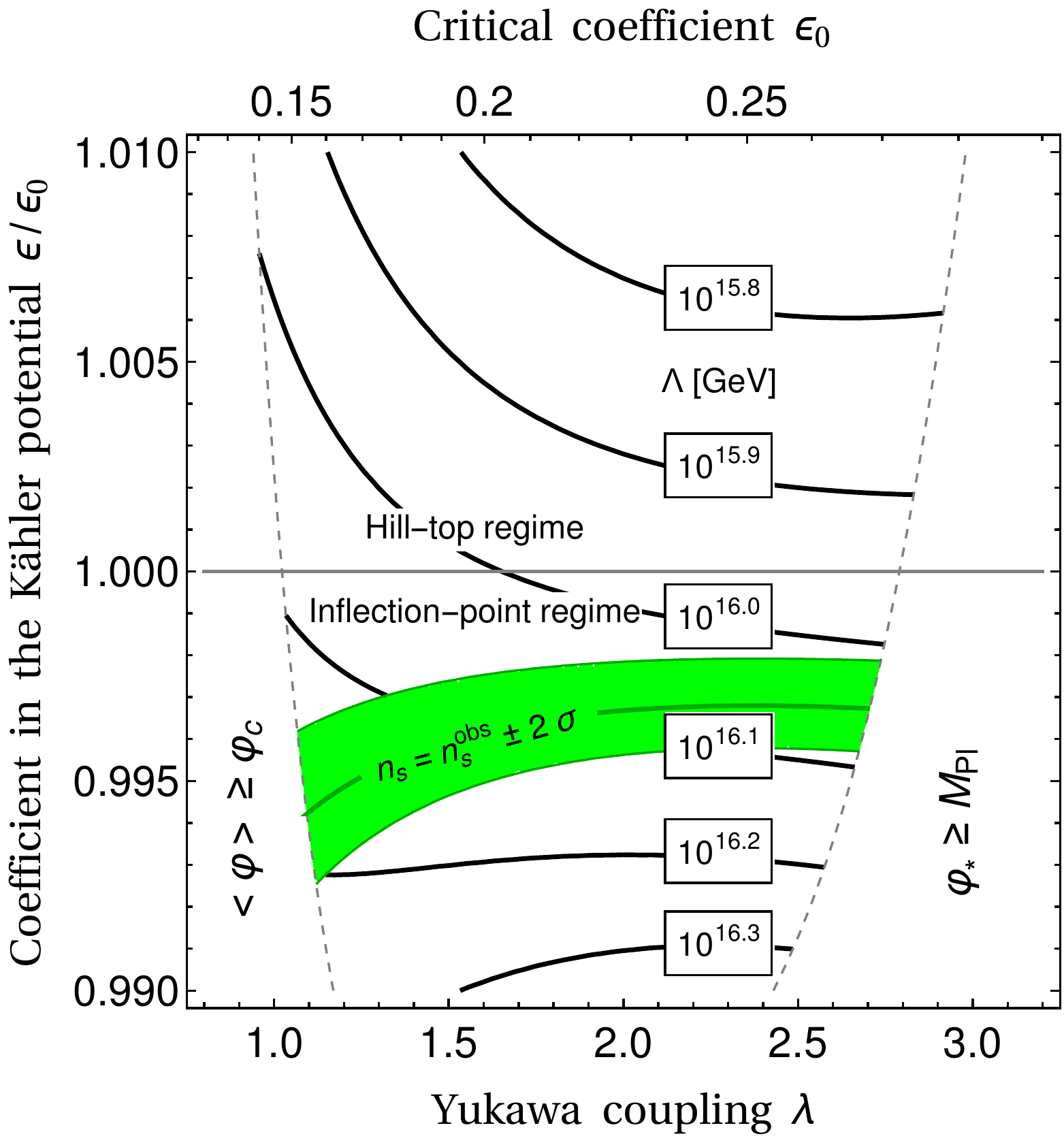}
\hfill
\includegraphics[width=0.48\textwidth]{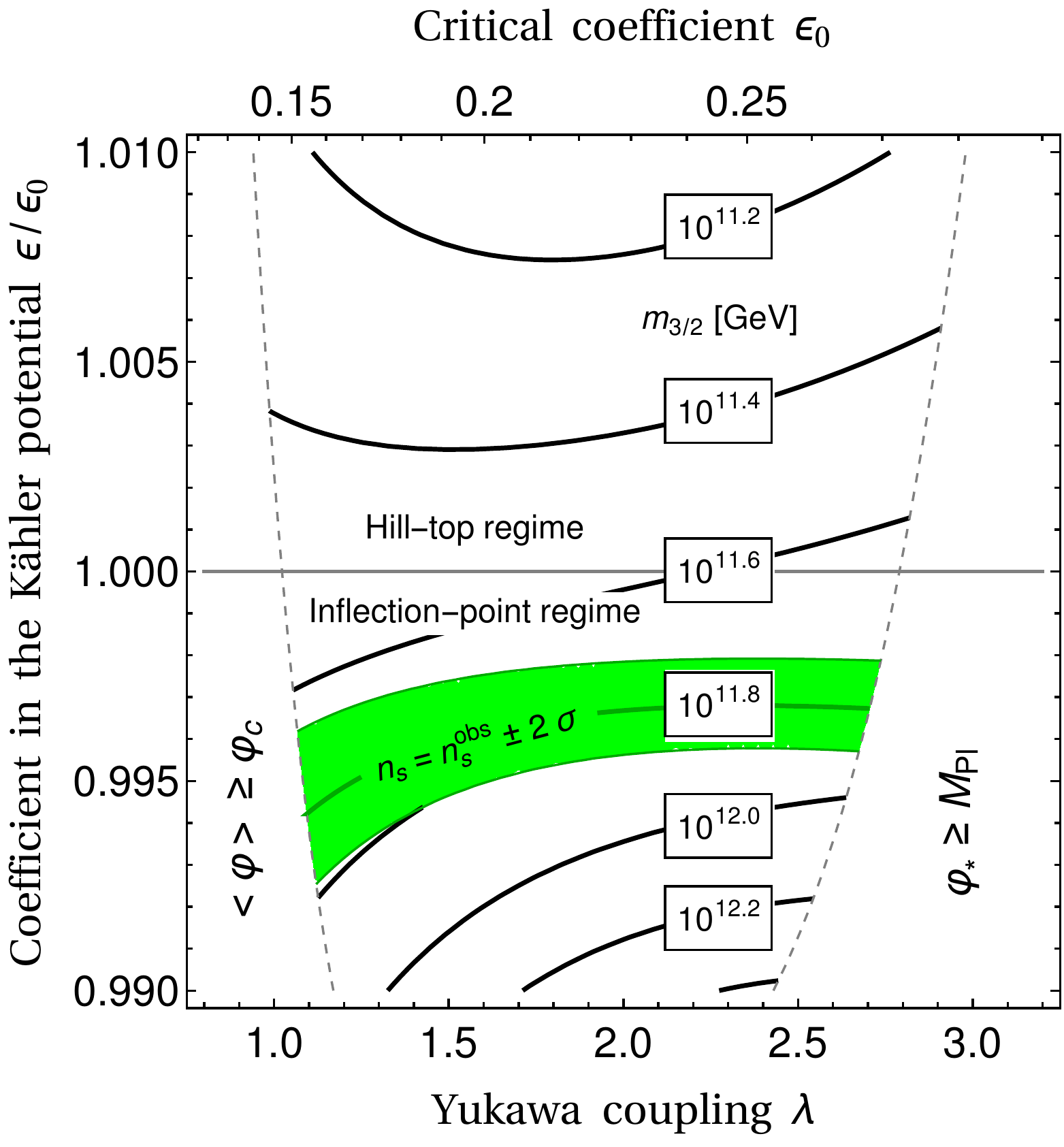}
\caption{Viable region in the parameter space spanned by the Yukawa coupling $\lambda$
and the coefficient in the noncanonical K\"ahler potential, $\epsilon$.
Here, the $\epsilon$ values on the vertical axis are normalized by the critical
values $\epsilon_0$, see Eq.~\eqref{eq:epsilon0}.
For $\lambda \lesssim 1$, the VEV of the Polonyi field begins
to exceed the critical field value $\varphi_c$, see Eq.~\eqref{eq:lambdamin},
whereas for $\lambda \gtrsim 3$, the Polonyi field begins
to take super-Planckian values during inflation (this follows from Eq.~\eqref{eq:Ne}).
The black contours respectively indicate the values of $\Lambda$ \textbf{(left panel)}
and $m_{3/2}$ \textbf{(right panel)} required to obtain the correct scalar
spectral amplitude, $A_s \simeq A_s^{\rm obs}$.}
\label{fig:scan}
\end{figure}


The outcome of our analysis is shown in Fig.~\ref{fig:scan}.
Both panels of Fig.~\ref{fig:scan} indicate where in parameter space our
prediction for $n_s$ falls into the observed $2\,\sigma$ range.
Interestingly enough, we are able to reproduce the observed value $n_s^{\rm obs}$
only in the ``inflection-point regime'' ($\epsilon < \epsilon_0$), in which
the scalar potential does not exhibit any local extrema close to
the inflection point at $\varphi = \varphi_{\rm flex}$.
In the ``hill-top regime'' ($\epsilon > \epsilon_0$), on the other hand,
the scalar spectral index always turns out to be small, $n_s \lesssim 0.94$.
If we adopt the notion that the critical coefficient $\epsilon_0$ should, in fact,
be regarded as an upper bound on $\epsilon$ (because the inflaton
would otherwise get trapped in a false vacuum), this result entails
the following interesting observation:
Maybe the coefficient $\epsilon$ has, for one reason or another, the tendency
to saturate its upper bound from below, $\epsilon \simeq \epsilon_0$.
The coefficient $\epsilon$ might, e.g., be intrinsically of $\mathcal{O}(1)$;
and it only ends up being slightly suppressed, $\epsilon \simeq 0.2$, because it
is bounded from above.
Alternatively, $\epsilon$ might tend to be close to $\epsilon_0$, because
such a parameter choice leads to a particularly flat potential and, hence,
to a particularly long period of inflation (i.e., to a large spatial volume
in which $\epsilon \simeq \epsilon_0$).
But anyhow, no matter what the origin of this fine-tuning is, the point is:
once we suppose that there is a physical reason that singles out $\epsilon$
values close to $\epsilon_0$, we are automatically led to
values of $n_s$ close to the observed value.
Or put differently, it is an intriguing coincidence that 
the observed value of $n_s$ is reproduced just in the vicinity
of the only special point on the $\epsilon$ axis.
One could have expected that successful inflation would
require $\epsilon\simeq\epsilon_0$.
But then it would not have been guaranteed that the resulting values of $n_s$
would end up being close to $n_s^{\rm obs}$.


Another interesting coincidence pertains to the required value
of the dynamical scale $\Lambda$.
As evident from the left panel of Fig.~\ref{fig:scan}, the scale $\Lambda$
needs to take values very close to the scale of grand unification,
\begin{align}
\Lambda \sim 10^{16}\,\textrm{GeV} \,.
\end{align}
This suggest the fascinating possibility that the scales of supersymmetry breaking,
inflation, and grand unification might, in fact, all be unified,
\begin{align}
\Lambda_{\rm SUSY} \equiv \Lambda_{\rm inf} \equiv \Lambda_{\rm GUT} \,.
\end{align}
A trivial condition for this kind of ``triple unification'' 
is that the energy scales of inflation and supersymmetry breaking are identical
to each other.
The unification of the three scales $\Lambda_{\rm SUSY}$, $\Lambda_{\rm inf}$,
and $\Lambda_{\rm GUT}$ can, thus, only be realized in the context of
Polonyi inflation and in no other inflation model.
In this paper, we have not made any attempt to embed our model into a larger
GUT framework.
More work on the possible connection between Polonyi inflation and grand unification
is, therefore, certainly needed.
As evident from the right panel of Fig.~\ref{fig:scan}, the gravitino turns out
to be superheavy in our model,
\begin{align}
m_{3/2} \sim 10^{12}\,\textrm{GeV} \,.
\end{align}
As already pointed out in the introduction (see Eq.~\eqref{eq:m32r}),
the gravitino mass is directly related to our prediction
for the tensor-to-scalar ratio $r$ (this relation readily follows
from $m_{3/2} \simeq H_{\rm inf}$ in our model),
\begin{align}
r \simeq 2 \times 10^{-5} \:\left(\frac{m_{3/2}}{10^{12}\,\textrm{GeV}}\right)^2 \,.
\end{align}
The experimental confirmation of this relation is certainly challenging---but
\textit{if} it should be accomplished, it would represent an unequivocal
smoking-gun signal for Polonyi inflation, i.e., for a close link between the
dynamics of supersymmetry breaking and inflation.


\subsection{Neutralino dark matter and thermal leptogenesis}


In the minimal framework of pure gravity mediation,
gravitino masses of $\mathcal{O}\left(10^{12}\right)\,\textrm{GeV}$ typically
result in MSSM gaugino masses $M_{\tilde{g},\tilde{w},\tilde{b}}$
of $\mathcal{O}\left(10^{10}\right)\,\textrm{GeV}$,
\begin{align}
M_{\tilde{g},\tilde{w},\tilde{b}} \sim \frac{m_{3/2}}{16\pi^2} \sim 10^{10}\,\textrm{GeV} \,,
\label{eq:Mgaugino}
\end{align}
where the suppression by the loop factor $1/\left(16\pi^2\right)$ is a
consequence of anomaly mediation~\cite{Giudice:1998xp,Dine:1992yw}.
Here, the wino $\tilde{w}$ typically ends up being the lightest gaugino.
At tree level, the gaugino masses vanish in our model.
To see this, recall that the Polonyi field $\Phi$ carries $R$ charge $2$.
Couplings between the Polonyi field and the SM gauge fields
of the form $\Phi\,\mathcal{W}_\alpha\mathcal{W}^\alpha$
are, thus, forbidden in the superpotential.
After inflation, the Polonyi field predominantly decays into gravitinos
(via the $\left|\Phi\right|^4$ term in the effective K\"ahler
potential~\cite{Kawasaki:2006gs}),
so that inflation is followed by a phase of gravitino domination
in our scenario~\cite{Jeong:2012en}.
The nonthermal wino production in gravitino decays then easily 
overcloses the universe~\cite{Moroi:1999zb}.
To achieve an acceptable present-day wino abundance, we, therefore,
have to assume that the wino mass is somehow suppressed.
Fortunately, this is possible in the context of pure gravity mediation,
where the gaugino masses in Eq.~\eqref{eq:Mgaugino} also receive threshold
corrections from Higgsino loops~\cite{Ibe:2006de,Ibe:2012hu}.
These loop corrections are potentially of the same order of magnitude as the
anomaly-mediated masses.
The different contributions to the wino mass can, in particular,
cancel, so that the mass $M_{\tilde{w}}$ is reduced down to%
\footnote{Alternatively, the wino mass may be reduced
via gauge mediation~\cite{Jeong:2012en}.
We also mention that, in split-SUSY spectra~\cite{ArkaniHamed:2004fb},
anomaly-mediated gaugino masses are absent altogether.
In this case, gaugino masses may be as small as
$m_{3/2}^3/M_{\rm Pl}^2\sim 1\,\textrm{GeV}$, which would even require
some extra mass contributions.
A split-SUSY spectrum may, e.g., be achieved, if supersymmetry breaking
is a pure SUGRA effect~\cite{Izawa:2010ym}.
It is an interesting open question how to realize a split-SUSY spectrum
in the context of DSB models \`a la IYIT.
We leave this question for future work. For now, we simply assume
that the contribution to $M_{\tilde{w}}$ from anomaly mediation
ends up being canceled by another mass contribution
of equal magnitude and opposite sign.}
\begin{align}
M_{\tilde{w}} \sim 3 \,\textrm{TeV} \,.
\end{align}
We shall assume that this is case for anthropic reasons.
In case the wino mass is in the TeV range, the wino reaches thermal equilibrium
after reheating and eventually represents an excellent candidate for
dark matter in the form of \textit{weakly interacting massive particles} or WIMPs.
Larger wino masses, on the other hand, result in an overabundance of
dark matter, thereby making our Universe hostile for life.


The phase of gravitino domination after Polonyi inflation ends once the gravitino
begins to decay into the massless fields of the MSSM.
Here, the corresponding gravitino decay rate is given as~\cite{Jeong:2012en}
\begin{align}
\Gamma_{3/2} \simeq \frac{193}{384\pi} \frac{m_{3/2}^3}{M_{\rm Pl}^2} \simeq
30 \,\textrm{MeV} \left(\frac{m_{3/2}}{10^{12}\,\textrm{GeV}}\right)^3 \,.
\end{align}
The reheating temperature reached during gravitino decay, $T_{\rm rh}$,
then takes the following value,
\begin{align}
T_{\rm rh} \simeq \left(\frac{90}{\pi^2\,g_*}\right)^{1/4} \sqrt{\Gamma_{3/2} M_{\rm Pl}}
\simeq 1 \times 10^8 \,\textrm{GeV} \left(\frac{m_{3/2}}{10^{12}\,\textrm{GeV}}\right)^{3/2} \,,
\end{align}
where $g_* = 915/4$ denotes the effective number of relativistic DOFs in the MSSM
at high temperatures.
Contrary to many other inflation models, the reheating temperature after Polonyi
inflation is, therefore, not a free parameter, which is controlled by the unknown strength
of the inflaton couplings to matter.
$T_{\rm rh}$ rather follows directly from the universal decay rate of the
gravitino, which only depends on the gravitino mass.
As the gravitino mass is, in turn, more or less fixed by the amplitude
of the scalar power spectrum in Polonyi inflation (see Fig.~\ref{fig:scan}),
we arrive at the interesting result that Polonyi inflation makes
a definite prediction for the reheating temperature: $T_{\rm rh} \sim 10^8 \,\textrm{GeV}$.
Remarkably enough, this is not far away from the temperature that is needed
for the successful realization of thermal leptogenesis~\cite{Fukugita:1986hr}.
In its simplest form, thermal leptogenesis requires the reheating temperature
to be at least of $\mathcal{O}\left(10^9\right)\,\textrm{GeV}$.
But resonance effects in the case of mildly degenerate heavy-neutrino mass
spectrum easily increase the efficiency of thermal leptogenesis
also at lower temperatures~\cite{Pilaftsis:1997jf}.
The generation of the baryon asymmetry of the universe after Polonyi
inflation may, thus, very well be accounted for by thermal leptogenesis.


\subsection{Benchmark scenario}
\label{subsec:bench}


\begin{table}
\centering
\begin{tabular}{llll}
Quantity & Symbol & Value [GeV] & Reference\\\hline\hline
\multicolumn{4}{c}{$\vphantom{\bigg(}$Energy and mass scales
in the SUSY-breaking sector} \\\hline
Dynamical scale                                        & $\Lambda$
& $1.3 \times 10^{16}$                   & Eq.~\eqref{eq:Mij}   \\
Effective supersymmetric mass scale                    & $\overline{m}$
& $1.3 \times 10^{16}$                   & Eq.~\eqref{eq:x}     \\
Soft SUSY-breaking mass scale                          & $m$
& $1.7 \times 10^{15}$                   & Eq.~\eqref{eq:mara}  \\
SUSY breaking scale                                    & $\mu$
& $1.5 \times 10^{15}$                   & Eq.~\eqref{eq:mu}    \\
Energy scale of inflation                              & $V_*^{1/4}$
& $1.5 \times 10^{15}$                   & Eq.~\eqref{eq:CMBobs} \\
Effective potential energy scale at large field values & $\Lambda_{\rm HE}$
& $7.4 \times 10^{14}$                   & Eq.~\eqref{eq:LambdaLEHE} \\
Effective potential energy scale at small field values & $\Lambda_{\rm LE}$
& $5.9 \times 10^{14}$                   & Eq.~\eqref{eq:LambdaLEHE} \\
Effective Polonyi mass around the origin               & $m_{\rm eff}$
& $3.2 \times 10^{13}$                   & Eq.~\eqref{eq:meff2} \\
Gravitino mass                                         & $m_{3/2}$
& $5.6 \times 10^{11}$                   & Eq.~\eqref{eq:m32} \\
Inflationary Hubble rate                               & $H_{\rm inf}$
& $5.6 \times 10^{11}$                   & Eq.~\eqref{eq:bubbles} \\\hline
\multicolumn{4}{c}{$\vphantom{\bigg(}$Energy and mass scales
in the $R$ symmetry-breaking sector} \\\hline
VEV of the field $P$ in the true vacuum$^*$            & $\left<P\right>$
& $2.9 \times 10^{17}$                   & Eq.~\eqref{eq:VEVSUGRA}   \\
Characteristic value for the scale $v_P$               & $\bar{v}_P$
& $2.4 \times 10^{17}$                   & Eq.~\eqref{eq:vPbench}   \\
Gaugino condensation scale                             & $\tilde\Lambda_{\rm eff}$
& $6.4 \times 10^{16}$                   & Eq.~\eqref{eq:GCscale}   \\
VEV of the field $Y$ in the true vacuum$^*$            & $\left<Y\right>$
& $1.2 \times 10^{17}$                   & Eq.~\eqref{eq:VEVSUGRA}   \\
Dynamical scale$^*$                                    & $\tilde\Lambda$
& $3.0 \times 10^{16}$                   & Eq.~\eqref{eq:Lambdatilde} \\
Constant term in the superpotential                    & $w^{1/3}$
& $1.5 \times 10^{16}$                   & Eq.~\eqref{eq:WeffR}   \\
Mass of the field $P$ in the true vacuum$^*$           & $m_p$
& $1.5 \times 10^{12}$                   & Eq.~\eqref{eq:wm32mphiyp}   \\
Mass of the field $Y$ in the true vacuum$^*$           & $m_y$
& $1.5 \times 10^{12}$                   & Eq.~\eqref{eq:wm32mphiyp}    \\\hline
\multicolumn{4}{c}{$\vphantom{\bigg(}$Important Polonyi field values during and after inflation} \\\hline
Beginning of slow-roll inflation                               & $\varphi_{\rm sr}^-$
& $1.04\,M_{\rm Pl}$                                           & Eq.~\eqref{eq:phiSR} \\
Location of the inflection point                               & $\varphi_{\rm flex}$
& $0.72\,M_{\rm Pl}$                                           & Eq.~\eqref{eq:phiflex} \\
Field value $N_e^* = 55$ $e$-folds before the end of inflation & $\varphi_*$
& $0.70\,M_{\rm Pl}$                                           & Eq.~\eqref{eq:Ne} \\
End of slow-roll inflation                                     & $\varphi_{\rm sr}^+$
& $0.43\,M_{\rm Pl}$                                           & Eq.~\eqref{eq:phiSR} \\
Critical field value                                           & $\varphi_c$
& $1.5 \times 10^{16}$                           & Eq.~\eqref{eq:phic} \\
Polonyi VEV in the true vacuum                                 & $\left<\varphi\right>$
& $3.6 \times 10^{15}$                           & Eq.~\eqref{eq:VEVSUGRA} \\\hline\hline
\end{tabular}
\caption{Numerical results for a number of important energy scales, masses and field
values in the context of the benchmark scenario discussed in Sec.~\ref{subsec:bench}.
The input parameter values used in this scenario are listed in
Eqs.~\eqref{eq:point}, \eqref{eq:vPbench}, and \eqref{eq:alphabetabench}.
Quantities labeled with a star ($^*$) are evaluated for $v_P = \bar{v}_P$.
Their scaling with $v_P$ is indicated in Eqs.~\eqref{eq:PYLbench} and
\eqref{eq:alphabetabench}.}
\label{tab:benchmark}
\end{table}


Finally, we shall illustrate our findings by means of a
concrete example.
To this end, we will now consider the following point in parameter space,
which may be regarded as a representative benchmark scenario,
\begin{align}
\Lambda \simeq 1.26 \times 10^{16}\,\textrm{GeV} \,, \quad
\lambda \simeq 1.66 \,, \quad \epsilon \simeq 0.204 \,.
\label{eq:point}
\end{align}
For this choice of $\lambda$, the critical coefficient
$\epsilon_0$ is given by $\epsilon_0 \simeq 0.205$ (see Eq.~\eqref{eq:epsilon0}).
Our choice of $\epsilon$ is, therefore, fine-tuned in the sense that it
deviates from $\epsilon_0$ only by half a per cent or so.
For the parameter values in Eq.~\eqref{eq:point}, we then obtain the following
predictions for the inflationary CMB observables,
\begin{align}
A_s \simeq 2.14 \times 10^{-9} \,, \quad n_s \simeq 0.968 \,, \quad r \simeq 6.01 \times 10^{-6} \,,
\end{align}
where the values for $A_s$ and $n_s$ coincide with the experimental best-fit values
by construction.
Next to the parameters $\Lambda$, $\lambda$, and $\epsilon$ (which represent the free
parameters in the SUSY-breaking sector), we also need to specify the parameters in
the $R$ symmetry-breaking sector, i.e., the coefficients $\alpha$ and $\beta$ as well
as the energy scale $v_P$ (see Eq.~\eqref{eq:WP}).
Here, the scale $v_P$ is required to take a value within a finite interval,
\begin{align}
5.3 \times 10^{16}\,\textrm{GeV} \lesssim v_P \lesssim 1.1 \times 10^{18}\,\textrm{GeV} \,.
\label{eq:vPinterval}
\end{align}
We recall that the lower bound on $v_P$ is a consequence of the requirement that the heavy-quark
mass scale in the $R$ symmetry-breaking sector should always exceed the dynamical scale,
$\left<P\right> \gtrsim \tilde\Lambda$ (see Eq.~\eqref{eq:mubound}), while
the upper bound on $v_P$ follows from the requirement that the dynamics of the 
$R$ symmetry-breaking sector should have no noticeable effect on inflation,
$v_P \ll \epsilon^{1/2}\,M_{\rm Pl}$ (see Eq.~\eqref{eq:vmax12}).
In the following, we shall use the geometric mean of the two boundary values
in Eq.~\eqref{eq:vPinterval} as a characteristic value for $v_P$,
\begin{align}
\bar{v}_P \simeq 2.41 \times 10^{17}\,\textrm{GeV} \,.
\label{eq:vPbench}
\end{align}
For the VEVs $\left<P\right>$ and $\left<Y\right>$ as well as for the dynamical scale $\tilde\Lambda$,
we then obtain (see Eqs.~\eqref{eq:Lambdatilde} and \eqref{eq:VEVSUGRA})
\begin{align}
\left<P\right> \simeq 2.9 \times 10^{17}\,\textrm{GeV}\left(\frac{v_P}{\bar{v}_P}\right) \,, \quad
\left<Y\right> \simeq 1.2 \times 10^{17}\,\textrm{GeV}\left(\frac{v_P}{\bar{v}_P}\right) \,, \quad
\tilde\Lambda \simeq 3.0 \times 10^{16}\,\textrm{GeV}\left(\frac{\bar{v}_P}{v_P}\right)^{1/2} \,.
\label{eq:PYLbench}
\end{align}
Similarly, we find the following relations for the two reference values $\alpha_0$
and $\beta_0$ (see Eqs.~\eqref{eq:alpha0} and \eqref{eq:beta0}),
\begin{align}
\alpha_0 \simeq 4.0 \times 10^{-6} \left(\frac{\bar{v}_P}{v_P}\right) \,, \quad
\beta_0  \simeq 4.0 \times 10^{-6} \left(\frac{\bar{v}_P}{v_P}\right) \,.
\label{eq:alphabetabench}
\end{align}
The actual value of $\alpha$ is given by $a_1^{1/2}\left(\varphi_0\right)\, \alpha_0$
(see Eq.~\eqref{eq:a12}), where $a_1\left(\varphi_0\right)$ varies as a function of
$\varphi_0$, i.e, the Polonyi field value at the onset of gaugino condensation
in the $R$ symmetry-breaking sector.
Allowing for $\varphi_0$ values in the range $0\lesssim \varphi_0 \lesssim \varphi_c$,
the function $a_1\left(\varphi_0\right)$ roughly takes values between $1.00$ and $1.02$.
Up to deviations of the order of one per cent or so, the coefficient $\alpha$, therefore,
coincides with the reference value $\alpha_0$.
Meanwhile, the coefficient $\beta$ can be freely chosen, as long as we respect the
constraint $\left|\beta\right| \lesssim \beta_0$.


The parameter values in Eqs.~\eqref{eq:point}, \eqref{eq:vPbench}, and \eqref{eq:alphabetabench}
now completely fix the numerical properties of our benchmark scenario.
In Tab.~\ref{tab:benchmark}, we give an overview of all the resulting numerical
values for the various energy scales and field values of interest in our model.
Tab.~\ref{tab:benchmark}, thus, provides an example for a possible realization of our model,
which illustrates how to achieve successful Polonyi inflation as well as spontaneous
$R$ symmetry breaking at late times in the context of strongly coupled gauge theories.


\section{Conclusions and outlook}
\label{sec:conclusions}


The large value of the SM Higgs boson mass as well as the null result
of SUSY searches at the LHC call for a paradigm shift
in our expectations towards the role of supersymmetry in nature.
From the perspective of string theory, dark matter, and grand unification,
supersymmetry still represents a well-motivated extension of the standard
model.
But in view of the recent experimental data, we now begin to realize
that supersymmetry's main purpose may actually not lie in stabilizing the
electroweak scale.
Instead, it now appears more likely that supersymmetry is,
in fact, spontaneously broken at a scale much higher than the electroweak scale.
This opens up a whole range of new phenomenological possibilities.


As we have demonstrated in this paper, the spontaneous breaking of
supersymmetry at a high scale, $\Lambda \sim 10^{16}\,\textrm{GeV}$, may, e.g.,
offer a dynamical explanation for the occurrence of cosmic inflation in
the early universe.
We have dubbed the ensuing inflationary scenario \textit{Polonyi inflation},
as it identifies cosmic inflation as a natural by-product of spontaneous
supersymmetry breaking in an effective Polonyi model.
Generally speaking, the key idea of our proposal is to obtain inflation
from the Polonyi superpotential,
\begin{align}
W = \mu^2 \,\Phi + w \,,
\label{eq:PolonyiW}
\end{align}
with the Polonyi superfield $\Phi$ breaking supersymmetry via its nonvanishing
F-term, $\left<\left|F_\Phi\right|\right> = \mu^2$, and acting as the chiral inflaton
field at the same time.
In the context of Polonyi inflation, spontaneous supersymmetry breaking and
cosmic inflation are nothing but two sides of the same coin:
Inflation is driven by the vacuum energy density associated with the
spontaneous breaking of supersymmetry and the scalar inflaton is identified
as the complex Polonyi field, i.e., the pseudoflat direction in the scalar
potential of the SUSY-breaking sector.
This connection between supersymmetry breaking and inflation
results in several characteristic parameter relations
that cast a new light on well-known quantities.
In Polonyi inflation, the Hubble rate during inflation, e.g., is
equal to the gravitino mass at low energies,
\begin{align}
H_{\rm inf} \simeq m_{3/2} \sim 10^{12}\,\textrm{GeV} \,.
\end{align}
This implies that the amplitude of the
primordial density fluctuations, $\delta\rho/\rho$, scales with $m_{3/2}$,
\begin{align}
\frac{\delta \rho}{\rho} = A_s^{1/2} = \frac{\sqrt{2}}{\pi} \frac{1}{r^{1/2}}
\frac{m_{3/2}}{M_{\rm Pl}} \sim 10^{-5} \: \left(\frac{10^{-4}}{r}\right)^{1/2}
\bigg(\frac{m_{3/2}}{10^{12}\,\textrm{GeV}}\bigg) \,.
\end{align}
There are indications that the observed value of $\delta\rho/\rho$
might be the result of anthropic selection~\cite{Tegmark:1997in}.
Given typical values of the tensor-to-scalar ratio $r$ in
slow-roll inflation, the above relation then points towards
very large values of the gravitino mass.
In Polonyi inflation, the ultimate reason why supersymmetry is
broken at a high scale, therefore, consists in the necessity 
of reproducing the anthropic value of $\delta\rho/\rho$.
The general concept of Polonyi inflation, thus, not only yields an
answer to the question \textit{``Why inflation?''}, it also explains
why supersymmetry is necessarily broken at a very high scale.
Moreover, we point out that $m_{3/2} \sim 10^{12}\,\textrm{GeV}$
is an interesting result in view of the
stability of the SM Higgs potential.
Recall that, solely within the standard model, the Higgs quartic coupling
(most likely) turns negative at energies
around $\mathcal{O}\left(10^{11\cdots12}\right)\,\textrm{GeV}$~\cite{Degrassi:2012ry}.
This instability can be remedied by supersymmetry---as long as the
soft sparticle masses are at most of $\mathcal{O}\left(10^{12}\right)\,\textrm{GeV}$.
In this sense, our result for $m_{3/2}$ turns out to saturate the upper bound
on the gravitino mass implied by vacuum stability!
This may or may not be a coincidence.


In this paper, we have constructed a minimal model that illustrates
how the idea of Polonyi inflation may be realized in the context of strongly
coupled supersymmetric gauge theories.
Our construction consists of two separate hidden sectors that
respectively account for the dynamical origin of the parameters
$\mu$ and $w$ in Eq.~\eqref{eq:PolonyiW}.
Here, one hidden sector is identical to the simplest version of the IYIT model
of dynamical supersymmetry breaking.
This sector contains the Polonyi field $\Phi$ and is responsible
for the generation of the SUSY-breaking parameter $\mu$ via strong dynamics.
The other hidden sector features the same matter content as the
SUSY-breaking sector, but contains less singlet fields.
The masses of the matter fields in this sector are controlled
by the VEV of a singlet field $P$.
At early times, $\left<P\right>$ vanishes.
At this time, the constant in the superpotential, $W \supset w$,
is zero.
As we are able to show, the effective scalar potential for the Polonyi
field at this stage then takes the same form as the inflaton potential of ordinary
F-term hybrid inflation in the limit of a vanishing gravitino mass.
It is this potential that gives rise to the actual period of Polonyi inflation.%
\footnote{It is amusing to note that the scalar potential of Polonyi inflation
is identical to the potential of F-term hybrid inflation in the special
limit $m_{3/2} \rightarrow 0$---albeit, in the true vacuum at low energies,
$m_{3/2}$ is exceptionally large in our model.}
At small Polonyi field values after the end of slow-roll inflation,
the Hubble-induced mass for the field $P$ decreases---until, at a certain field
value, $P$ becomes tachyonically unstable.
This instability results in a waterfall transition in the $R$ symmetry-breaking
sector, such that $\left<P\right> \neq 0$.
The second hidden sector then turns into a pure SYM theory, in which $R$
symmetry is broken via gaugino condensation.
In consequence of that, the constant term $w$ appears in the superpotential.
Here, the final value of $w$ needs to be fine-tuned, such that the CC in the true
vacuum vanishes.


To sum up, we conclude that our effective Polonyi model results in an inflationary
scenario similar to ordinary F-term hybrid inflation, apart from a few key differences:
(i) All corrections to the scalar potential
proportional to $m_{3/2}$ are missing in our model.
(ii) This means, in particular, that the scalar potential
does not contain any odd powers of the inflaton field.
Because of that, the inflaton potential does not depend on the
complex phase of the inflaton field, as it is usually the
case in F-term hybrid inflation.
(iii) Moreover, in our scenario, the low-energy value of the gravitino mass,\
$m_{3/2}$, is not bounded from above by the inflationary Hubble rate.
In Polonyi inflation, these two scales are, in fact, equal to each other,
$m_{3/2} \simeq H_{\rm inf}$, whereas
in F-term hybrid inflation we have to demand that
$m_{3/2} \lesssim 10^{-3} H_{\rm inf}$, so as to ensure
that the slow-roll conditions do not get violated.
(iv) In Polonyi inflation, the inflaton sector does not undergo a
waterfall phase transition at the end of inflation.
Instead, the inflationary vacuum energy density continues to act as
the vacuum energy density associated with the spontaneous breaking 
of supersymmetry after inflation.
(v) Meanwhile, small inflaton field values trigger a (harmless)
waterfall phase transition in a separate hidden sector, in the course of
which only an approximate global $Z_2$ symmetry becomes broken.
In contrast to F-term hybrid inflation, Polonyi inflation, therefore,
does not suffer from the production of dangerous topological defects
at the end of inflation.


In the present paper, we have only touched on the phenomenological
implications of our model and more work is certainly needed.
For one thing, the study of the inflationary phase may still be
further refined. 
As it turns out, inflation takes place at field values only shortly
below the Planck scale in our model. 
Inflaton terms of $\mathcal{O}\left(\varphi^6\right)$ in the scalar potential
may, therefore, have a noticeable effect on the predictions for the inflationary CMB
observables.
We do not expect these terms to change our conclusions qualitatively.
But quantitatively, they may be relevant.
For another thing, the end of inflation and the subsequent reheating
phase require a closer examination.
Here, it would be interesting to study the implications of
different interactions between the two hidden sectors in the K\"ahler
potential more comprehensively.
Also, the actual dynamics of reheating deserve
further investigation.
Our model does not suffer from the usual Polonyi
problem, as the decay of the Polonyi field is identified with the
first stage of the reheating process itself.
But besides that, a dedicated study of reheating tracking the
expansion of the universe during the transition to the radiation-dominated era
as well as the oscillatory motion of all scalar fields is needed.
Here, particular attention should be paid to the study of 
all possible decay modes of the Polonyi field as well as of the gravitino.
A better understanding of reheating would then enable us to better
assess the prospects of successful thermal leptogenesis.
Finally, the phenomenological implications of our model at low energies
need to be explored in more detail.
As examples of observational signatures at low energies, we merely
mention two interesting possibilities:
(i) In our scenario, a wino LSP with an anthropically selected mass
around 3 TeV accounts for dark matter.
This wino can be searched for in direct and indirect detection experiments.
In addition, the almost mass-degenerate chargino may be seen in the form of
macroscopic charged tracks in collider experiments.
All in all, our model predicts that the neutral and the charged wino are
the only sparticles that should show up at low energies.
All other sparticles have masses of at least
$\mathcal{O}\left(10^{10}\right)\,\textrm{GeV}$ and are, thus, expected
to be decoupled from low-energy phenomenology for the most part.
(ii) Moreover, the discrete $Z_4^R$ symmetry that we use to forbid the
constant term in the superpotential at early times predicts
the existence of vector-like matter fields charged under the SM gauge group,
which cancel the SM contributions to the $Z_4^R$ gauge anomalies.
These vector-like matter fields may have masses within the reach of collider
experiments, which would allow to probe our assumption of an anomaly-free
$R$ symmetry at accessible energies.


Last but not least, one should attempt to implement our idea of
Polonyi inflation into other dynamical models.
It would be interesting to assess which other
DSB models apart from the IYIT model might serve
as a UV completion of the inflationary dynamics.
In addition to that, a more systematic study of the Polonyi K\"ahler
potential---beyond near-canonical and approximately shift-symmetric K\"ahler
potentials---would be desirable.
Maybe there are particular (exotic) choices for the K\"ahler potential that
render it unnecessary to supplement the tree-level Polonyi model with
radiative corrections, as we have done it in this paper.
Likewise, our model of late-time $R$ symmetry breaking may be
extended or supplemented by alternative means to generate the
constant term in the superpotential at the end of inflation.
As an alternative to our scenario, which employs the waterfall
field $P$, one might consider a direct coupling of the Polonyi
field to the gauge fields of a strongly coupled pure SYM theory
via the gauge-kinetic function,
\begin{align}
\mathcal{L} \subset \int d^2\theta \,
\frac{1}{4}\left(\frac{1}{g_0^2} + \frac{\Phi^n}{\phi_0^n}\right) 
\mathcal{W}_\alpha \mathcal{W}^\alpha + \textrm{h.c.} \,, \quad n = 1,2, \cdots\,.
\end{align}
For appropriately chosen values of the parameters $g_0$ and $\phi_0$ and
the integer $n$, the SYM theory is weakly coupled during inflation, returning to the
strongly coupled regime only at the end of inflation.
This results in gaugino condensation and, hence,
late-time $R$ symmetry breaking at $\left|\phi\right| \lesssim \phi_0$.
We will give a more careful discussion of this mechanism in the context of
Polonyi inflation elsewhere.
Furthermore, gaugino condensation in a pure SYM theory is not
the only way to break $R$ symmetry and other possibilities should
be considered as well.
Eventually, the model presented in this paper may only be regarded
as a first step towards a new understanding of the close relation
between inflation and spontaneous supersymmetry breaking.
The concept of Polonyi inflation leads us into uncharted territory
and we are convinced that, exploring these new avenues,
we will encounter both big surprises and valuable insights.


\subsubsection*{Acknowledgements}


At the early stages of this project, one of us (K.\,S.) was still employed
as a postdoctoral researcher at Kavli IPMU. 
After the completion of roughly half the project, K.\,S.\ then transferred
to MPIK.
This work has been supported in part by Grants-in-Aid for Scientific Research
from the Ministry of Education, Science, Sports, and Culture (MEXT), Japan,
No.\ 26104009 and No.\ 26287039 (T.\,T.\,Y.) as well as by the World
Premier International Research Center Initiative (WPI), MEXT, Japan
(K.\,S.\ and T.\,T.\,Y.).


\newpage

\appendix


\section{Effective scalar potential for the Polonyi/inflaton field}
\label{app:potential}


In this appendix, we give the details of our derivation of
the effective one-loop potential, $V_{\rm 1-loop}$, for
the Polonyi field $\phi = \varphi/\sqrt{2}\,e^{i\theta}$,
i.e., the pseudoflat scalar direction in the IYIT DSB model,
see Eqs.~\eqref{eq:V1loopLEHE}, \eqref{eq:meff2},
\eqref{eq:LambdaLEHE}, and \eqref{eq:V1loop} in Sec.~\ref{subsec:potential}.
The starting point of our computation is the Coleman-Weinberg formula for
the effective one-loop potential~\cite{Coleman:1973jx},
which takes the following form in supersymmetric theories,
\begin{align}
V_{\rm 1-loop}\left(\varphi\right) = \frac{1}{64\pi^2}\,
\textrm{STr} \left[\mathcal{M}^4\left(\varphi\right)
\left(\ln\left(\frac{\mathcal{M}^2\left(\varphi\right)}{Q^2}\right)
+ c \right)\right] \,.
\label{eq:VCW}
\end{align}
Here, $Q$ is the renormalization scale and $c$ denotes a dimensionless constant
that is introduced for notational convenience.
We note that $Q$ and $c$ are only defined up to the following transformation,
\begin{align}
c \rightarrow c + \Delta c \,, \quad Q \rightarrow e^{\Delta c/2}\,Q \,,
\end{align}
so that the constant $c$ can always be removed by a finite renormalization.
Eq.~\eqref{eq:VCW} now tells us that the potential $V_{\rm 1-loop}$ is
given as the supertrace (STr) of a specific function
of the mass matrix $\mathcal{M}$, which encompasses all
bosonic and fermionic mass eigenvalues in the presence of
a nonvanishing Polonyi field background.
In a first step, we, therefore, need to determine the full mass spectrum
of the IYIT sector.


\subsection{Mass spectrum of the low-energy effective theory}
\label{subsec:massspectrum}


Apart from the Polonyi field $\Phi$ (which is massless at tree level in global supersymmetry),
the IYIT sector of our model consists of six meson flavors, $\Xi^0$ and $X^n$,
and six singlet fields, $\Sigma$ and $S_n$.
The scalar and fermionic masses for these chiral multiplets follow
from the effective tree-level superpotential in Eq.~\eqref{eq:Weff}.
For an arbitrary background value of the Polonyi field $\varphi$,
we find the following scalar mass eigenvalues,
\begin{align}
M_a^2\left(\varphi;p,q\right) =
m_a^2 + \frac{1}{2}\left(M^2\left(\varphi\right)+q\, m^2\right) + \frac{p}{2}
\left[\left(M^2\left(\varphi\right)+q\, m^2\right)^2 +
\left(2\,m_a\, M\left(\varphi\right)\right)^2\right]^{1/2} \,,
\label{eq:M2bos}
\end{align}
where $a = 0$ refers to the zeroth flavor, i.e., to the fields $\Xi^0$ and $\Sigma$,
while $a = 1,\cdots5$ refers to the $n$ other flavors, i.e., to the fields $X^n$ and $S_n$.
Moreover, the discrete parameter $p=\pm1$ distinguishes
between the two different types of particles involved,
i.e., between mesons ($p=+1$) and singlets ($p=-1$), while the
parameter $q=\pm1$ accounts for the soft
mass splitting within each complex scalar, $\pm m^2$, in consequence of
spontaneous supersymmetry breaking.
In this respect, Eq.~\eqref{eq:M2bos} serves as an illustration of
our statement in Sec.~\ref{subsec:SUSY}, where we note that the mass parameter
$m$ plays the role of the soft SUSY-breaking
mass scale in the IYIT model, see the discussion below Eq.~\eqref{eq:mara}.
By setting the soft mass $m$ to zero, we then readily obtain the mass eigenvalues of
the fermionic components in the meson and singlet multiplets,
\begin{align}
m = 0 \quad\Rightarrow\quad
\widetilde{M}_a^2\left(\varphi;p\right) = m_a^2 + \frac{1}{2} M^2\left(\varphi\right)
+ \frac{p}{2} \left[M^4\left(\varphi\right) + \left(2\,m_a\,
M\left(\varphi\right)\right)^2\right]^{1/2} \,.
\label{eq:M2fer}
\end{align}
For $\varphi \neq 0$, all scalar and fermionic masses are different from each other
(as long as none of the Dirac masses $m_a$ are identical).
Away from the origin, the meson and singlet fields, therefore, give
rise to, in total, 24 real scalars and 12 Majorana fermions.
At the origin, $M_a^2\left(\varphi;p,q\right)$ and
$\widetilde{M}_a^2\left(\varphi;p\right)$ reduce to
\begin{align}
M\left(\varphi\right) = 0 \quad\Rightarrow\quad
M_a^2\left(0;p,q\right) = 
m_a^2 + \frac{1}{2}\left(p+q\right)m^2 \,, \quad 
M_a^2\left(0;p,q\right) = m_a^2 \,,
\label{eq:Ma20}
\end{align}
which illustrates that, for $\varphi = 0$, half of the real scalars and all of
the Majorana fermions pair up to form complex scalars and Dirac fermions, respectively.
At $\varphi = 0$, we then end up with 12 real and 6 complex scalars
as well as with 6 Dirac fermions.
Here, the complex scalars and the Dirac
fermions are, in particular, composed half of mesonic and half of singlet DOFs.


Furthermore, we note that none of the masses
$M_a^2\left(\varphi;p,q\right)$ and
$\widetilde{M}_a^2\left(\varphi;p\right)$ ever turns tachyonic,%
\footnote{A necessary and sufficient condition for $M_a^2\left(\varphi;p,q\right)\geq0$
is that the soft SUSY-breaking mass scale $m$ does not exceed any of the
supersymmetric Dirac masses: $m_a \geq m$ for all flavors $a$, which is always
satisfied in our model, see Eq.~\eqref{eq:rbound}.}
which means that the meson and singlet fields always remain stabilized at the origin.
This should be compared with the situation in ordinary F-term hybrid inflation,
where the masses of the FHI waterfall fields correspond to (the nonzero values
of) $M_a^2\left(\varphi;p,q\right)$ and
$\widetilde{M}_a^2\left(\varphi;p\right)$ in the limit of zero Dirac masses,
\begin{align}
m_a = 0 \quad\Rightarrow\quad
M_a^2\left(\varphi;p,q\right) & = \frac{1}{2}\left(M^2\left(\varphi\right)+q\, m^2\right)
+ \frac{p}{2} \left|M^2\left(\varphi\right)+q\, m^2\right|
\rightarrow M^2\left(\varphi\right)+q\, m^2 \,,
\label{eq:Ma2FHI}\\ \nonumber
\widetilde{M}_a^2\left(0;p\right) & = \frac{1}{2}\left(1+p\right)M^2\left(\varphi\right)
\rightarrow M^2\left(\varphi\right) \,.
\end{align}
In this case, the scalar mass eigenstates corresponding to $q = -1$
are tachyonically unstable, once $M\left(\varphi\right) < m$.
This instability is absent in our ``waterfall transition-free'' version of
F-term hybrid inflation.


Finally, before moving on and presenting our results for the effective Polonyi potential,
let us state approximate expressions for $M_a^2\left(\varphi;p,q\right)$ and
$\widetilde{M}_a^2\left(\varphi;p\right)$ that are valid at small and large
values of the Polonyi field, respectively, and which will become useful later on.
In the low-energy regime close to the origin in field space, i.e., at small
values of the order parameter, $x\left(\varphi\right) = \varphi / \varphi_c \lesssim 1$
(see Eq.~\eqref{eq:x}), we find,
\begin{align}
x\left(\varphi\right) \lesssim 1 \quad\Rightarrow\quad
M_a^2\left(\varphi;p,q\right) & =
m_a^2\left(1 + p\,\frac{M^2\left(\varphi\right)}{m^2}\right) +
\frac{q}{2}\left(p+q\right)\left(M^2\left(\varphi\right)+ q\,m^2\right)
+ \mathcal{O}\left(x^4\right) \,, \label{eq:Ma2LE} \\ \nonumber
\widetilde{M}_a^2\left(\varphi;p\right) & =
m_a^2 + \frac{1}{2} M^2\left(\varphi\right) +
p\,m_a\, M\left(\varphi\right) + \mathcal{O}\left(x^3\right) \,,
\end{align}
which reduces to the expressions in Eq.~\eqref{eq:Ma20} in the limit $\varphi\rightarrow0$.
Conversely, in the high-energy regime, at large values of
$x\left(\varphi\right)$, the masses $M_a^2\left(\varphi;p,q\right)$ and
$\widetilde{M}_a^2\left(\varphi;p\right)$ may be approximated as follows,
\begin{align}
x\left(\varphi\right) \gtrsim 1 \:\:\Rightarrow\:\:
M_a^2\left(\varphi;p,q\right) & = \frac{1}{2}\left(1+p\right)\left(M^2\left(\varphi\right)
+ 2\,m_a^2 + q\, m^2 \right) - p\, \frac{m_a^2\left(m_a^2 + q\, m^2\right)}{M^2\left(\varphi\right)}
+ \mathcal{O}\left(x^{-4}\right) \,, \label{eq:Ma2HE} \\ \nonumber 
\widetilde{M}_a^2\left(\varphi;p\right) & = \frac{1}{2}\left(1+p\right)
\left(M^2\left(\varphi\right) + 2\, m_a^2\right) - p\,\frac{m_a^4}{M^2\left(\varphi\right)}
+ \mathcal{O}\left(x^{-4}\right) \,.
\end{align}
In this regime, the meson masses ($p=+1$) asymptotically
approach the usual expressions from F-term hybrid inflation,
see Eq.~\eqref{eq:Ma2FHI}, while the the singlet masses ($p=-1$) only
acquire suppressed masses of $\mathcal{O}\left(m_a^2/M\right)$.
Here, note that the small singlet masses in the large-field regime
are the result of a seesaw mechanism of a sort:
For a large Majorana mass $M\left(\varphi\right)$,
the Dirac masses $m_a$ become suppressed by a factor of
$\mathcal{O}\left(m_a/M\right)$.
This is exactly what happens in the usual seesaw mechanism, which
explains the small masses of the SM neutrinos
as the outcome of large Majorana and small Dirac neutrino masses~\cite{seesaw}.


\subsection{Coleman-Weinberg one-loop effective potential}
\label{subsec:CW}


We are now in the position to calculate the effective one-loop potential
for the scalar Polonyi field.
The full result for $V_{\rm 1-loop}$ follows from Eqs.~\eqref{eq:VCW},
\eqref{eq:M2bos}, and \eqref{eq:M2fer} and takes a rather complicated form.
For this reason, we refrain from explicitly writing down the full expression for $V_{\rm 1-loop}$,
and merely refer to Fig.~\ref{fig:effpot}, where we plot the exact result 
for $V_{\rm 1-loop}$ as a function of the order parameter $x\left(\varphi\right)$.
Instead, let us now evaluate $V_{\rm 1-loop}$ for specific field values of interest.
For instance, at the origin, we find
\begin{align}
V_{\rm 1-loop}\left(0\right) = \frac{m^4}{32\pi^2}
\left[N_X \left(\ln\left(\frac{\overline{m}^2}{Q^2}\right) + c\right)
+ \sum_a L\left(r_a\right)\right] \,,
\end{align}
where $N_X = 6$ counts the number of meson flavors in the IYIT model,
$\overline{m} = \Lambda$ denotes the effective supersymmetric
mass scale in the IYIT sector (see Eqs.~\eqref{eq:x} and \eqref{eq:mbar}),
and where $L$ is a loop function that needs to be evaluated at the
respective mass ratios $r_a$ (see Eq.~\eqref{eq:mara}),
\begin{align}
L\left(r_a\right) = 
\frac{1}{2}\left(1+\frac{1}{r_a^2}\right)^2 \ln\left(1+r_a^2\right) +
\frac{1}{2}\left(1-\frac{1}{r_a^2}\right)^2 \ln\left(1-r_a^2\right) =
\frac{3}{2} + \mathcal{O}\left(r_a^4\right) \,.
\end{align}
Note that this result for $V_{\rm 1-loop}\left(0\right)$ can also be obtained by plugging
the masses in Eq.~\eqref{eq:Ma20} into the CW formula.
The function $L$ is nearly constant over its entire domain,
$2\ln2 =  L\left(1\right)\leq L\left(r_a\right) \leq  L\left(0\right) = 3/2$, and thus,
well approximated by $L\left(r_a\right) \approx 3/2$.
This motivates the following choices for $Q$ and $c$,
\begin{align}
Q = \overline{m} \,, \quad 
c = -\frac{3}{2} \,.
\end{align}
For these values of the renormalization scale $Q$ and the constant $c$,
the tree-level vacuum energy density of the IYIT model,
$V_0 = \left<\left|F_\Phi\right|\right>^2 = \mu^4$, receives a small negative shift
$\Delta V_0$ of $\mathcal{O}\left(m^8/m_a^4\right)$,
\begin{align}
\Delta V_0 = V_{\rm 1-loop}\left(0\right) = 
- N_X \frac{m^4}{32\pi^2}
\left(\frac{3}{2}-\frac{1}{N_X}\sum_a L\left(r_a\right)\right) \,.
\label{eq:DeltaV0}
\end{align}
This correction to the vacuum energy density and, hence,
the ratio $\left|\Delta V_0\right|/V_0$, are bounded from above,
\begin{align}
\left|\Delta V_0\right| \leq N_X \,\frac{m^4}{32\pi^2}\left(\frac{3}{2}-2\ln2\right) \,, \quad
\frac{\left|\Delta V_0\right|}{V_0} \lesssim 10^{-3} \left(1+r_0^2\right) \lambda^2 \,.
\end{align}
Here, we have used that $L$ takes its smallest value
in the limit $r_a \rightarrow 1$, where $L \rightarrow 2\ln2$.
For perturbative values of the Yukawa coupling, $\lambda \lesssim 4$,
we can, therefore, safely neglect the radiative correction $\Delta V_0$.


Having fixed $Q$ and $c$,
let us now study the effective potential for extreme values
of the order parameter, i.e., for $x\left(\varphi\right) \ll 1$
and $x\left(\varphi\right) \gg 1$.
Below the critical field value, the inflaton-induced mass $M\left(\varphi\right)$
is smaller than the supersymmetric Dirac masses $m_a$.
At energies below $m_a$, the ``heavy'' meson flavors can then be integrated out,
which results in a quadratic potential for $\varphi$ around the origin,
\begin{align}
V_{\rm 1-loop}^{\rm LE}\left(\varphi\right) = \frac{1}{2} m_{\rm eff}^2\,\varphi^2
+ \mathcal{O}\left(x^4\right) \,,
\end{align}
where $m_{\rm eff}$ denotes the effective Polonyi mass at one-loop level
(see also Eq.~\eqref{eq:meff2} and Footnote~\ref{fn:meff}),
\begin{align}
m_{\rm eff}^2 = \left(2\ln2-1\right) N_X^{\rm eff}\left(r_a\right)
\frac{\kappa_\Phi^2}{16\pi^2}\, m^2 \,, \quad 
N_X^{\rm eff}\left(r_a\right) =  \sum_a \omega\left(r_a\right) \,.
\end{align}
Here, $\omega$ represents a loop function that acts as a normalized weight,
$0\leq\omega\leq1$, for each meson flavor,
\begin{align}
\omega\left(r_a\right) = \frac{\ell\left(r_a\right)}{2\ln2-1} \,, \quad
\ell\left(r_a\right) = 
\frac{1}{2}\left(1+\frac{1}{r_a^2}\right)^2
\ln\left(1+r_a^2\right) - \frac{1}{2}\left(1-\frac{1}{r_a^2}\right)^2
\ln\left(1-r_a^2\right) - \frac{1}{r_a^2} \,.
\label{eq:omega}
\end{align}
Evaluated at the mass ratio $r_a$, the function $\omega$ quantifies
the contribution from the respective meson flavor to $m_{\rm eff}$.
Correspondingly, the sum over all factors $\omega\left(r_a\right)$ yields the effective
number of flavors, $N_X^{\rm eff}$, that contribute to the effective Polonyi mass.
Moreover, we note that, over its entire domain, $0\leq r_a\leq 1$, the function $\omega$ is
well approximated by $\omega\left(r_a\right) \approx r_a^2$.
This allows us to rewrite $V_{\rm 1-loop}$ in the following way,
\begin{align}
V_{\rm 1-loop}^{\rm LE}\left(\varphi\right) & = 
\frac{m^2}{16\pi^2}\, M^2\left(\varphi\right)
\sum_a \ell\left(r_a\right) + \mathcal{O}\left(x^4\right)
\label{eq:V1loopLE}\\\nonumber
& \approx \left(2\ln2-1\right)
\frac{m^2}{16\pi^2}\, M^2\left(\varphi\right)
\sum_a r_a^2 + \mathcal{O}\left(x^4\right) \\\nonumber
& = \left(2\ln2-1\right)
\frac{m^4}{16\pi^2} \sum_a R_a^2\left(\varphi\right)
+ \mathcal{O}\left(x^4\right) \,,
\end{align}
which provides us with a couple of alternative expressions for
$V_{\rm 1-loop}$.
For instance, the first line of Eq.~\eqref{eq:V1loopLE} makes explicit
the dependence of the effective potential on the two mass scales $m$ and
$M\left(\varphi\right)$ at small field values,
$V_{\rm 1-loop}\left(\varphi\right) \propto m^2 M^2\left(\varphi\right)$.
It is interesting to note that this result for the effective potential
can also be obtained by plugging the approximate expressions for the
scalar and fermionic masses in Eq.~\eqref{eq:Ma2LE}
into the CW formula in Eq.~\eqref{eq:VCW}.
Meanwhile, the third line of Eq.~\eqref{eq:V1loopLE} illustrates
how $V_{\rm 1-loop}$ may be expressed as a function of the ratios
$R_a\left(\varphi\right) = M\left(\varphi\right)/m_a$
(see Eq.~\eqref{eq:Ra}).
In this case, the effective potential turns out to be given by the
fourth power of the soft SUSY-breaking mass, $m^4$, times a function of
the SUSY-preserving mass parameters $M\left(\varphi\right)$ and $m_a$,
$V_{\rm 1-loop}\left(\varphi\right) \propto m^4 \,f\left(R_a\right)$.
The expressions for $V_{\rm 1-loop}$ in Eq.~\eqref{eq:V1loopLE},
therefore, comply with the expectation that, in supersymmetric theories,
the effective potential ought to be proportional to the soft
SUSY-breaking mass scale $m$~\cite{Ellis:1982ed}.
Moreover, we can write $V_{\rm 1-loop}$ in the small-field regime
directly as a function of the order parameter,
\begin{align}
V_{\rm 1-loop}^{\rm LE}\left(\varphi\right) =
\Lambda_{\rm LE}^4 \, x^2\left(\varphi\right) + \mathcal{O}\left(x^4\right) \,,
\end{align}
where we have introduced $\Lambda_{\rm LE}$ as the effective potential energy scale
at small field values,
\begin{align}
\Lambda_{\rm LE}^4 = \frac{m^2\,\overline{m}^2}{16\pi^2}
\sum_a \ell\left(r_a\right) 
\approx \left(2\ln2-1\right) \frac{m^4}{16\pi^2}
\sum_a \left(\frac{\overline{m}}{m_a}\right)^2 \,.
 \end{align}


Finally, let us examine the effective Polonyi potential at small field values
in the limit of small supersymmetry breaking, i.e., for a small soft
SUSY-breaking mass $m$ in comparison to large SUSY-preserving Dirac
masses, $m \ll m_a$.
This limit is best quantified in terms of the geometric mean 
of all mass ratios $r_a$, which we will refer to as the hierarchy
parameter $y$ in the following (see Eq.~\eqref{eq:mara}),
\begin{align}
y = \Big(\prod_a r_a\Big)^{1/N_X} = \frac{m}{\overline{m}} \,,
\end{align}
which can take values between $0$ and $1$ (see Eq.~\eqref{eq:rbound}).
Here, $y=0$ corresponds the SUSY-preserving limit, while $y=1$
represents the case of maximal supersymmetry breaking.
From Eqs.~\eqref{eq:mara} and \eqref{eq:mbar}, it immediately
follows that the hierarchy parameter $y$ is, in fact, nothing but
an alternative measure for the strength of the Yukawa coupling
of the Polonyi field to the IYIT matter fields,
\begin{align}
y = \frac{\lambda}{\eta} \,.
\end{align}
That is, for perturbative values of the Yukawa coupling $\lambda$,
supersymmetry is only mildly broken in the meson and singlet multiplets
in the IYIT sector, while in the strongly coupled limit, i.e., for
nonperturbative values of the Yukawa coupling $\lambda$, the effect
of supersymmetry breaking is maximal.
In passing, we mention that the parameter $y$ also quantifies
the ratio between  the order parameter of our model, $x\left(\varphi\right)$,
and the order parameter of the waterfall transition in ordinary
F-term hybrid inflation, $R\left(\varphi\right)$,
\begin{align}
x\left(\varphi\right) = y\, R\left(\varphi\right) \,, \quad 
R\left(\varphi\right) = \frac{M\left(\varphi\right)}{m} =
\frac{\left|\phi\right|}{v} = \frac{\varphi}{\sqrt{2}\,v}\,.
\end{align}
In the limit of maximal supersymmetry breaking, both
order parameters, therefore, coincide with each other.
In all other cases, we have $R\left(\varphi\right) > x\left(\varphi\right)$.
This implies that, for $y<1$, the mesonic phase transition
in the IYIT model (see Fig.~\ref{fig:effpot}) takes place around a critical
field value, $\varphi \sim \varphi_c$, that is parametrically larger than the
critical field value associated with the waterfall transition
in F-term hybrid inflation, $\varphi = \sqrt{2}\,v$.
Let us now turn to the limit of small supersymmetry breaking.
For small values of $y$, i.e., for small mass ratios $r_a$,
the weight function $\omega$ in Eq.~\eqref{eq:omega} can be approximated as follows,
\begin{align}
\omega\left(r_a\right) = \frac{r_a^2}{3 \left(2\ln2-1\right)} + 
\mathcal{O}\left(r_a^6\right) \,.
\end{align}
The effective potential $V_{\rm 1-loop}$
and the potential energy scale $\Lambda_{\rm LE}$
at small field values then reduce to
\begin{align}
V_{\rm 1-loop}^{\rm LE}\left(\varphi\right) =
\frac{m^4}{48\pi^2} \sum_a R_a^2\left(\varphi\right)
+ \mathcal{O}\left(x^4,y^8\right) \,, \quad 
\Lambda_{\rm LE}^4 =  \frac{m^4}{48\pi^2}
\sum_a \left(\frac{\overline{m}}{m_a}\right)^2
+ \mathcal{O}\left(y^8\right) \,.
\label{eq:V1loopLEy}
\end{align}
We will come back to these results in Sec.~\ref{subsec:kahler}.


Next, we shall examine the effective Polonyi potential for large values of the
order parameter $x\left(\varphi\right)$.
In the large-field regime, $x\left(\varphi\right) \gtrsim 1$, the IYIT matter fields
acquire a large field-dependent mass, such that their ``bare'' supersymmetric masses 
$m_a$ become irrelevant, $M\left(\varphi\right) \gtrsim m_a$.
Integrating out the matter fields then results in the usual logarithmic effective potential
known from ordinary F-term hybrid inflation,
\begin{align}
V_{\rm 1-loop}^{\rm HE}\left(\varphi\right) =  N_X\frac{m^4}{16\pi^2}
\ln x\left(\varphi\right) + \mathcal{O}\left(x^{-4}\right) \,, \quad
x\left(\varphi\right) = \frac{\varphi}{\varphi_c} =
\frac{M\left(\varphi\right)}{\overline{m}} \,.
\label{eq:V1loopHE}
\end{align}
At large field values, the dependence of the effective potential 
on the mass scales $m$ and $M\left(\varphi\right)$ is, therefore, of
the form $V_{\rm 1-loop}\left(\varphi\right) \propto m^4 \,\ln M\left(\varphi\right)$.
Up to corrections of $\mathcal{O}\left(x^{-2}\right)$, we are able to
reproduce this result by inserting the approximate
masses in Eq.~\eqref{eq:Ma2HE} into the CW formula in Eq.~\eqref{eq:VCW}.
Moreover, we can read off the effective potential energy scale in the large-field
regime, $\Lambda_{\rm HE}$, from Eq.~\eqref{eq:V1loopHE},
\begin{align}
V_{\rm 1-loop}^{\rm HE}\left(\varphi\right) = \Lambda_{\rm HE}^4  \ln x\left(\varphi\right)
+ \mathcal{O}\left(x^{-4}\right) \,, \quad
\Lambda_{\rm HE}^4 = N_X\frac{m^4}{16\pi^2} \,,
\label{eq:V1loopHELambda}
\end{align}
which is again of $\mathcal{O}\left(m^4\right)$.
Last but not least, we can rewrite the effective potential as a function
of the mass ratios $R_a\left(\varphi\right)$,
which provides us with a large-field counterpart to the small-field
result in Eq.~\eqref{eq:V1loopLEy},
\begin{align}
V_{\rm 1-loop}^{\rm HE}\left(\varphi\right) =  \frac{m^4}{16\pi^2}\sum_a
\ln R_a\left(\varphi\right) + \mathcal{O}\left(x^{-4}\right) \,.
\label{eq:V1loopHEy}
\end{align}


\subsection{Reformulation in terms of an effective K\"ahler potential}
\label{subsec:kahler}


At tree level, the F-term scalar potential is determined by two input functions,
the superpotential $W$ and the K\"ahler potential $K$.
Here, the superpotential does not receive any radiative corrections in perturbation theory
according to the SUSY nonrenormalization theorem~\cite{Grisaru:1979wc}.
The K\"ahler potential, on the other hand, \textit{is} renormalized, which allows one
to rewrite parts of the effective action in the form of an effective K\"ahler potential.
We shall now demonstrate how this applies to our results in
the previous section.
First, we note that, schematically, the full result for $V_{\rm 1-loop}$ in Eq.~\eqref{eq:VCW}
is of the following form,
\begin{align}
V_{\rm 1-loop}\left(\varphi\right) = \frac{m^4}{16\pi^2}
\sum_{n=0}^\infty \frac{f_n \left(R_a\right)}{x^{4n}\left(\varphi\right)}\,
y^{4n} = \frac{m^4}{16\pi^2}\,
f_0 \left(R_a\right) + \mathcal{O}\left(y^8\right) \,,
\label{eq:V1loopxy}
\end{align}
where the order parameter $x\left(\varphi\right) = M\left(\varphi\right)/\overline{m}$
and the mass ratios $R_a = M\left(\varphi\right)/m_a$ denote the ratios of supersymmetric
mass scales, while the hierarchy parameter $y = m/\overline{m}$ quantifies the ratio 
between the soft SUSY-breaking mass scale $m$ and the effective supersymmetric mass scale
$\overline{m}$.
The above expression for $V_{\rm 1-loop}$, thus, illustrates that the effective potential
represents a power series in $y^4$, the coefficients of which are functions of
SUSY-preserving mass parameters.
From our results in Eqs.~\eqref{eq:V1loopLEy} and \eqref{eq:V1loopHEy}, we can
read off the coefficient function $f_0\left(R_a\right)$ belonging to
the leading-order contribution to $V_{\rm 1-loop}$,
\begin{align}
f_0\left(R_a\right) = 
\begin{cases}
\sum_a\frac{1}{3}\, R_a^2\left(\varphi\right) + \mathcal{O}\left(x^4\right) & ; \: x \ll 1 \\ 
\sum_a\ln R_a\left(\varphi\right) + \mathcal{O}\left(x^{-4}\right)          & ; \: x \gg 1 
\end{cases} \,.
\end{align}


As we are now going to show, this result for the function $f_0\left(R_a\right)$ can be
reproduced by supplementing the tree-level K\"ahler potential, $K_{\rm tree}$, by an
effective one-loop K\"ahler potential, $K_{\rm 1-loop}$,
\begin{align}
K_{\rm eff} = K_{\rm tree} +  K_{\rm 1-loop} \,,
\label{eq:Keff}
\end{align}
where $K_{\rm tree}$ and $K_{\rm 1-loop}$ are respectively given
as follows~\cite{Gaillard:1993es,Grisaru:1996ve},
\begin{align}
K_{\rm tree} = \Phi^\dagger \Phi \,, \quad 
K_{\rm 1-loop} = - \frac{1}{32\pi^2}\, \textrm{Tr}\left[
\widetilde{\mathcal{M}}^\dagger\widetilde{\mathcal{M}}\left(
\ln\left(\frac{\widetilde{\mathcal{M}}^\dagger
\widetilde{\mathcal{M}}^{\phantom{\dagger}}}{Q^2}\right)-c'\right)\right] \,.
\label{eq:K1loop}
\end{align}
Here, $\widetilde{\mathcal{M}}$ denotes the fermionic mass matrix, which is identical to
the Hessian of the superpotential, and $c'$ is a numerical constant
that we will determine shortly.
In global supersymmetry, the effective K\"ahler potential in 
Eq.~\eqref{eq:Keff} results in the following F-term scalar potential
for the Polonyi field $\phi = \varphi/\sqrt{2}\,e^{i\theta}$,
\begin{align}
V_{\rm eff}^{\rm approx}\left(\varphi\right) = \left(\frac{\partial^2 K_{\rm eff}\left(\phi,\phi^*\right)}
{\partial \phi\, \partial \phi^*}\right)^{-1}
\left|\frac{\partial W_{\rm eff}\left(\phi\right)}{\partial \phi}\right|^2 \,,
\label{eq:Veffapprox}
\end{align}
where the effective superpotential consists of the Polonyi tadpole term,
$W_{\rm eff} \simeq \mu^2\, \Phi$, such that
\begin{align}
\left|\frac{\partial W_{\rm eff}\left(\phi\right)}{\partial \phi}\right|^2 = \mu^4 = V_0  \,.
\end{align}
It is important to note that the scalar potential in Eq.~\eqref{eq:Veffapprox} is
only an approximation of the full effective potential.
To see this, let us expand the inverse K\"ahler metric in powers of
the loop factor $\kappa_\Phi^2/\left(16\pi^2\right)$,
\begin{align}
\left(\frac{\partial^2 K_{\rm eff}\left(\phi,\phi^*\right)}
{\partial \phi\, \partial \phi^*}\right)^{-1} = 1 -
\frac{\partial^2 K_{\rm 1-loop}\left(\phi,\phi^*\right)}{\partial \phi\, \partial \phi^*}
+ \mathcal{O}\left(\Big(\frac{\kappa_\Phi^2}{16\pi^2}\Big)^2\right) \,.
\end{align}
The leading contribution to $V_{\rm eff}^{\rm approx}$ then yields what we shall
refer to as the \textit{truncated effective potential},
\begin{align}
V_{\rm eff}^{\rm trunc}\left(\varphi\right) = \left(1 -
\frac{\partial^2 K_{\rm 1-loop}\left(\phi,\phi^*\right)}{\partial \phi\, \partial \phi^*}\right)
\left|\frac{\partial W_{\rm eff}\left(\phi\right)}{\partial \phi}\right|^2 \,.
\end{align}
This truncated effective potential coincides with the full effective potential
up to corrections of $\mathcal{O}\left(y^8\right)$,
\begin{align}
V_0 + V_{\rm 1-loop}\left(\varphi\right) = V_{\rm eff}^{\rm trunc}\left(\varphi\right)
+ \mathcal{O}\left(y^8\right) \,.
\label{eq:Vefftrunc}
\end{align}
An explicit proof of this relation is given by Intriligator, Seiberg, and Shih
in Appendix~A.5 of~\cite{Intriligator:2006dd}.
For an earlier discussion of the effective action in supersymmetric theories,
see~\cite{Buchbinder:1994iw}. 
From the relation in Eq.~\eqref{eq:Vefftrunc}, it is evident that the truncated potential
$V_{\rm eff}^{\rm trunc}$ is only a good approximation of the actual (true)
effective potential, as long as the effect of spontaneous supersymmetry breaking is small,
i.e., as long as $y \ll 1$.
Otherwise, the $\mathcal{O}\left(y^8\right)$ corrections in Eq.~\eqref{eq:Vefftrunc}
become important.
These radiative corrections---while present in the full expression for $V_{\rm 1-loop}$
in Eq.~\eqref{eq:VCW}---cannot be captured by the effective K\"ahler potential.
One example for such a correction is, e.g., 
the shift in the tree-level vacuum energy density, $\Delta V_0$ (see Eq.~\eqref{eq:DeltaV0}),
which is proportional to $m^4 r_a^4 = m^8 / m_a^4 \sim m_a^4\, y^8$ and which cannot be explained
in terms of an effective K\"ahler potential.
Instead, the shift in the vacuum energy density $\Delta V_0$ and
all other  $\mathcal{O}\left(y^8\right)$ corrections follow from an effective potential
for the auxiliary component of the Polonyi multiplet, $\left.\Phi\right|_{\theta^2} = F_\Phi$.
This \textit{effective auxiliary field potential} contains higher-order terms in $F_\Phi$ and
is only negligible, once the effect of supersymmetry breaking is nothing but
a small correction to the otherwise supersymmetric dynamics~\cite{Buchbinder:1994iw}.
Fortunately, this is exactly the case in our model, since the soft SUSY-breaking mass scale
$m$ is always the smallest mass parameter in the IYIT sector (see Eq.~\eqref{eq:rbound}).


Let us now derive an explicit expression for the one-loop K\"ahler potential $K_{\rm 1-loop}$.
To do so, we first need to promote the fermionic masses in Eq.~\eqref{eq:M2fer}
to functions of the chiral Polonyi superfield $\Phi$,
\begin{align}
\widetilde{M}_a^2\left(\varphi;p\right) \rightarrow \frac{1}{2}
\left(\mathcal{A}_a + p\, \mathcal{B}_a\right) \,, \quad
\mathcal{A}_a = \kappa_\Phi^2\, \Phi^\dagger\Phi + 2\, m_a^2 \,, \quad
\mathcal{B}_a = \left[\kappa_\Phi^2\, \Phi^\dagger\Phi \left(
\kappa_\Phi^2\, \Phi^\dagger\Phi + 4\, m_a^2\right)\right]^{1/2} \,.
\end{align}
These masses then need to be inserted into the general expression
for $K_{\rm 1-loop}$ in Eq.~\eqref{eq:K1loop},
\begin{align}
K_{\rm 1-loop} = - \frac{1}{32\pi^2} \sum_a \left[
\mathcal{A}_a \left(\ln\left(\frac{m_a^2}{Q^2}\right) + c'\right)
+ \mathcal{B}_a\left(\frac{1}{2}\ln\left(1+\frac{\mathcal{B}_a}{\mathcal{A}_a}\right)
- \frac{1}{2}\ln\left(1-\frac{\mathcal{B}_a}{\mathcal{A}_a}\right)\right)\right] \,.
\end{align}
Note that this result for $K_{\rm 1-loop}$ is independent of the SUSY-breaking
mass scale $m$.
Or put differently, the effective K\"ahler potential ``does not know anything about
supersymmetry breaking''. 
It is, thus, clear that $V_{\rm eff}^{\rm trunc}$ (or $V_{\rm eff}^{\rm approx}$ for
that purpose) can approximate the full effective potential only up to
$\mathcal{O}\left(y^4\right)$.
We shall now study $K_{\rm 1-loop}$ as a function of $\mathcal{X}\left(\Phi^\dagger\Phi\right)$,
the superfield analog of the order parameter $x\left(\varphi\right)$,
\begin{align}
\mathcal{X}\big(\Phi^\dagger\Phi\big) =
\left(\frac{\Phi^\dagger\Phi}{\varphi_c^2/2}\right)^{1/2} \,.
\end{align}
Close to the origin, $\mathcal{X}\left(\Phi^\dagger\Phi\right) \ll 1$, the effective
one-loop K\"ahler potential consists of the following terms,
\begin{align}
K_{\rm 1-loop} = \Delta K_0 - N_X \frac{\kappa_\Phi^2}{32\pi^2}
\left(\ln\left(\frac{\overline{m}^2}{Q^2}\right) + c' + 2\right)\Phi^\dagger\Phi 
+ \mathcal{O}\left(\mathcal{X}^2\right) \,,
\label{eq:K1loop0}
\end{align}
where the first term, $\Delta K_0$, denotes a constant shift
in the VEV of K\"ahler potential that does not have any consequences
in global supersymmetry and that is negligible in supergravity,
$\Delta K_0 \ll M_{\rm Pl}^2$,
\begin{align}
\Delta K_0 = - \frac{1}{16\pi^2} \sum_a m_a^2 
\left(\ln\left(\frac{m_a^2}{\overline{m}^2}\right) - 2\right) \,.
\end{align}
Meanwhile, the second term in Eq.~\eqref{eq:K1loop0} represents a
shift in the canonical kinetic term of the Polonyi field.
In the following, we will choose the renormalization scale $Q$
and the constant $c'$ as follows,
\begin{align}
Q = \overline{m} \,, \quad c' = -2 \,,
\end{align}
so that this term exactly vanishes (i.e., so that the Polonyi field $\Phi$
remains canonically normalized).


The first relevant correction to $K_{\rm tree}$ in
the small-field regime, $\mathcal{X}\left(\Phi^\dagger\Phi\right) \lesssim 1$,
is then of $\mathcal{O}\left(\mathcal{X}^2\right)$,
\begin{align}
K_{\rm 1-loop}^{\rm LE} = - \frac{1}{64\pi^2} \sum_a \frac{1}{3}
\left(\frac{\kappa_\Phi^2 \Phi^\dagger\Phi}{m_a}\right)^2
+ \mathcal{O}\left(\mathcal{X}^4\right) \,.
\label{eq:K1loopLE}
\end{align}
This form of the K\"ahler potential nicely reflects the fact, at small
field values, the radiative corrections to the Polonyi potential follow
from integrating out ``heavy'' meson fields with masses $m_a$.
In terms of the order parameter superfield $\mathcal{X}\left(\Phi^\dagger\Phi\right)$,
the one-loop K\"ahler potential in Eq.~\eqref{eq:K1loopLE} reads as follows,
\begin{align}
K_{\rm 1-loop}^{\rm LE} = -\frac{1}{4\, V_0} 
\Lambda_{\rm LE}^4\, \mathcal{X}^2\big(\Phi^\dagger\Phi\big)\, \Phi^\dagger\Phi
+ \mathcal{O}\left(\mathcal{X}^4\right)
\,, \quad \Lambda_{\rm LE}^4 =  \frac{m^4}{48\pi^2}
\sum_a \left(\frac{\overline{m}}{m_a}\right)^2 \,, 
\end{align}
where the expression for the effective potential energy scale $\Lambda_{\rm LE}$
is the identical to the one that we obtained in Sec.~\ref{subsec:CW} in the
limit of small supersymmetry breaking, see Eq.~\eqref{eq:V1loopLEy}.
In this sense, $K_{\rm 1-loop}^{\rm LE}$
indeed reproduces the effective one-loop potential
in the small-field regime up to corrections of $\mathcal{O}\left(y^8\right)$,
\begin{align}
V_{\rm eff}^{\rm trunc}\left(\varphi\right) =
V_0 + \Lambda_{\rm LE}^4\, x^2\left(\varphi\right) + 
\mathcal{O}\left(x^4\right) \,.
\end{align}
Meanwhile, for large values of the Polonyi field, 
$\mathcal{X}\left(\Phi^\dagger\Phi\right) \gtrsim 1$, the one-loop K\"ahler potential
reduces to
\begin{align}
K_{\rm 1-loop}^{\rm HE} = - N_X \frac{\kappa_\Phi^2}{32\pi^2}
\left(\ln\left(\frac{\kappa_\Phi^2 \Phi^\dagger \Phi}{\overline{m}^2}\right)-2\right)
\Phi^\dagger\Phi + \Delta K_0' + \mathcal{O}\left(\mathcal{X}^{-4}\right) \,,
\label{eq:K1loopHE}
\end{align}
which represents a logarithmic renormalization of the kinetic term.
Here, $\Delta K_0'$ is a correction to the VEV of the K\"ahler potential,
which is meaningless in global supersymmetry and irrelevant in supergravity,
\begin{align}
\Delta K_0' = - \frac{1}{16\pi^2}
\left(\ln\left(\frac{\kappa_\Phi^2\Phi^\dagger\Phi}{\overline{m}^2}\right)-1\right) \sum_a m_a^2 \,.
\end{align}
Again, we may rewrite the one-loop K\"ahler potential
in terms of the order parameter superfield $\mathcal{X}\left(\Phi^\dagger\Phi\right)$,
\begin{align}
K_{\rm 1-loop}^{\rm HE} = - \frac{1}{V_0}
\Lambda_{\rm HE}^4 \left(\ln\mathcal{X}\big(\Phi^\dagger\Phi\big) - 1\right)
\Phi^\dagger\Phi + \mathcal{O}\left(\mathcal{X}^{-4}\right) 
\,, \quad \Lambda_{\rm HE}^4 =  N_X\frac{m^4}{16\pi^2}
\end{align}
which reproduces $V_{\rm 1-loop}$ and the potential energy scale $\Lambda_{\rm HE}$
in the large-field regime, see Eq.~\eqref{eq:V1loopHELambda},
\begin{align}
V_{\rm eff}^{\rm trunc}\left(\varphi\right) =
V_0 + \Lambda_{\rm HE}^4 \ln x\left(\varphi\right) + 
\mathcal{O}\left(x^{-4}\right) \,.
\end{align}
We, therefore, conclude that our results for $K_{\rm 1-loop}^{\rm LE}$
and $K_{\rm 1-loop}^{\rm HE}$ in Eqs.~\eqref{eq:K1loopLE}
and \eqref{eq:K1loopHE} indeed suffice to describe the radiative corrections to the Polonyi
potential, as long as the effect of supersymmetry breaking remains small, $y\ll1$.
In the limit of large supersymmetry breaking, $y\rightarrow1$, the full information on
the radiative corrections is, however, only contained in the effective one-loop
potential in Eq.~\eqref{eq:VCW}.


\section{Viability of inflation in the case of early \boldmath{$R$} symmetry breaking}
\label{app:viability}


In the main text, we study inflation in an effective Polonyi model that is based
on the effective superpotential $W_{\rm eff} \simeq \mu^2\,\Phi$, a near-canonical K\"ahler potential
and logarithmic radiative corrections.
In particular, we assume that the constant term in the superpotential,
$W \supset w$, which allows one to tune the CC in the true vacuum to zero,
is generated only after inflation.
In this appendix, we are now going to show that (for all K\"ahler potentials of interest)
this assumption is, indeed, inevitable.
That is, in contrast to the main text, we are now going to assume that the constant $w$ takes
its low-energy value already from the very beginning, $w=w_0$. 
We then study the prospects of realizing successful inflation in the Polonyi
model for different choices of the K\"ahler potential (near-canonical and
approximately shift-symmetric), finding that, for $w=w_0$ during inflation,
Polonyi inflation is always bound to fail.
Either the Polonyi potential turns out to be too steep to support slow-roll inflation
or there does not even exist a global Minkowski vacuum.
Of course, our analysis does not represent a general no-go theorem,
as some intricate choice for the K\"ahler potential may render inflation
in the Polonyi model viable, after all~\cite{Izawa:2007qa}.%
\footnote{In single-field helical-phase inflation, the phase of
the Polonyi field, $\theta = \arg \phi$, may play the role of the inflaton~\cite{Ketov:2015tpa}.
This scenario requires a particular tuning of the K\"ahler potential, in order to stabilize
the radial component of the complex Polonyi field, $\varphi = \sqrt{2}\left|\phi\right|$,
and eventually results in inflationary predictions that are equivalent to those of
natural inflation~\cite{Freese:1990rb}.}
But for our purposes, as we intend to focus on \textit{simple} forms of the K\"ahler
potential, these results suffice to convince us that successful Polonyi inflation
is better off, if $R$ symmetry is broken at late times.


\subsection{Canonical K\"ahler potential plus higher-dimensional corrections}
\label{subsec:canonical}


We first consider the full Polonyi superpotential in combination with
a near-canonical K\"ahler potential,
\begin{align}
W = \mu^2\, \Phi + w \,, \quad
K = \Phi^\dagger \Phi + \frac{\epsilon}{\left(2!\right)^2}
\left(\frac{\Phi^\dagger \Phi}{M_{\rm Pl}}\right)^2 + 
\mathcal{O}\left(\epsilon^2,M_{\rm Pl}^{-4}\right) \,, \quad \epsilon \lesssim 1 \,.
\end{align}
Apart from the subdominant higher-dimensional correction (parametrized
in terms of the parameter $\epsilon$), this is nothing but the
original Polonyi model~\cite{Polonyi:1977pj}.
For now, let us assume that this model does not receive any radiative
corrections from Yukawa couplings between $\Phi$ and some (heavy) matter
fields.
Given the above K\"ahler potential, the complex scalar contained 
in the chiral Polonyi field, $\tilde\phi \subset \Phi$,
\begin{align}
\tilde\phi = \frac{\tilde\varphi}{\sqrt{2}}\,e^{i\tilde\theta} \,,
\end{align}
is not canonically normalized.
The Polonyi phase, $\tilde\theta = \arg \tilde\phi$, is always stabilized at $\tilde\theta = 0$,
so that we can neglect it in the following.
Meanwhile, the properly normalized radial component $\varphi$ is given as follows,
\begin{align}
\varphi = \int d\tilde\varphi \left(\frac{\partial^2 K}{\partial\tilde\phi\,\partial\tilde\phi^*}\right)^{1/2}
= \tilde\varphi \left[1 + \frac{\epsilon}{12}\left(\frac{\tilde\varphi}{M_{\rm Pl}}\right)^2 +
\mathcal{O}\left(\epsilon^2\right)\right] \,.
\label{eq:phit}
\end{align}
The vacuum energy density in the global minimum vanishes,
once the constant $w$ is fine-tuned to
\begin{align}
w_0 = \left(2-\sqrt{3}\right) \left(1 - \epsilon + \mathcal{O}\left(\epsilon^2\right)
\right) \mu^2 M_{\rm Pl} \,.
\label{eq:w0can}
\end{align}
For this value of $w$, the real Polonyi field $\varphi$ takes the following
value in the true vacuum,
\begin{align}
\left<\varphi\right> = \sqrt{2}\left[\left(\sqrt{3}-1\right)\left(1+\epsilon\right)
- \frac{\epsilon}{6} + \mathcal{O}\left(\epsilon^2\right)\right] M_{\rm Pl} \,.
\label{eq:phican}
\end{align}
Around this vacuum, the Polonyi field has a tree-level mass of the order of the
gravitino mass,
\begin{align}
m_\varphi^2 = 2\sqrt{3}\left[1-\left(3-\sqrt{3}\right)\epsilon +
\mathcal{O}\left(\epsilon^2\right)\right] m_{3/2}^2 \,,
\end{align}
where the gravitino mass follows from inserting Eqs.~\eqref{eq:w0can}
and \eqref{eq:phican} into the general formula in Eq.~\eqref{eq:m32},
\begin{align}
m_{3/2} = e^{2-\sqrt{3}}\left[1 + \frac{1}{2}\left(3-\sqrt{3}\right)\epsilon
+ \mathcal{O}\left(\epsilon^2\right) \right] \frac{\mu^2}{M_{\rm Pl}} \,.
\end{align}


As it turns out, the scalar potential around the true vacuum is
too steep to support slow-roll inflation.
This can be seen by inspecting the slow-roll parameters
$\varepsilon$ and $\eta$ in the vicinity of $\varphi = \left<\varphi\right>$,
\begin{align}
\varepsilon\left(\varphi\right) & = 2 \left[\frac{1}{\delta\left(\varphi\right)} 
+ \frac{2\sqrt{3}-1}{2\sqrt{2}}
- \frac{2\sqrt{3}-3}{4\sqrt{2}} \,\epsilon\right]^{2}
+ \frac{\sqrt{3}+1}{\sqrt{3}} - \frac{5}{3}\,\epsilon
+ \mathcal{O}\left(\epsilon^2,\delta\right) \,,
\label{eq:SRparacan}\\ \nonumber
\eta\left(\varphi\right) & = 2 \left[\frac{1}{\delta\left(\varphi\right)} 
+ \frac{2\sqrt{3}-1}{\sqrt{2}}
- \frac{2\sqrt{3}-3}{2\sqrt{2}} \,\epsilon\right]^{2}
+ \frac{17}{2\sqrt{3}} - 4 - \frac{2}{3}\left(6\sqrt{3}-5\right)\epsilon
+ \mathcal{O}\left(\epsilon^2,\delta\right) \,,
\end{align}
where $\delta\left(\varphi\right)$ parametrizes the displacement from
the true vacuum in units of the Planck scale,
\begin{align}
\delta\left(\varphi\right) = \frac{\varphi-\left<\varphi\right>}{M_{\rm Pl}} \,.
\end{align}
The expressions in Eq.~\eqref{eq:SRparacan}
imply that we only have a chance of sufficiently suppressing $\varepsilon$ and $\eta$,
if we displace the inflaton field by some amount of
$\mathcal{O}\left(M_{\rm Pl}\right)$ from its VEV.
In the case of such large field displacements, the scalar potential is, however,
severely steepened by SUGRA corrections.
In fact, scanning over the entire Polonyi field range and
evaluating the slow-roll parameters $\varepsilon$ and $\eta$
numerically, we find that nowhere in field space
the scalar potential is sufficiently flat for slow-roll inflation,
\begin{align}
\min \max\left\{\varepsilon,\left|\eta\right|\right\} \sim 0.3 \,.
\end{align}
In view of this result, we then also have little hope that radiative corrections
may improve the situation.
Instead, we expect that radiative correction would, overall,
rather increase the magnitude of the slow-roll parameters $\varepsilon$ and $\eta$
even further.
We, thus, arrive at the same conclusion as the analysis in~\cite{Ovrut:1983my},
namely that inflation in the original Polonyi model, with $w=w_0$
during inflation, does not work.


\subsection{Approximate shift symmetry along the real axis}


Next, we assume that the K\"ahler potential exhibits an approximate shift
symmetry along the real axis,
\begin{align}
W = \mu^2\, \Phi + w \,, \quad
K = \frac{\epsilon}{2} \left(\Phi + \Phi^\dagger\right)^2
- \frac{1}{2} \left(\Phi - \Phi^\dagger\right)^2 +
\mathcal{O}\left(\epsilon^2,M_{\rm Pl}^{-2}\right) \,, \quad \epsilon \ll 1 \,.
\end{align}
The complex scalar contained in the chiral Polonyi field, $\tilde\phi \subset \Phi$,
then decomposes as follows,
\begin{align}
\tilde\phi = \frac{1}{\sqrt{2}} \left(\tilde\varphi + i\,\tilde\sigma\right) \,,
\end{align}
where $\tilde\varphi$ and $\tilde\sigma$ denote the (not canonically normalized)
inflaton field and its scalar partner (i.e., the ``sinflaton''), respectively.
First, we note that, in the limit of an exact shift symmetry,
this theory does not admit a Minkowski vacuum.
For $\epsilon = 0$, the only stationary point of the scalar
potential is located at
\begin{align}
\left<\tilde\varphi\right> = - \sqrt{2}\,\frac{w}{\mu^2} \,, \quad 
\left<\tilde\sigma\right> = 0 \,,
\label{eq:phishiRePre}
\end{align}
so that the Polonyi tadpole term ``eats up'' the constant $w$.
At this stationary point, we, therefore, have $\left<W\right> = 0$,
which implies unbroken $R$ symmetry and, hence, zero gravitino mass.
But more importantly, this point in field space corresponds
to a dS state, as the vacuum energy density does not vanish,
$\left<V\right> = \mu^4$.
On top of that, it is not even stable, as the $\tilde\varphi$
direction acquires a tachyonic mass, $m_\varphi^2 = -3\,\mu^4/M_{\rm Pl}^2$.


We can remedy this situation and distort the scalar potential by slightly breaking
the shift symmetry in the $\tilde\varphi$ direction, i.e., by allowing for
small nonzero values of $\epsilon$.
This stabilizes the stationary point in Eq.~\eqref{eq:phishiRePre}
and allows for the possibility of a global Minkowski vacuum.
In addition, a slightly broken shift symmetry in the K\"ahler potential
appears to be more natural, anyway, as the superpotential already breaks
the shift symmetry explicitly.
The canonically normalized fields $\varphi$ and $\sigma$ are then given as follows,
\begin{align}
\varphi = \left(1+\epsilon\right)^{1/2} \tilde\varphi \,, \quad
\sigma = \left(1+\epsilon\right)^{1/2} \tilde\sigma \,.
\end{align}
For $\epsilon \neq 0$, the model exhibits a global Minkowski vacuum
at a super-Planckian value of the inflaton field,
\begin{align}
\left<\varphi\right> = \frac{1+\epsilon}{\epsilon} \left[\frac{\sqrt{3}}{\sqrt{2}} -
\left(\frac{\epsilon}{1+\epsilon}\right)^{1/2}\right] M_{\rm Pl} \,, \quad
\left<\sigma\right> = 0 \,.
\label{eq:phishiRe}
\end{align}
We note that this solution is only consistent as long as $\epsilon$ is positive.
Moreover, it becomes unphysical in the limit of an exact shift symmetry,
i.e., for $\epsilon\rightarrow 0$.
This is also the reason why, in contrast to Sec.~\ref{subsec:canonical},
we now refrain from expanding our results in powers of the small parameter $\epsilon$.
In order to tune the vacuum energy density in the vacuum to zero,
the constant $w$ needs to take the following value,
\begin{align}
w_0 = -\frac{1}{\sqrt{2}\,\epsilon} \left[\frac{\sqrt{3}}{\sqrt{2}}\,
\left(1+\epsilon\right)^{1/2} - 2\, \epsilon^{1/2}\right] \mu^2 M_{\rm Pl} \,,
\label{eq:wshiRe}
\end{align}
which also diverges for $\epsilon\rightarrow 0$.
As we are able to read off from Eqs.~\eqref{eq:phishiRe} in \eqref{eq:wshiRe},
the relation between $\left<\tilde\varphi\right>$ and $w$ in
Eq.~\eqref{eq:phishiRePre} now no longer applies.
Thanks to the slightly broken shift symmetry, we now have
$\left<\tilde\varphi\right> = \left(1+\epsilon\right)^{-1/2} \left<\varphi\right>
= -\sqrt{2}\, w/\mu^2 + \epsilon^{-1/2}$, which results in
a nonzero gravitino mass,
\begin{align}
m_{3/2} = \frac{1}{\sqrt{2}\,\epsilon^{1/2}}
\exp\left[\frac{\epsilon}{2\left(1+\epsilon\right)}
\frac{\left<\varphi\right>^2}{M_{\rm Pl}^2}\right]
\frac{\mu^2}{M_{\rm Pl}} \,.
\end{align}
The field $\varphi$ then acquires a non-tachyonic mass of the order of the gravitino mass,
\begin{align}
m_\varphi^2 = 2\sqrt{3}
\left(\frac{2\,\epsilon}{1+\epsilon}\right)^{3/2} m_{3/2}^2 \,.
\end{align}


Again, the scalar potential around the true vacuum is too steep for slow-roll inflation.
In the vicinity of $\varphi = \left<\varphi\right>$, the slow-roll parameters
$\varepsilon$ and $\eta$ (as functions of
$\delta\left(\varphi\right) = \left(\varphi-\left<\varphi\right>\right)/M_{\rm Pl}^2$)
are now given as,
\begin{align}
\varepsilon\left(\varphi\right) & = 2\left[\frac{1}{\delta\left(\varphi\right)} + 
\frac{\sqrt{3}}{\sqrt{2}}-\frac{\epsilon^{1/2}}{2}\right]^2 +
\mathcal{O}\left(\epsilon\right) \,, \\ \nonumber
\eta\left(\varphi\right) & = 2\left[\frac{1}{\delta\left(\varphi\right)}
+ \sqrt{6} -\epsilon^{1/2}\right]^2
- \sqrt{6}\left(\sqrt{6}-2\,\epsilon^{1/2}\right) +
\mathcal{O}\left(\epsilon\right) \,.
\end{align}
A numerical scan of the full expressions for $\varepsilon$ and $\eta$ over the entire
field range reveals that both parameters are always of $\mathcal{O}(1)$ or larger
and, hence, always too large for slow-roll inflation.
Similarly as in the case of a near-canonical K\"ahler potential, we do not
expect that radiative corrections could improve this situation.
We, therefore, conclude that, assuming an approximate shift symmetry
along the real axis in the K\"ahler potential, inflation based on the full
Polonyi superpotential is unfortunately not an option.


\subsection{Approximate shift symmetry along the imaginary axis}


Finally, let us examine what happens if we replace the approximate shift symmetry
along the real axis by an approximate shift symmetry along the imaginary axis.
As we will see shortly, in this case, we find a stationary point at $\left<\varphi\right> = 0$,
similarly as in Sec.~\ref{subsec:SUGRA}.
For that reason, our analysis in this section will resemble the discussion in
Sec.~\ref{subsec:SUGRA} much more closely than our analysis in the previous section.
But, first things first.
Now, we supplement the Polonyi potential by a K\"ahler potential of the following
form,
\begin{align}
W = \mu^2\, \Phi + w \,, \quad
K = \frac{1}{2} \left(\Phi + \Phi^\dagger\right)^2
- \frac{\epsilon}{2} \left(\Phi - \Phi^\dagger\right)^2 +
\mathcal{O}\left(\epsilon^2,M_{\rm Pl}^{-2}\right) \,.
\label{eq:KshiIm}
\end{align}
The (not canonically normalized) inflaton field $\tilde\varphi$ and the
(not canonically normalized) sinflaton field $\tilde\sigma$ then
correspond to the imaginary and the real part of
the complex scalar $\tilde\phi \subset \Phi$, respectively,
\begin{align}
\tilde\phi = \frac{1}{\sqrt{2}} \left(\tilde\sigma + i\,\tilde\varphi\right) \,,
\end{align}
As in the previous section, the canonically normalized fields $\varphi$
and $\sigma$ are related to $\tilde\varphi$ and $\tilde\sigma$ as follows,
\begin{align}
\varphi = \left(1+\epsilon\right)^{1/2} \tilde\varphi \,, \quad
\sigma = \left(1+\epsilon\right)^{1/2} \tilde\sigma \,.
\end{align}
Now we find a stationary point at a Planckian value of
the sinflaton field rather than the inflaton field,
\begin{align}
\left<\varphi\right> = 0 \,, \quad
\left<\sigma\right> = \left(1+\epsilon\right)^{1/2}
\left[1-\frac{\sqrt{3}}{\sqrt{2}}\,
\left(1+\epsilon\right)^{1/2}\right] M_{\rm Pl} \,.
\end{align}
The vacuum energy density at this stationary point vanishes, once
$w$ takes the following value,
\begin{align}
w_0 = - \left[\sqrt{2} - \frac{\sqrt{3}}{2} \left(1+\epsilon\right)^{1/2}\right] \mu^2 M_{\rm Pl} \,.
\end{align}
We then obtain for the gravitino mass as well as for the inflaton mass
around this stationary point,
\begin{align}
m_{3/2} = \frac{1}{\sqrt{2} }\exp\left[\frac{1}{2\left(1+\epsilon\right)}
\frac{\left<\sigma\right>^2}{M_{\rm Pl}^2}\right]
\frac{\mu^2}{M_{\rm Pl}} \,, \quad
m_\varphi^2 = \left[4-2\sqrt{3} \left(\frac{2}{1+\epsilon}\right)^{3/2}\right] m_{3/2}^2 \,.
\label{eq:m32mphi}
\end{align}


None of the above quantities diverges in the limit of an exact shift
symmetry.
In contrast to the previous section, the limit $\epsilon\rightarrow0$,
therefore, does not take us out of the physical regime.
At the same time, independent of whether $\epsilon$ is exactly zero
or just small, $\epsilon \ll 1$, we find that the inflaton direction
in field space turns out to be tachyonic, $m_\varphi^2 < 0$.
The stationary point at $\varphi = 0$, therefore, does not
represent the global minimum of the scalar potential.
Instead, the scalar potential is either unbounded from below
(for $\epsilon = 0$) or it exhibits a global AdS vacuum
(for $\epsilon \neq 0$) at a large inflaton field value
$\varphi_{\rm min}\propto M_{\rm Pl}/\epsilon$.
It is obvious that the former case does not represent a viable
setting for inflation.
For this reason, we shall focus on the latter case
in the following and argue that in this case, too, inflation
cannot be successfully realized.
Here, our argument will closely follow the discussion in Sec.~\ref{subsec:SUGRA}.
First, let us consider the scalar potential for the inflaton field $\varphi$
with the sinflaton field $\sigma$ being fixed at $\sigma = \left<\sigma\right>$,
\begin{align}
V\left(\varphi\right) = \exp\left[\frac{\epsilon}{1+\epsilon}
\frac{\varphi^2}{M_{\rm Pl}^2}\right]
\left(\frac{1}{2}m_\varphi^2\varphi^2
+ \frac{\lambda_\varphi}{4!}\, \varphi^4\right) \,, \quad 
\lambda_\varphi = \frac{48\,\epsilon^2}{\left(1+\epsilon\right)^3}
\frac{m_{3/2}^2}{M_{\rm Pl}^2}
\end{align}
We assume that the shift symmetry the K\"ahler potential in Eq.~\eqref{eq:KshiIm} is only
slightly broken, $\epsilon \ll 6^{1/3}-1$, so that $m_\varphi^2 < 0$;
and we assume that the potential is not unbounded from below,
$\epsilon \neq 0$, so that $\lambda_\varphi >0$.
Under these assumptions, the inflaton potential exhibits a global AdS vacuum at
\begin{align}
\varphi_{\rm min} = \left(\frac{1+\epsilon}{\epsilon}\right)^{1/2}
\left[\left(1+a^2\right)^{1/2}-\left(1-a\right)\right]^{1/2} M_{\rm Pl} \,, \quad
a = -\frac{\epsilon}{1+\epsilon}\frac{6}{\lambda_\varphi}\frac{m_\varphi^2}{M_{\rm Pl}^2} \,.
\end{align}
At this field value, the scalar potential takes the following value,
\begin{align}
V_{\rm min} = - \exp\left[\left(1+a^2\right)^{1/2}-\left(1-a\right)\right]
\left[\left(1+a^2\right)^{1/2}-1\right] \frac{4}{1+\epsilon} \,
m_{3/2}^2\, M_{\rm Pl}^2 \,,
\end{align}
which represents the vacuum energy density in the global AdS minimum.
Similarly as in Sec.~\ref{subsec:SUGRA}, we may hope that there might be a chance to
lift this AdS vacuum by radiative corrections.
But as it turns out, this attempt to stabilize the inflaton potential
fails for the same reason as in Sec.~\ref{subsec:SUGRA}.
Within the parameter ranges of interest, $\left|\epsilon\right| \lesssim 10^{-0.5}$
and $\lambda /\eta \lesssim 10^{-0.5}$, we always find a hierarchy among the critical
field value $\varphi_c$ and the position of the AdS vacuum,
$\varphi_c / \varphi_{\rm min} \lesssim 0.1$. 
The quadratic radiative corrections close to the origin, therefore, fail to
stabilize the inflaton potential all the way up to $\varphi = \varphi_{\rm min}$.
In consequence of that, we always end up with a local minimum
in between $\varphi_c$ and $\varphi_{\rm min}$.
Inflation taking place in the logarithmic part of the
effective potential then always ``gets stuck'' and the
inflaton has no chance of reaching the Minkowski
vacuum at the origin (see also our discussion in Sec.~\ref{subsec:SUGRA}).


This completes our argument that inflation based on the Polonyi
superpotential $W = \mu^2\,\Phi + w$---with the constant $w$ being set to
its final value $w=w_0$ already from the very beginning---neither works for
a near-canonical nor for an approximately shift-symmetric
K\"ahler potential.
It is this conclusion that leads us to resort to studying inflation
in combination with late-time $R$ symmetry breaking in the main text.
And indeed, successful Polonyi inflation turns out to be feasible,
once $w$ vanishes during inflation.



\end{document}